\DeclareRobustCommand{\ionn}[2]{%
\relax\ifmmode
\ifx\testbx\f@series
{\mathbf{#1\,\mathsc{#2}}}\else
{\mathrm{#1\,\mathsc{#2}}}\fi
\else\textup{#1\,{\mdseries\textsc{#2}}}%
\fi}

\documentclass[onecolumn,12pt]{aastex61}

\usepackage{amsmath} 
\usepackage{xcolor}

\begin{document}

\title{White Paper on MAAT@GTC}


\author{Francisco Prada}
\affil{Instituto de Astrof\'isica de Andaluc\'ia (CSIC), Glorieta de la Astronom\'ia, E-18008, Granada, Spain}
\correspondingauthor{Francisco Prada}
\email{f.prada@csic.es}
\author{Robert Content}
\affil{Australian Astronomical Optics—Macquarie, Macquarie University, NSW 2109, Sydney, Australia}
\author{Ariel Goobar}
\affil{Oskar Klein Centre, Department of Physics, Stockholm University, SE 106 91 Stockholm, Sweden}
\author{Luca Izzo}
\affil{DARK, Niels Bohr Institute, Copenhagen University, Lyngbyvej 2, 2100, Copenhagen, Denmark}
\author{Enrique P\'erez}
\affil{Instituto de Astrof\'isica de Andaluc\'ia (CSIC), Glorieta de la Astronom\'ia, E-18008, Granada, Spain}


\author{Adriano Agnello}
\affil{DARK, Niels Bohr Institute, Copenhagen University, Lyngbyvej 2, 2100, Copenhagen, Denmark}
\author{Carlos del Burgo}
\affil{Instituto Nacional de Astrofísica, Óptica y Electrónica, Luis Enrique Erro 1, Sta. Ma. Tonantzintla, Puebla, Mexico}
\author{Vik Dhillon}
\affil{Department of Physics and Astronomy, University of Sheffield, Sheffield S3 7RH, UK}
\author{Jose M. Diego}
\affil{Instituto de F\'isica de Cantabria (CSIC-UC), Avda. Los Castros s/n, 39005, Santander, Spain}
\author{Llu\'is Galbany}
\affil{Departamento de F\'isica Te\'orica y del Cosmos, Universidad de Granada, E-18071, Granada, Spain}
\author{Jorge Garc\'ia-Rojas}
\affil{Instituto de Astrof\'isica de Canarias, E-38205 La Laguna, Spain}
\affil{Departamento de Astrof\'isica, Universidad de La Laguna, E-38206 La Laguna, Spain}
\author{David Jones}
\affil{Instituto de Astrof\'isica de Canarias, E-38205 La Laguna, Spain}
\affil{Departamento de Astrof\'isica, Universidad de La Laguna, E-38206 La Laguna, Spain}
\author{Jon Lawrence}
\affil{Australian Astronomical Optics—Macquarie, Macquarie University, NSW 2109, Sydney, Australia}
\author{Eduardo Mart\'in}
\affil{Instituto de Astrof\'isica de Canarias, E-38205 La Laguna, Spain}
\affil{Departamento de Astrof\'isica, Universidad de La Laguna, E-38206 La Laguna, Spain}
\affil{Consejo Superior de Investigaciones Cient\'ificas, Madrid, Spain}
\author{Evencio Mediavilla}
\affil{Instituto de Astrof\'isica de Canarias, E-38205 La Laguna, Spain}
\affil{Departamento de Astrof\'isica, Universidad de La Laguna, E-38206 La Laguna, Spain}
\author{M. \'Angeles P\'erez~Garc\'ia}
\affil{Departamento de F\'isica Fundamental, Universidad de Salamanca, E-37008, Salamanca, Spain}
\author{Jorge S\'anchez~Almeida}
\affil{Instituto de Astrof\'isica de Canarias, E-38205 La Laguna, Spain}
\affil{Departamento de Astrof\'isica, Universidad de La Laguna, E-38206 La Laguna, Spain}


\author{Jos\'e A. Acosta~Pulido}
\affil{Instituto de Astrof\'isica de Canarias, E-38205 La Laguna, Spain}
\affil{Departamento de Astrof\'isica, Universidad de La Laguna, E-38206 La Laguna, Spain}
\author{\'Angel R. L\'opez-S\'anchez}
\affil{Australian Astronomical Optics—Macquarie, Macquarie University, NSW 2109, Sydney, Australia}
\affil{Department of Physics and Astronomy, Macquarie University; Macquarie University Research Centre for Astronomy; ARC Centre of Excellence for All Sky Astrophysics in 3 Dimensions (ASTRO-3D), Australia}
\author{Santiago Arribas}
\affil{Centro de Astrobiolog\'ia, Carretera Ajalvir km 4, E-28850 Torrej\'on de Ardoz, Madrid, Spain}
\author{Francisco J. Carrera}
\affil{Instituto de F\'isica de Cantabria (CSIC-UC), Avda. Los Castros s/n, 39005, Santander, Spain}
\author{Amalia Corral}
\affil{Instituto de F\'isica de Cantabria (CSIC-UC), Avda. Los Castros s/n, 39005, Santander, Spain}
\author{Inmaculada Dom\'inguez}
\affil{Departamento de F\'isica Te\'orica y del Cosmos, Universidad de Granada, E-18071, Granada, Spain}
\author{Silvia Mateos}
\affil{Instituto de F\'isica de Cantabria (CSIC-UC), Avda. Los Castros s/n, 39005, Santander, Spain}
\author{Silvia Mart\'inez~Nu\~nez}
\affil{Instituto de F\'isica de Cantabria (CSIC-UC), Avda. Los Castros s/n, 39005, Santander, Spain}
\author{Eva Villaver}
\affil{Departamento de F\'isica Te\'orica, Universidad Aut\'onoma de Madrid, Spain}
\author{Mar\'ia Rosa Zapatero~Osorio}
\affil{Centro de Astrobiolog\'ia, Carretera Ajalvir km 4, E-28850 Torrej\'on de Ardoz, Madrid, Spain}


\author{Conrado Albertus}
\affil{Departamento de F\'isica Fundamental, Universidad de Salamanca, E-37008, Salamanca, Spain}
\author{Fabrizio Arrigoni Battaia}
\affil{Max Planck Institut f\"ur Astrophysik, Karl-Schwarzschild-Str. 1, D-85741 Garching bei München, Germany}
\author{David Barrado}
\affil{Centro de Astrobiolog\'ia, Carretera Ajalvir km 4, E-28850 Torrej\'on de Ardoz, Madrid, Spain}
\author{V\'ictor J.~S. B\'ejar}
\affil{Instituto de Astrof\'isica de Canarias, E-38205 La Laguna, Spain}
\affil{Departamento de Astrof\'isica, Universidad de La Laguna, E-38206 La Laguna, Spain}
\author{Henri M.~J. Boffin}
\affil{European Southern Observatory, Karl-Schwarzschild-str. 2, D-85748 Garching, Germany}
\author{Herv\'e Bouy}
\affil{Laboratoire d'Astrophysique de Bordeaux, Univ. Bordeaux, CNRS, B18N, Allée Geoffroy Saint-Hillaire, F-33615 Pessac, France}
\author{Adam Burgasser}
\affil{Center for Astrophysics and Space Science, University of California San Diego, La Jolla, CA 92093, USA}
\author{C\'esar Esteban}
\affil{Instituto de Astrof\'isica de Canarias, E-38205 La Laguna, Spain}
\affil{Departamento de Astrof\'isica, Universidad de La Laguna, E-38206 La Laguna, Spain}
\author{Nicolas Lodieu}
\affil{Instituto de Astrof\'isica de Canarias, E-38205 La Laguna, Spain}
\affil{Departamento de Astrof\'isica, Universidad de La Laguna, E-38206 La Laguna, Spain}
\author{Mar\'ia Morales~Calder\'on}
\affil{Centro de Astrobiolog\'ia, Carretera Ajalvir km 4, E-28850 Torrej\'on de Ardoz, Madrid, Spain}
\author{Antonio P\'erez~Garrido}
\affil{Dpto. F\'isica Aplicada, Universidad Polit\'ecnica de Cartagena, 30202, Cartagena, Murcia, Spain}
\author{Pablo Rodr\'iguez~Gil}
\affil{Instituto de Astrof\'isica de Canarias, E-38205 La Laguna, Spain}
\affil{Departamento de Astrof\'isica, Universidad de La Laguna, E-38206 La Laguna, Spain}
\author{Ana Sagués~Carracedo}
\affil{Oskar Klein Centre, Department of Physics, Stockholm University, SE 106 91 Stockholm, Sweden}
\author{Miguel Santander~Garc\'ia}
\affil{Observatorio Astron\'omico Nacional (OAN-IGN), Alfonso XII, 3, 28014, Madrid, Spain}
\author{Enrique Solano}
\affil{Centro de Astrobiolog\'ia, Carretera Ajalvir km 4, E-28850 Torrej\'on de Ardoz, Madrid, Spain}
\author{Manuel A.~P. Torres}
\affil{Instituto de Astrof\'isica de Canarias, E-38205 La Laguna, Spain}
\affil{Departamento de Astrof\'isica, Universidad de La Laguna, E-38206 La Laguna, Spain}
\author{Roger Wesson}
\affil{Department of Physics and Astronomy, University College London, Gower Street, London WC1E 6BT, UK}

\submitjournal{arXiv, a white paper for the astronomical community}

\vspace*{4cm}

\begin{abstract}
MAAT is proposed as a visitor mirror-slicer optical system that will allow the OSIRIS spectrograph on the 10.4-m Gran telescopio CANARIAS (GTC) the capability to perform Integral Field Spectroscopy (IFS) over a seeing-limited FoV $14.20''\times 10''$ with a slice width of $0.303''$. MAAT@GTC will enhance the resolution power of OSIRIS by 1.6 times as compared to its $0.6''$ wide long-slit. All the eleven OSIRIS grisms and volume-phase holographic gratings will be available to provide broad spectral coverage with moderate resolution (R=600 up to 4100) in the $3600 - 10000$ \AA\ wavelength range. MAAT unique observing capabilities will broaden its use to the needs of the GTC community to unveil the nature of most striking phenomena in the universe well beyond time-domain astronomy. The GTC equipped with OSIRIS+MAAT will also play a fundamental role in synergy with other facilities, some of them operating on the northern ORM at La Palma. This White Paper presents the different aspects of MAAT@GTC - including scientific and technical specifications, outstanding science cases, and an outline of the instrument concept.
\end{abstract}

\begingroup
\let\clearpage\relax
\tableofcontents
\endgroup

\newpage

\section{MAAT basic description}
\label{sec:maat}

\begin{figure}[htb]
\centering
\includegraphics[width=0.3\linewidth]{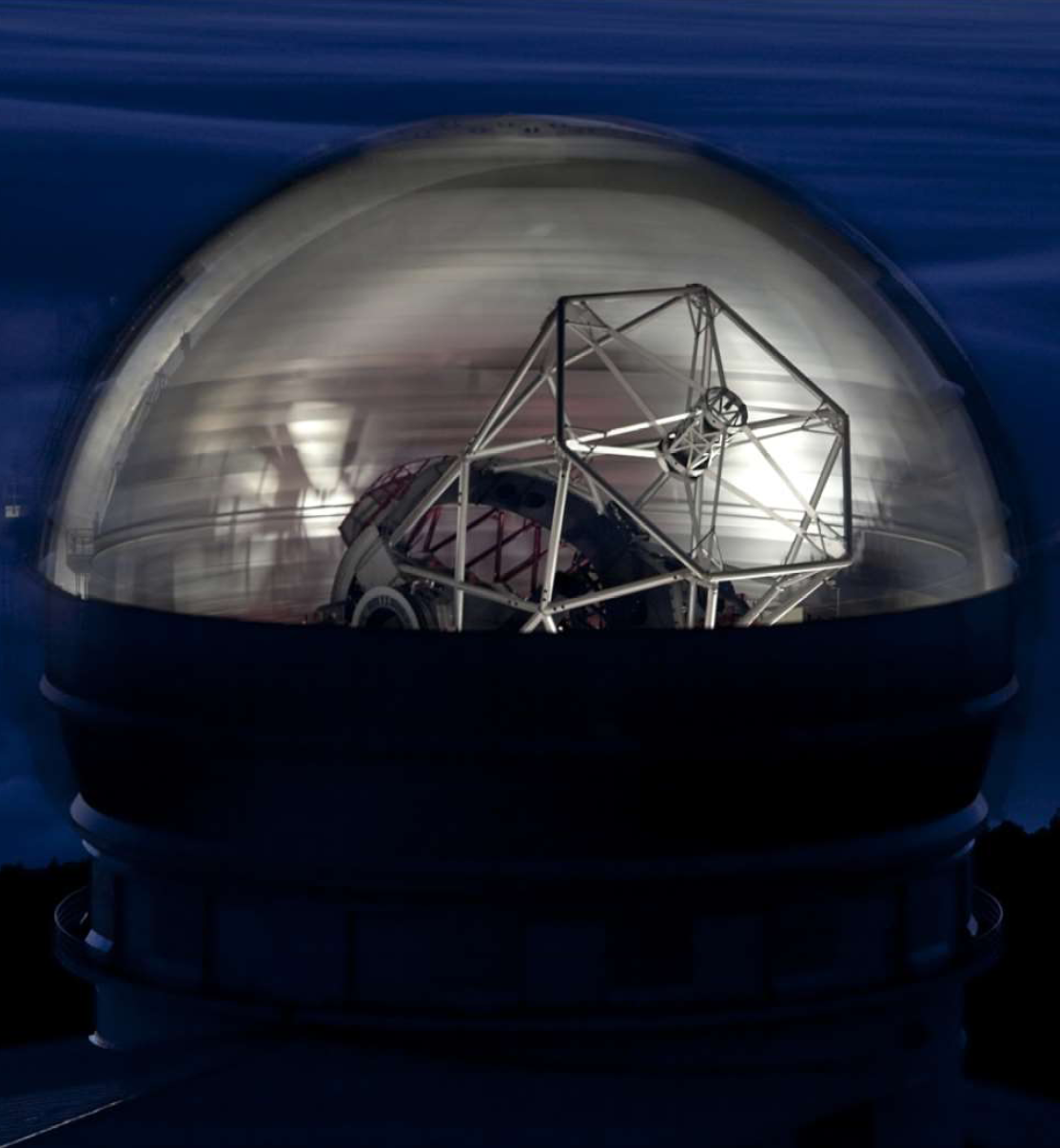}\quad\includegraphics[width=0.327\linewidth]{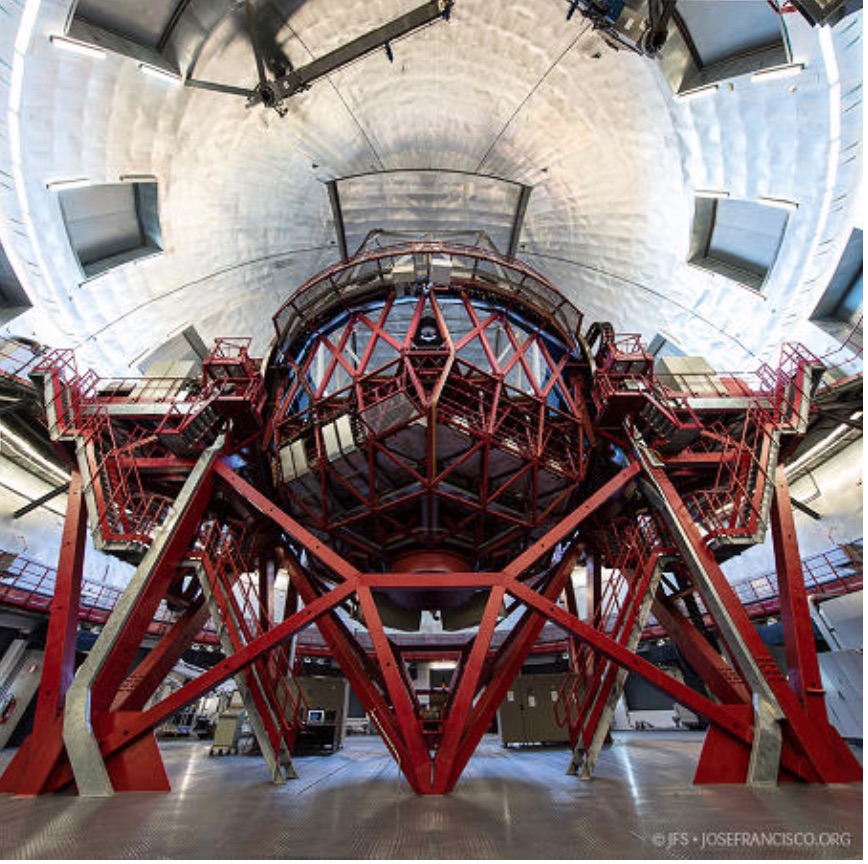}
\caption{Outside / inside view of the Gran Telescopio CANARIAS (GTC). Credits: ING and \textit{josefrancisco.org}.
}
\label{fig:gtcview}
\end{figure}

MAAT\footnote{MAAT refers to the ancient Egyptian concepts of truth, balance, order, harmony, law, morality, justice, and cosmic order.} (Mirror-slicer Array for Astronomical Transients) is proposed as a new mirror-slicer optical system that will add the OSIRIS\footnote{\href{http://www.gtc.iac.es/instruments/osiris/osiris.php}{http://www.gtc.iac.es/instruments/osiris/osiris.php}} spectrograph on the 10.4-m GTC telescope\footnote{\href{http://www.gtc.iac.es}{http://www.gtc.iac.es}} (see an outside / inside view of GTC in Figure~\ref{fig:gtcview}). The combination of MAAT and OSIRIS, the most demanded instrument on the GTC, will allow astronomers to perform integral-field spectroscopy (IFS) over a seeing-limited field-of-view $14.20'' \times 10.0''$, with an angular resolution of $0.303\arcsec \times 0.127\arcsec$. MAAT will enhance the resolution power of OSIRIS by 1.6 times as compared to its $0.6''$ wide long-slit. All the eleven OSIRIS grisms and volume-phase holographic (VPH) gratings will be available to provide broad spectral coverage with moderate resolution (600--4100) in the spectral range 360--1000 nm. The basic parameters of MAAT@GTC are listed in Table~\ref{tab:maatbasic}.

\begin{deluxetable}{rcc}[h!] 
\small
\tablecaption{The MAAT basic parameters\label{tab:maatbasic}}
\tablehead{
\colhead{Parameter} & \colhead{Value} & \colhead{Notes}
}
\startdata
Spectrograph & OSIRIS  &  Install at GTC Cassegrain focus \\
Module & Integral Field Unit &  \\
Field-of-View  & $14.20\arcsec \times 10.00\arcsec$ & IFU sky area is 142 arcsec$^2$ (141 arcsec$^2$ without vignetting)  \\
Field aspect ratio & 1.42 & The footprint can be rotated to match the target shape or multiple objects  \\
Slicer width & $0.303\arcsec$  &   \\
Spatial sampling   & $0.303\arcsec \times 0.127\arcsec$ & $0.303\arcsec \times 0.254\arcsec$ with $1\times2$ CCD binning  \\
Wavelength range & 360 to 1000 nm &   \\
Spectral resolution  & 600 to 4100 & Enhanced 1.6 times resolution power w.r.t. a $0.6\arcsec$ long-slit  \\
Detector  &  $4k \times 4k$ (15 $\mu$m pixel) &  New Teledyne-e2v CCD231-84 deep-depleted astro multi-2 \\
CCD plate scale &  $0.127\arcsec$ per pixel &  \\
\enddata
\end{deluxetable}

\begin{figure}[htb]
\centering
\includegraphics[width=0.7\linewidth]{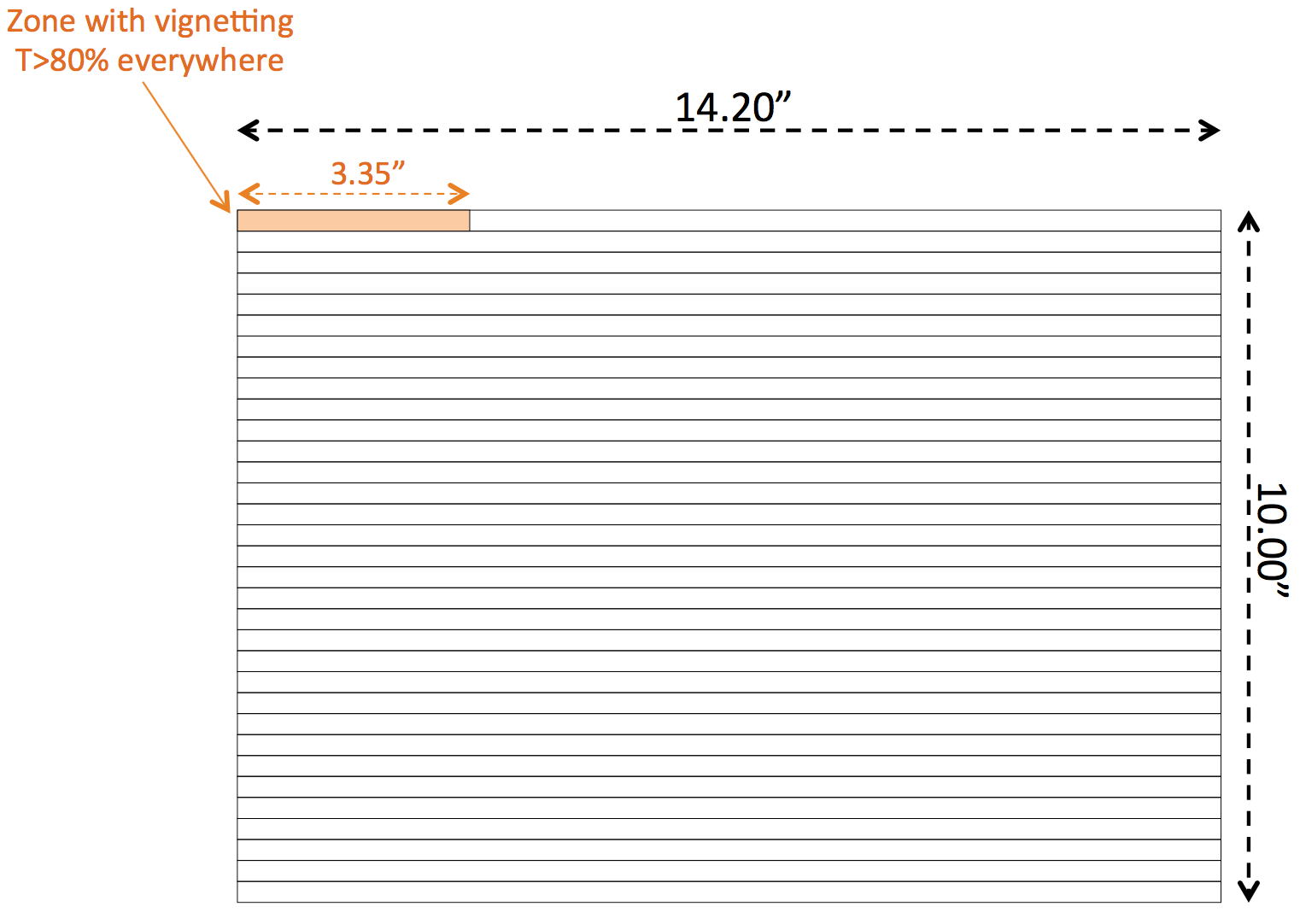}
\caption{Sky footprint of the MAAT mirror-slicer IFU.}
\label{fig:maatfootp}
\end{figure}

The integral-field unit (IFU) consists of an imaging slicer optical system with 33 slices (with 4 pixel on the CCD separattion between slices) each of $0.303'' \times 14.20''$. Figure~\ref{fig:maatfootp} shows the sky footprint of the MAAT mirror-slicer IFU. The IFS mode will take advantage of the expected significant increase in the overall OSIRIS efficiency due to its new e2v $4k \times 4k$ detector that will be installed after the relocation of OSIRIS at the Cassegrain focus of GTC. Table~\ref{tab:IFSs} compares MAAT with the other existing seeing-limited IFS on 10m-class telescopes located on both hemispheres. In terms of opportunity, of all these systems, MAAT provides the unique capability of a broad-band spectral coverage over the entire spectral range from the UV up to the near-IR (360--1000 nm).

\begin{deluxetable*}{rccccccc}[!] 
\small
\tablecaption{Seeing-limited Integral Field Spectrographs on 10m-class telescopes.\label{tab:IFSs}}
\tablehead{
\colhead{Sky} & \colhead{Telescope} & \colhead{Instrument} & \colhead{Spectral range} & \colhead{Resolution} & \colhead{Field of View} & \colhead{Spatial sampling} & \colhead{IFU}
}
\startdata
Southern & VLT & MUSE  & 480--930 nm & 1770--3590 & $59.9\arcsec \times 60.0\arcsec$ & $0.2\arcsec \times 0.2\arcsec$ & mirror slicer \\
Northern & Keck  & KCWI  & 350--560 nm & 3000--4000 & $8.25\arcsec \times 20.0\arcsec$ & $0.34\arcsec \times 0.147\arcsec$ & mirror slicer \\
N \& S & Gemini & GMOS-IFU  & 360--940 nm & 600--4400 & $5.0\arcsec \times 7.0\arcsec$ & $0.2\arcsec$ & lenslet/fibers \\
Northern & GTC & MAAT  & 360--1000 nm & 600--4100 & $10.0\arcsec \times 14.20\arcsec$ & $0.303\arcsec \times 0.127\arcsec$ & mirror slicer \\
\enddata
\end{deluxetable*}

\begin{figure}[htb]
\centering
\includegraphics[width=0.8\linewidth]{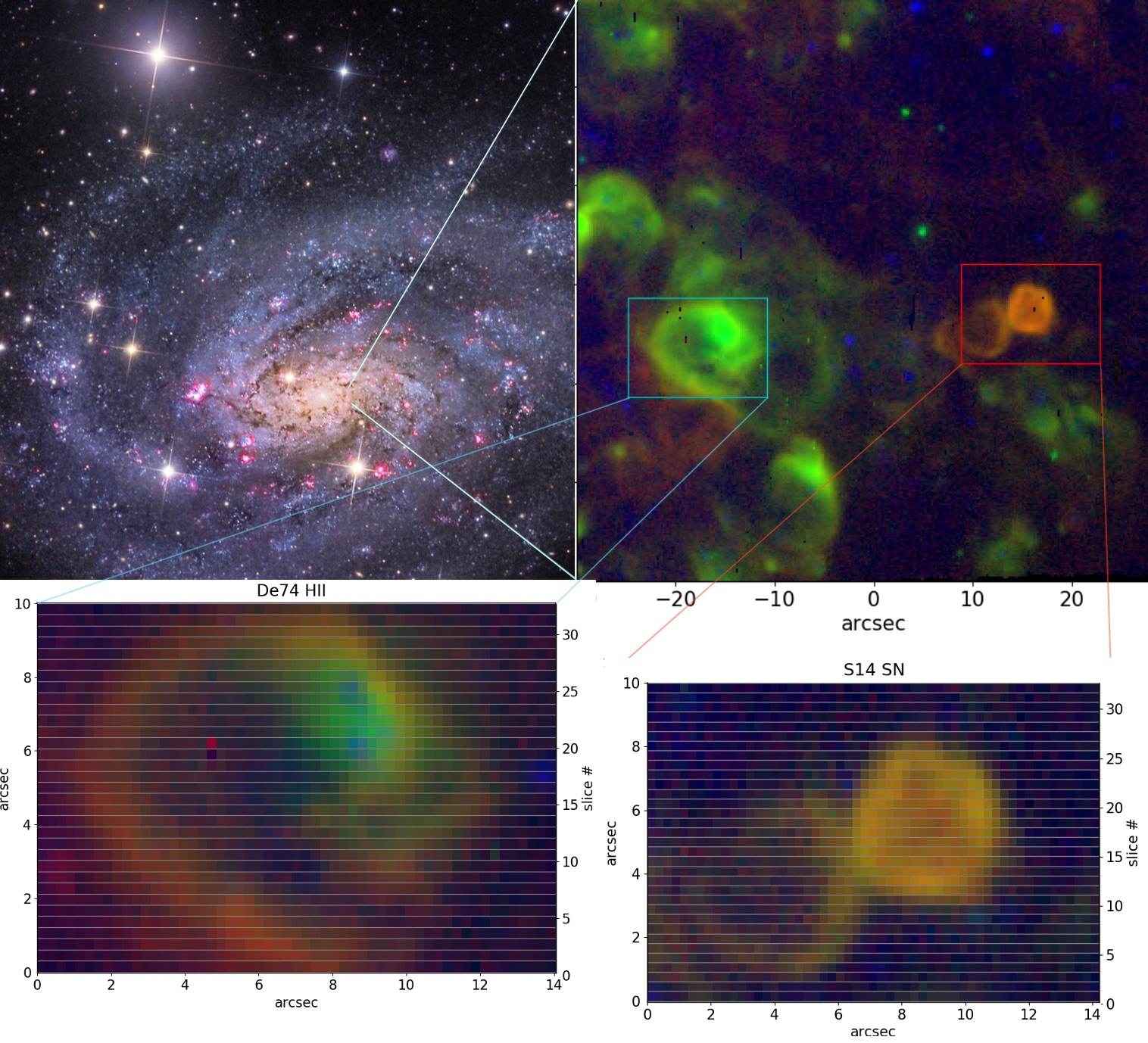}
\caption{MAAT simulation of two nebular objects in the spiral galaxy NGC~300.
The \ionn{H}{ii} region De74 (left) and the SN remnant S14 (right). 
The two bottom images is how the slicer sees each of the objects
that are then dispersed onto the CCD to obtain the frames as in Figure~\ref{NGC300_HIISNR}. The simulations are based on a MUSE \citep{Bacon1995} archival spectral cube of NGC~300.  
}
\label{NGC300MAAT}
\end{figure}

\begin{figure}[htb]
\centering
\includegraphics[width=0.75\linewidth]{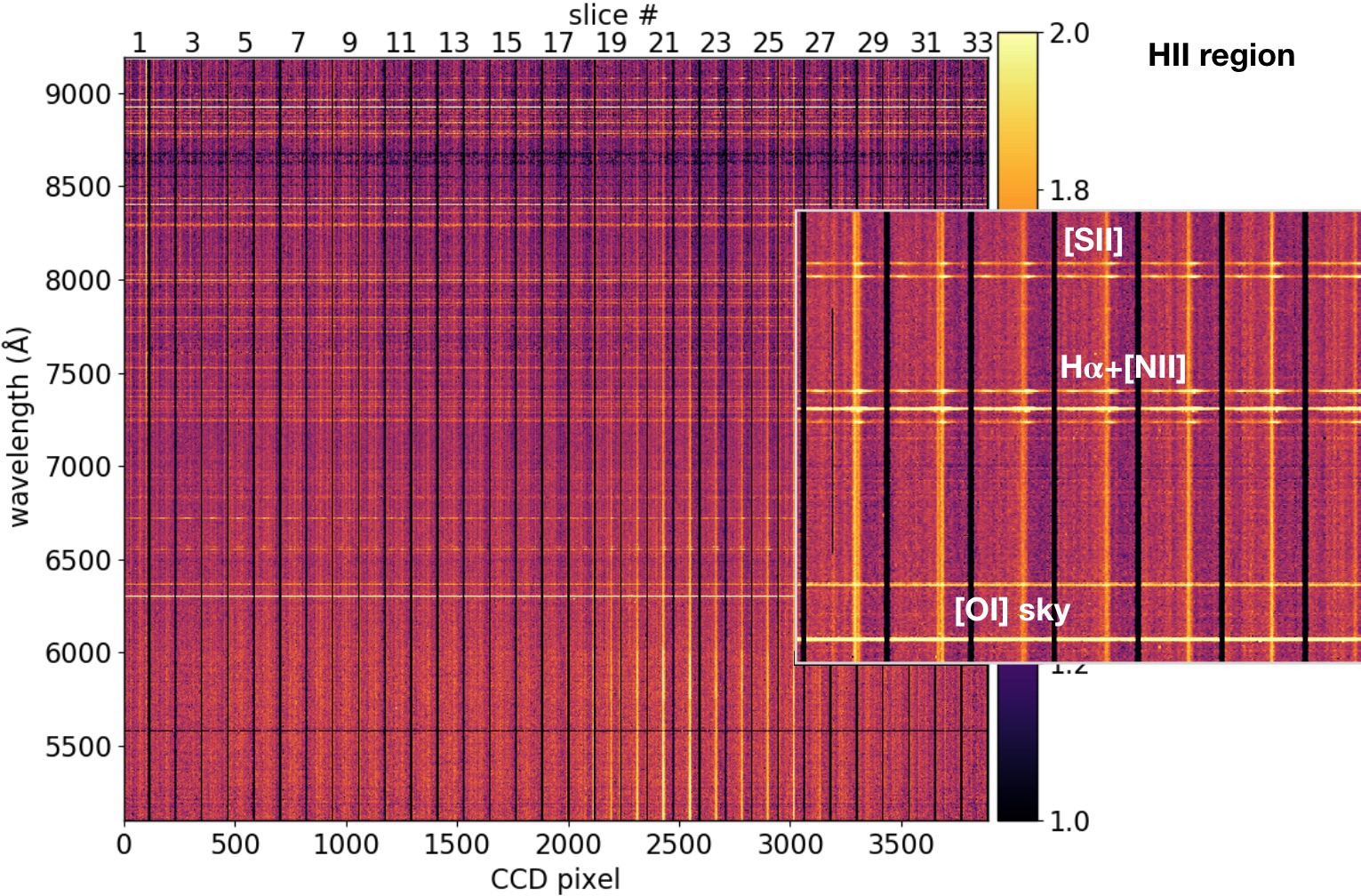}\quad\includegraphics[width=0.75\linewidth]{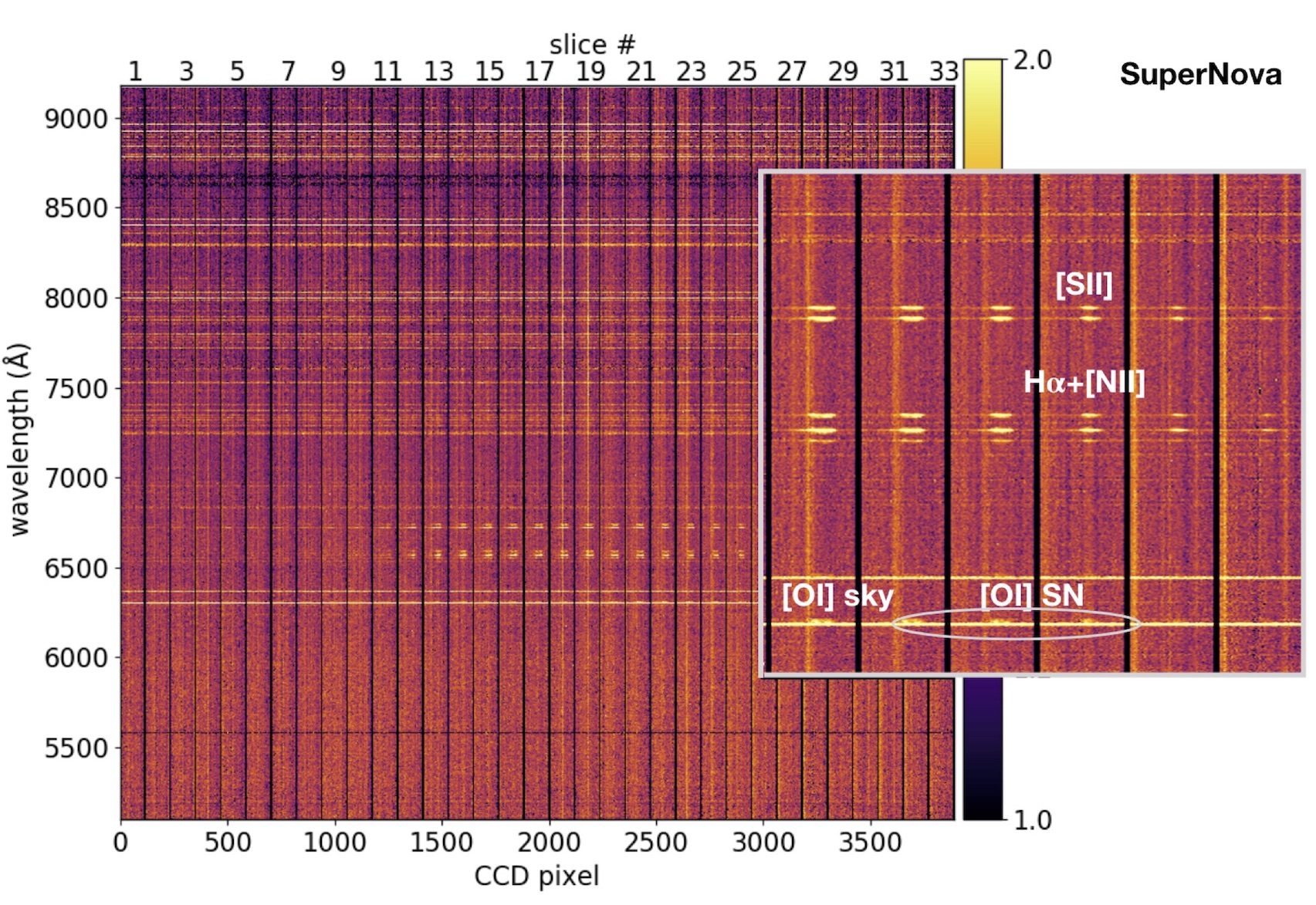}
\caption{Top: MAAT simulation of the giant \ionn{H}{ii} region De74 in the galaxy NGC~300.
The inset shows a zoom in around the [\ionn{O}{i}]6300 to [\ionn{S}{ii}]6731 spectral region.
Notice that the [OI]$\lambda\lambda6300,6364$ are Earth atmospheric emission lines,
while the other emission lines in the inset are nebular lines in De74.
Bottom: MAAT simulation of the SN remnant S14 in the galaxy NGC~300.
The inset shows a zoom in around the [\ionn{O}{i}]6300 to [\ionn{S}{ii}]6731 spectral region.
Notice that the strong [\ionn{O}{i}]$\lambda\lambda6300,6364$ are Earth atmospheric emission lines,
while the other emission lines in the inset are nebular lines in S14. 
An ellipse around [\ionn{O}{i}]6300 marks how a faint [\ionn{O}{i}]6300 emission is detected in the 
SNR just redshifted with respect to the bright sky lines. The simulations are based on a MUSE archival spectral cube of NGC~300.
}
\label{NGC300_HIISNR}
\end{figure}

Figure~\ref{NGC300MAAT} showcases the simulated MAAT view of the galaxy NGC~300 (see Section~\ref{sec:maatdata} for details). The top-right image zooms in on the circumnuclear region of NGC~300 (top-left)
around the giant \ionn{H}{ii} region De74 on the left (green as dominated by the [\ionn{O}{iii}] and H$\alpha$ emission lines) and the SN remnant S14 on the right (orange as dominated by the shock excited [\ionn{S}{ii}] emission lines). The two images below correspond to the giant \ionn{H}{ii} region De74 and the SN remnant S14 as seen by the slicer. The detector spectral images (Figure~\ref{NGC300_HIISNR}) show the wealth of emission lines typical of these type of objects; an expanded view around the [\ionn{O}{i}]6300 to [\ionn{S}{ii}]6731 spectral region includes the bright [OI] Earth atmospheric emission and it clearly shows the presence of [OI]6300 in the SN remnant but it is conspicuously absent in the \ionn{H}{ii} region.

MAAT is devised as an IFS-mode for OSIRIS devoted to unveiling the nature of most striking phenomena in the universe, see Section~\ref{sec:science}. Furthermore, MAAT top-level requirements will broaden its use to the needs of the GTC community for a wide range of competitive science topics that covers the entire astronomy given its unique observing capabilities.


\section{Why MAAT on the GTC?}
\label{sec:why}

Long-slit observations of a point source are affected by the atmosphere in at least two ways: the slit size should be adapted to match the seeing (with the subsequent undesired changes in spectral resolution), and the chromatic atmospheric dispersion spatially shifts and broadens the images depending on the wavelength, see below. Both drawbacks affect to the quality of the observed spectra and often lead to large observing overheads. Integral Field Spectroscopy records simultaneously complete spectral and spatial information of the source, a data cube, which can be corrected from chromatic atmospheric refraction. Consequently, basically no object acquisition is needed to obtain high quality spectroscopic data. Light losses and spectral bias are related to the ``slit effect", which is caused when the slit is illuminated asymmetrically \citep{Bacon1995}. In addition, from the data cube it is possible to obtain images at a constant wavelength. This allows one to perform spectro-astrometry, a technique to study the structure and kinematics of an astronomical source on scales much smaller than the diffraction limit of the telescope.
 
Most of the advantages of the IFS technique are direct consequences of the simultaneity when recording spatial and spectral information. The simultaneity not only implies a more efficient way of observing but, more importantly, it guarantees a great homogeneity in the data \citep{delburgo2000}. In addition, IFS has other advantages. With IFS systems there is no need for an accurate centering of the object in the slit or to adapt the slit width (spectral resolution) to the seeing conditions; neither are the data affected by 
``slit effects" when determining radial velocities \citep{Vanderriest1995}. Using IFS it is possible to determine and to correct in the spectra the effects of chromatic atmospheric refraction, using an a posteriori procedure. This is obviously important to preserve the spectro-photometric properties of the spectra without the need of an atmospheric dispersion corrector (ADC). Note that for long-slit observations the presence of chromatic atmospheric refraction imposes strong restrictions, which cannot be corrected by any means. Thus, the slit must be orientated along a direction defined by the parallactic angle (i.e. the spatial direction is predefined) if it is required to preserve the relative fluxes in the spectrum. If the slit is orientated at a desired position angle (not coincident with the one along which the differential atmospheric refraction takes place) special filters for the acquisition / guiding systems should be used to optimize the detection in a particular wavelength range, other spectral ranges being affected by light losses. To have a complete feel for these difficulties, we should note that the effects of differential atmospheric refraction are dynamic, in the sense that they change with time. Moreover, if trying to get a 2D map using the long-slit technique the resulting dataset suffers from inhomogeneity, problems in the spatial sampling, and observing time inefficiency. All these problems are strong drawbacks to obtain high quality spectra, in particular when monitoring of one object at different epochs with different atmospheric conditions is required. In summary, we can say that IFS can be the first choice technique to do spectroscopic monitoring of even point sources.

\begin{figure}[htb]
\centering
\includegraphics[width=0.5\linewidth]{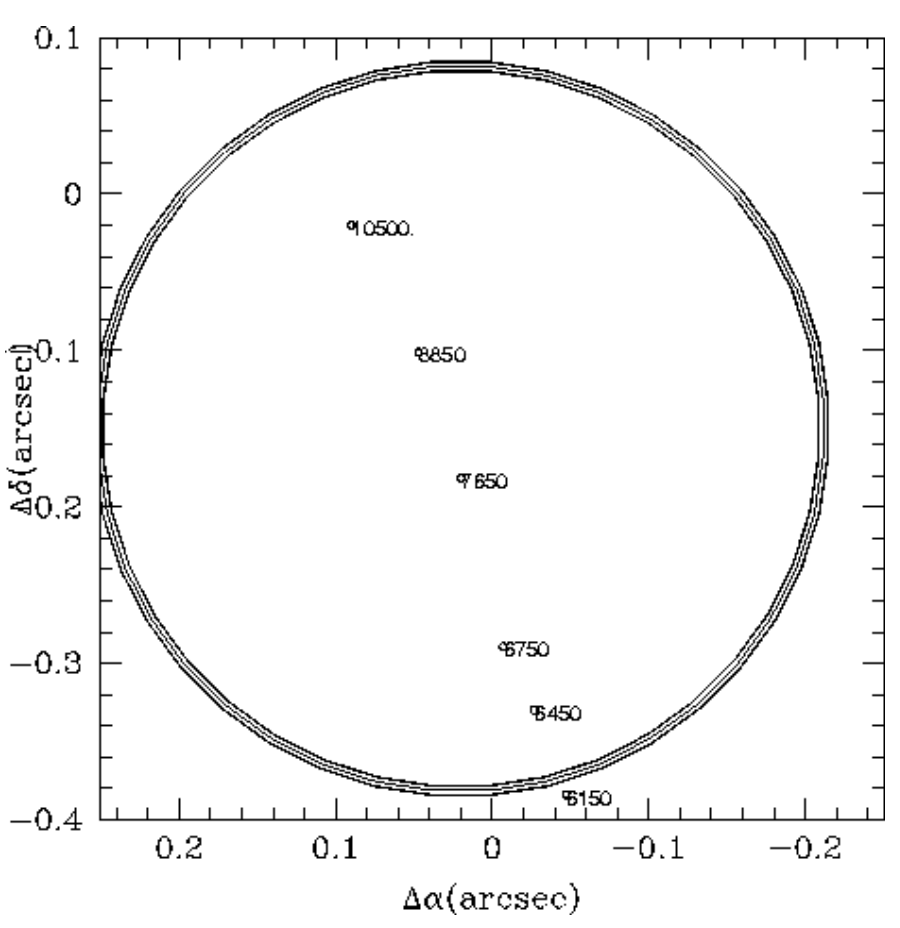}
\caption{Relative location of the photo-centers at different wavelength channels (indicated in \AA). The circle represents a fibre core in adequate scale. Notice the nice alignment of the photo-centers along parallactic angle.}
\label{fig:fiber}
\end{figure} 

The idea of measuring the wavelength dependence of the center of an astronomical source to study its structure and kinematics at scales much smaller than the diffraction limit \citep[spectro-astrometry,][]{Bailey1998} was initially developed from different experimental grounds (speckle interferometry, broad-band photometry, long-slit spectroscopy) and was based in specific and/or iterative observational strategies. This complexity is overtaken by Integral Field Spectroscopy that offers a simple way to obtain the spatial distribution of a given spectral feature \citep[see][for a long-slit vs. IFS comparison]{Gnerucci2010, Gnerucci2011, Gnerucci2011a}. 

An example of what can be obtained (without any specific observational strategy or methodological analysis) directly from IFS can be seen in Figure~\ref{fig:fiber} \citep{Arribas1999} where the photo-centers of a point-like source at different wavelengths trace chromatic atmospheric refraction with an accuracy better than $\sim5$ m.a.s., well below the spaxel size (a fiber face of 0.45 arcsec) and below the diffraction limit of the telescope (approx. 40 m.a.s. for the 4.2-m WHT telescope). Spectro-astrometry has been used or suggested  to study binaries \citep{Baines2004,Porter2005}, disks around stars \citep{Takami2003,Whelan2005}, supermassive black-holes \citep{Gnerucci2010, Gnerucci2011, Gnerucci2011a}, jets in AGNs, mergers and super massive black-hole binaries \citep{Unwin2008}, reverberation mapping \citep{Shen2012}, extrasolar planets, etc. We do not need, now, to focus on an specific scientific target, but it is convenient to give an estimate of the astrometric accuracy that can be reached with GTC. According to \citet{Lindegren1978,Lindegren2005}, assuming that the telescope diameter is 10.4 m and that we received $10^5$ photons from the source, we will have an astrometric accuracy in the determination of the photo-center at 6500 \AA\ of 13 microarcsecs (not far from the expected 8.5 microarcsec with the E-ELT at 1.6 $\mu m$).

All these advantages mentioned above among others are offered by MAAT@GTC, which  will provide to the GTC community with highly competitive unique observing capabilities (complementary to the existing and upcoming instrumentation), i.e.,

\begin{enumerate}
	\item Seeing-limited and wide-band IFS at low / moderate spectral resolution,
	\item All photons are collected, and a larger efficiency is obtained,
	\item MAAT can perform spectro-photometry and spectro-astrometry,
	\item Advantage on bad (any) seeing conditions. MAAT keeps its nominal spectral resolution regardless the seeing\footnote{This offers a significant advantage to accommodate the programs in queue observing according to their seeing requirements.},
	\item Target acquisition with no overheads. An entire FoV image could be generated from the 3D data cube.
\end{enumerate}

These unique capabilities are fundamental in order to develop the outstanding science objectives presented below. The scientific reach of MAAT@GTC is proving to be very substantial with a major impact on Astronomy and Cosmology.

\section{Science objectives}
\label{sec:science}

While the science potential of MAAT@GTC is essentially unlimited, this white paper will highlight the focus on a selected set of outstanding science topics enabled by the proposed instrument in different areas of expertise. These include:

\begin{itemize}
\item The nature of the diffuse universe: the intergalactic and circumgalactic mediums,
\item Strong galaxy lensing studies,
\item Time-domain cosmography with strongly lensed quasars and supernovae,
\item Identification and characterisation of EM-GW counterparts,
\item Exploration of the host galaxy environment of supernovae,
\item Binary masses and nebulae abundances,
\item Brown dwarfs and planetary mass objects,
\item Synergies with worldwide telescopes, and other facilities on La Palma;
\end{itemize}

Our view of the universe has changed dramatically over the last two decades: we have come to realize that $95\%$ of the cosmic composition is made of dark matter and dark energy instead of baryonic matter that most science had focused on for centuries.  This drastic development was originated from novel astronomical observations facilitated mainly by technological advances. One of the key pieces in the paradigm shift in cosmology was the detection of the accelerated expansion of the universe \citep{Riess1999,Perlmutter1999}, attributed to an exotic component named 
``dark energy". This evidence was first observed through transient astrophysical phenomena, Type Ia supernovae (SN), used as distance indicators in cosmology \citep{Phillips1993}.

As we look forward, we can identify exciting new avenues where breakthroughs can be expected, many of them also involving astrophysical transients. To begin with, we have just witnessed the dawn of the era of multi-messenger astronomy. Mergers of binary compact objects, generating gravitational wave (GW) signals along with electromagnetic waves (and possibly neutrinos), allow us to probe the densest states of matter and serve as laboratories for gravity at its most extreme conditions. At optical wavelengths, the resulting phenomena, dubbed \textit{kilonovae}, hold great promise for scientific explorations ranging from the origin of heavy elements through r-process reactions \citep{Smartt2017,Watson2019}, to the most accurate studies of the expansion of the universe \citep{Abbott2017,Hjorth2017,Dhawan2019}.

Gravitational lensing offers yet another way to study the power of gravity and the properties of curved space-time, thereby tracing the nature and distribution of dark matter and dark energy in an independent way. Gravitational lensing of transients, most notably quasars and supernovae, is emerging as a new precision tool in astronomy: in addition to the spatial information, these time-structured light beacons allow us to measure time-delays between light rays deflected by bodies in the line-of-sight \citep{Refsdal1964}. These deflectors come in many different shapes and mass-scales: black holes, stars, galaxies, and galaxy clusters. Gravitational lensing offers unique ways to weigh these structures, along with the measurement of global cosmological parameters, most notably the Hubble constant $H_0$ \citep{Treu2010}. 

\begin{figure}[htb]
\centering
\includegraphics[width=0.9\linewidth]{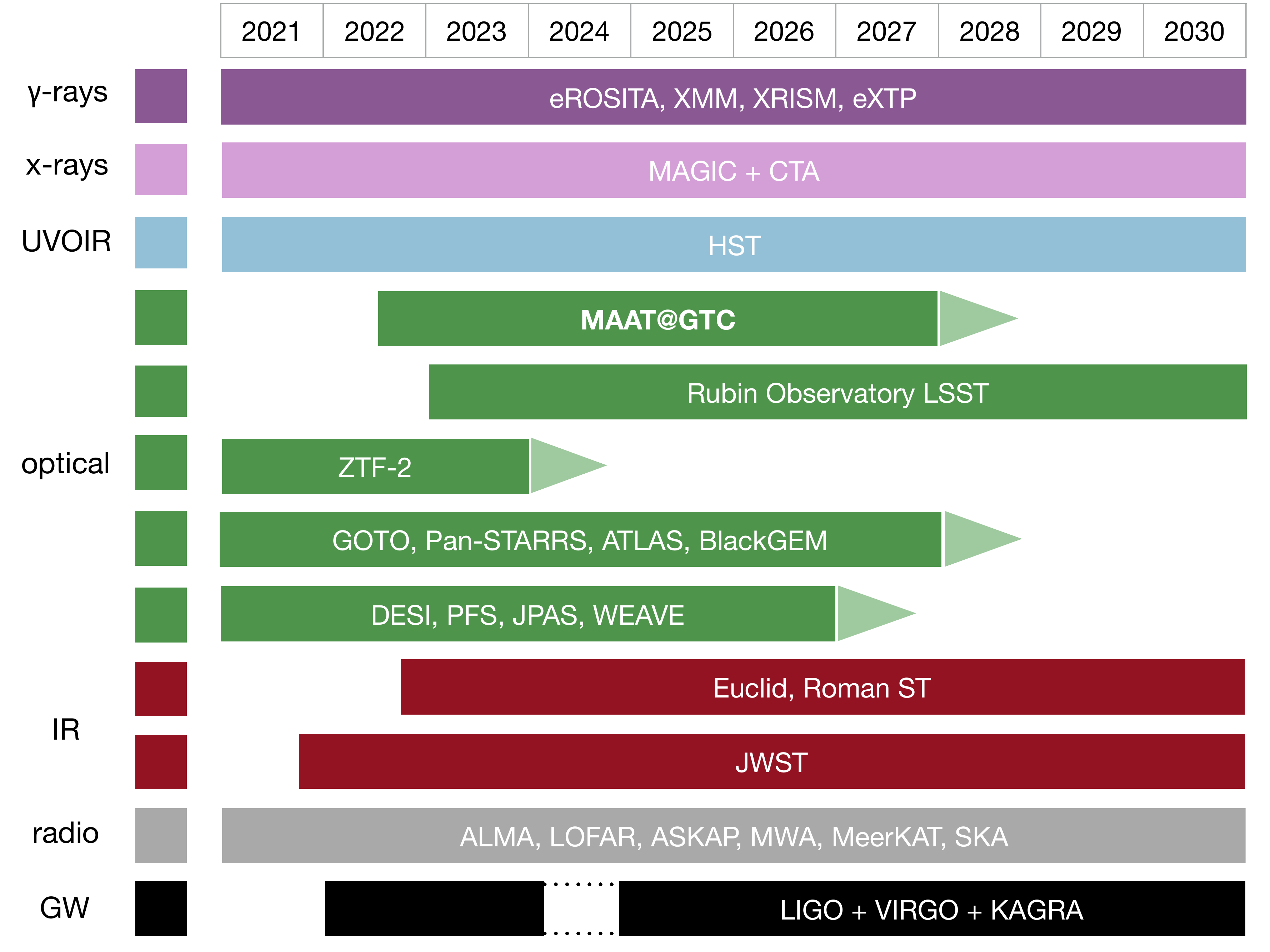}
\caption{Synergy chart of MAAT@GTC in the timeline context of upcoming major facilities. MAGIC, CTA, WEAVE, and GOTO are located at the ORM.}
\label{fig:synergies}
\end{figure} 

These intrinsically very rare phenomena can now be detected by large scale imaging surveys operating at optical wavelengths scanning the heavens with unprecedented speed and efficiency. Projects like ZTF \citep{2019PASP..131g8001G}, soon to become ZTF-2\footnote{\href{https://www.ztf.caltech.edu}{https://www.ztf.caltech.edu}}, Pan-STARRS\footnote{\href{https://www.ifa.hawaii.edu/research/Pan-STARRS.shtml}{https://www.ifa.hawaii.edu/research/Pan-STARRS.shtml}}, GOTO\footnote{\href{https://goto-observatory.org}{https://goto-observatory.org}}, ATLAS\footnote{\href{https://fallingstar.com/home.php}{https://fallingstar.com/home.php}}, BlackGEM\footnote{\href{https://astro.ru.nl/blackgem/}{https://astro.ru.nl/blackgem/}} and soon LSST\footnote{\href{https://www.lsst.org}{https://www.lsst.org}}, 
uncover the variable sky in ways that have not been possible until now. It is in this 
context, that the proposed instrument becomes the critical missing element. While imaging surveys are essential for the discovery of rare transients, timely identification of the nature and evolution of transients and their host galaxy environments requires spectroscopic screening in a telescope with great light collection power. Thus, the proposed IFU for OSIRIS on the 10.4-m Gran Telescopio CANARIAS, MAAT, presents us with unique opportunities to complete the time-domain revolution in astronomy.

The GTC equipped with OSIRIS+MAAT will play a fundamental role in synergy with other facilities operating in La Palma, opening a new era for transient studies at the Observatory of the Roque de Los Muchachos (see Figure~\ref{fig:synergies}). Furthermore, MAAT top-level requirements allow to broaden its use to the needs of the GTC community for a wide range of competitive science topics given its unique observing capabilities well beyond time-domain astronomy.

\subsection{Characterizing the CGM and IGM of galaxies with MAAT: two practical cases at two extreme redshifts}

\begin{figure}[htb]
\centering
\includegraphics[width=0.9\linewidth]{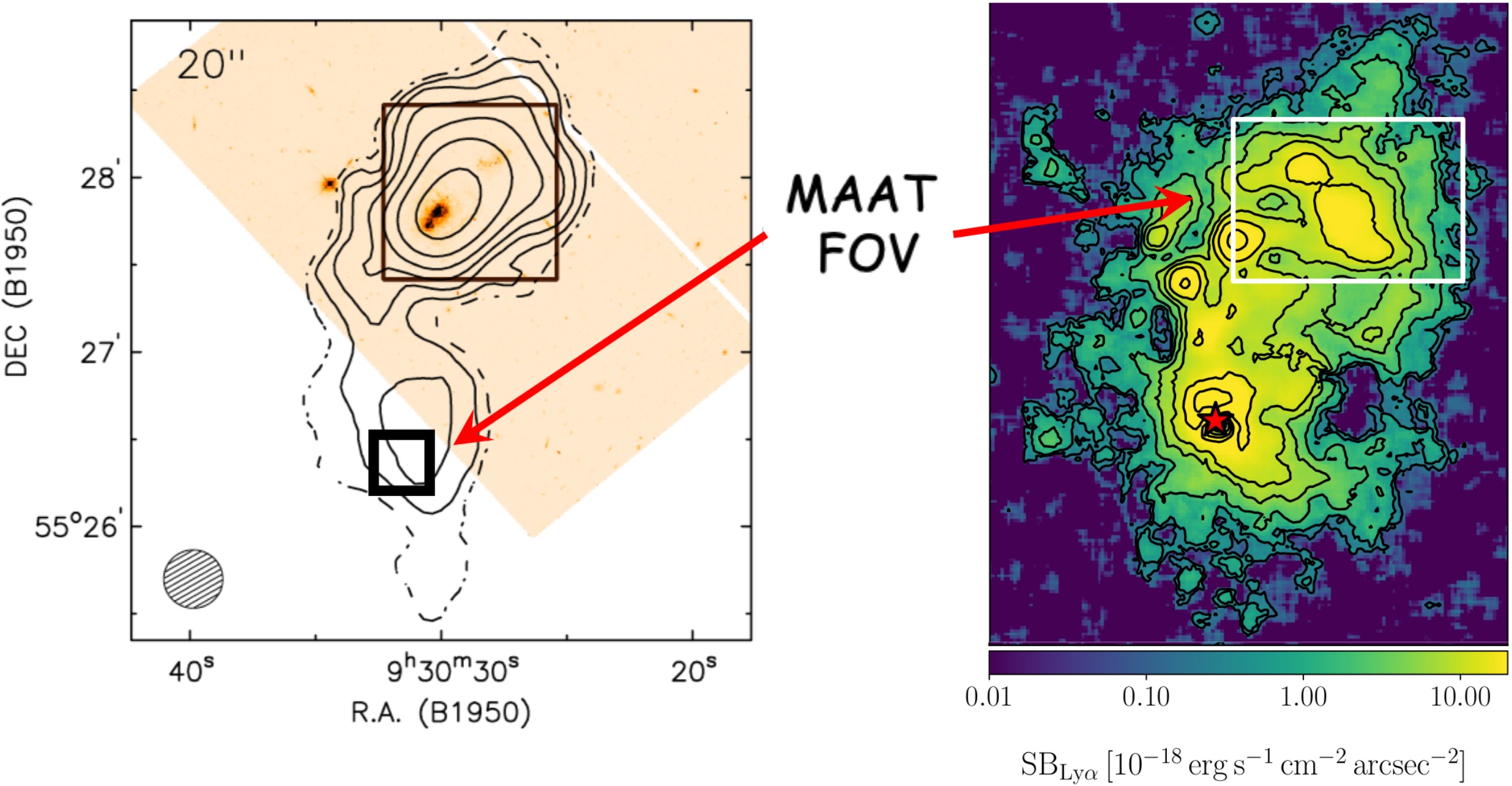}
\caption{\emph{Left:} \ionn{H}{i} emission around IZw18 \cite[the contours, from][]{2012A&A...537A..72L}. MAAT@GTC will allow us to discover whether the observed \ionn{H}{i} plume is pristine gas in process of being accreted onto the galaxy (the background image).
\emph{Right:} Ly$\alpha$ emission around the $z\sim3$ radio-quiet quasar SDSSJ1020+1040 (red star) \cite[figure adapted from][]{2018MNRAS.473.3907A}. MAAT@GTC will allow us to distinguish between inflows and outflows.}
\label{fig:fig1}
\end{figure}

Numerical simulations predict that gas accretion from the cosmic web drives star formation in disk galaxies \cite[e.g.,][]{2014A&ARv..22...71S}. The process is particularly important in the early universe, but it also occurs in isolated dwarf galaxies of the local universe \citep[e.g.,][]{angel2012}. The cosmic gas is expected to coexist with gas from galaxy-wide winds driven by feedback from  star-formation and AGNs. Although this cosmic gas accretion is a central ingredient of the current theory of galaxy formation it has not been observationally confirmed yet, and MAAT@GTC may play a fundamental role in this confirmation. MAAT FoV (14.2\arcsec$\times$10\arcsec) is insufficient to carry out blind searches for  diffuse gas in the
CGM (Circum-Galactic Medium) and IGM (Inter-Galactic Medium) of galaxies, however, its FoV, spectral coverage, and spectral resolution allow chemo-dynamical studies to distinguish metal-poor inflows from metal-rich outflows. Examples of two studies at extreme redshifts are: (Case 1) is the \ionn{H}{i} plume of IZw18 pristine gas in process of being accreted  \cite[Fig.~\ref{fig:fig1}, left;][]{2012A&A...537A..72L} (Case 2) What is happening in and around the $z\sim3$ enormous Ly$\alpha$ nebula SDSSJ1020+1040 \cite[Fig.~\ref{fig:fig1}, right;][]{2018MNRAS.473.3907A}. The shape of the Ly$\alpha$ line allows to distinguish between inflows and outflows.

\subsection{Improving strong lensing models with MAAT}

\begin{figure}[htb]
\centering
\includegraphics[width=0.9\linewidth]{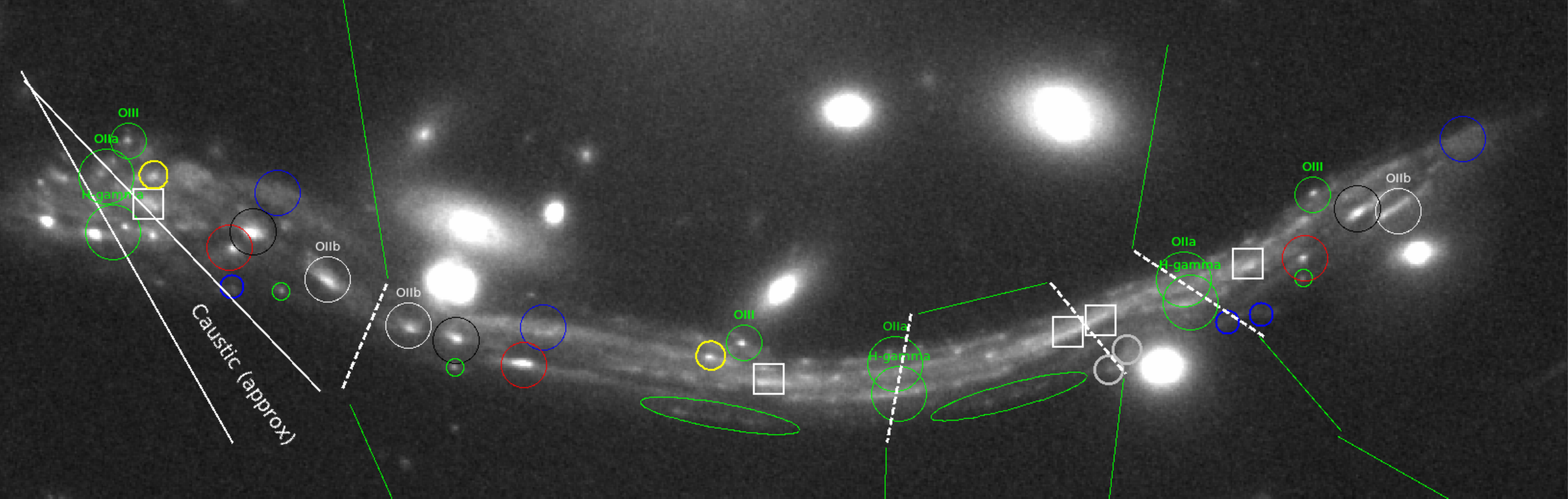}
\caption{Giant strongly lensed arc in the galaxy cluster A370 as seen by HST. 
    The arc corresponds to a foreground galaxy that is being multiply lensed 5 times. 
    Identification of multiply lensed features is possible with IFS data from MUSE, that can resolve individual compact regions in the foreground galaxy through their emission lines. The feature identification from IFS data can be used to build a model of the critical curves (dashed lines), that can later be confronted with actual lensing models. 
}
\label{Fig_A370}
\end{figure}

Strong gravitational lensing (SGL) allows to directly map the distribution of mass in massive clusters, where background galaxies can be observed as multiple images. Each one of these images carries information about changes in the potential along the line of sight. In most cases, tehse changes in the potential take place in a relatively small volume around the lens. Detailed maps of the lensing potential can reveal substructures in the dark matter, that can later be linked with models of dark matter, such as the standard cold dark matter scenario, or more alternative models such as self-interacting dark matter, axion-like particle dark matter, or even primordial black holes. To reach this level of detail, robust confirmation of the lensed systems, accurate measurements of the redshifts of the background galaxies, member identification of galaxies in the cluster, and if possible, independent mass estimates of the mass based on velocity dispersion, are needed.    
IFS has proven extremely valuable in the last years in confirming lensed systems in crowded fields, such as massive galaxy clusters. The Hubble Frontier Fields program (HFF) is the state of the art in galaxy cluster lensing. As part of the HFF program, six clusters were observed to unprecedented detail. These observations unveiled hundreds of background lensed galaxies. Spectroscopic confirmation of most of the lensed galaxies and their counter-images in these clusters has been possible mostly thanks to instruments equipped with IFS capabilities (such as MUSE). 
IFS is being used also to identify features, or knots, in giant arcs that are multiply lensed, as shown in Fig.~\ref{Fig_A370}. 
Very often, matching pairs of images that are multiply lensed within the same giant arc is not trivial given the rich substructure commonly present in these arcs. IFS can provide the fingerprint signature of each one of these signatures, facilitating their matching. The identification of these knots is of great interest to pin point the position of the critical curves, or to compute flux ratios, which can later be used to improve on the lensing models, or to identify substructures down to scales as small as thousands of solar masses. 
Also, IFS can be used to estimate the mass of individual member galaxies in the cluster. This is particularly important for the central Bright Cluster Galaxy (BCG), since lensing constraints are typically scarce near the BCG, and velocity dispersion measurements from IFS data can provide a much needed anchor point for the central mass in the lens model. 

IFS observations are also useful to identify regions in the lensed galaxies near the critical curves with very active star formation rates. These active regions are expected to harbor very bright, but short lived, stars, that can be observed after being magnified by factors of hundreds to thousands. These regions are ideal targets to search for transients produced by bright stars crossing the caustic of the cluster, or more likely micro-caustics produced by intervening microlenses in the cluster. These microlenses can be stars in the intracluster medium, but also dark matter candidates like primordial black holes, or small scale fluctuations in the dark matter distribution (that corrugate the critical curves) produced by interference in wave dark matter models. Such observations are already a reality after the discovery of the first such event \citep{Kelly2018,Diego2018}.

\subsection{Time-delay Cosmography}

\subsubsection{Strongly gravitational lensed quasars}

The time-delay distance inferred from strongly gravitational lensed quasars can be used as a powerful estimator of the Hubble constant and the main cosmological parameters. Quasars are very luminous astrophysical sources, so they can be observed from large distances. This makes them not only fascinating objects of study, but also useful as markers for studying Hubble-Lema\^itre’s Law. Indeed, light emitted from quasars fluctuates; when the follow-up of this variable luminosity is observed through multiple lensed images in gravitational lensed systems it can provide a direct measurement of distance \citep{Refsdal1964}, which is independent of local calibrators generally used in the cosmological distance scale ladder.  This method is based on the measurement of the time-delay between two or more images (typically spanning 20d-120d, see e.g. \citet{Millon2020}), which in turn are directly related to the distances to the deflectors and the sources, and on the a priory knowledge of the intervening lens mass. It is so powerful that just three lenses are needed to determine $H_0$ with a precision of $3.8\%$ \citep{Suyu2017,Bonvin2017,Shajib2019} in a way that is completely independent from any distance-ladder anchoring (e.g. Cepheids or TRGB). Increasing the number of lensed quasars, for which we can infer the value of the Hubble constant $H_0$, will then improve the accuracy on $H_0$ and this indicates the kind of contribution that MAAT@GTC will provide to fundamental cosmology in mediating the current tension between the Planck and local distance ladder estimates of $H_0$, even allow a first peek into possible new physics beyond the concordance paradigm \citep{Arendse2019}.

\begin{figure}[htb]
\centering
\includegraphics[width=0.8\linewidth]{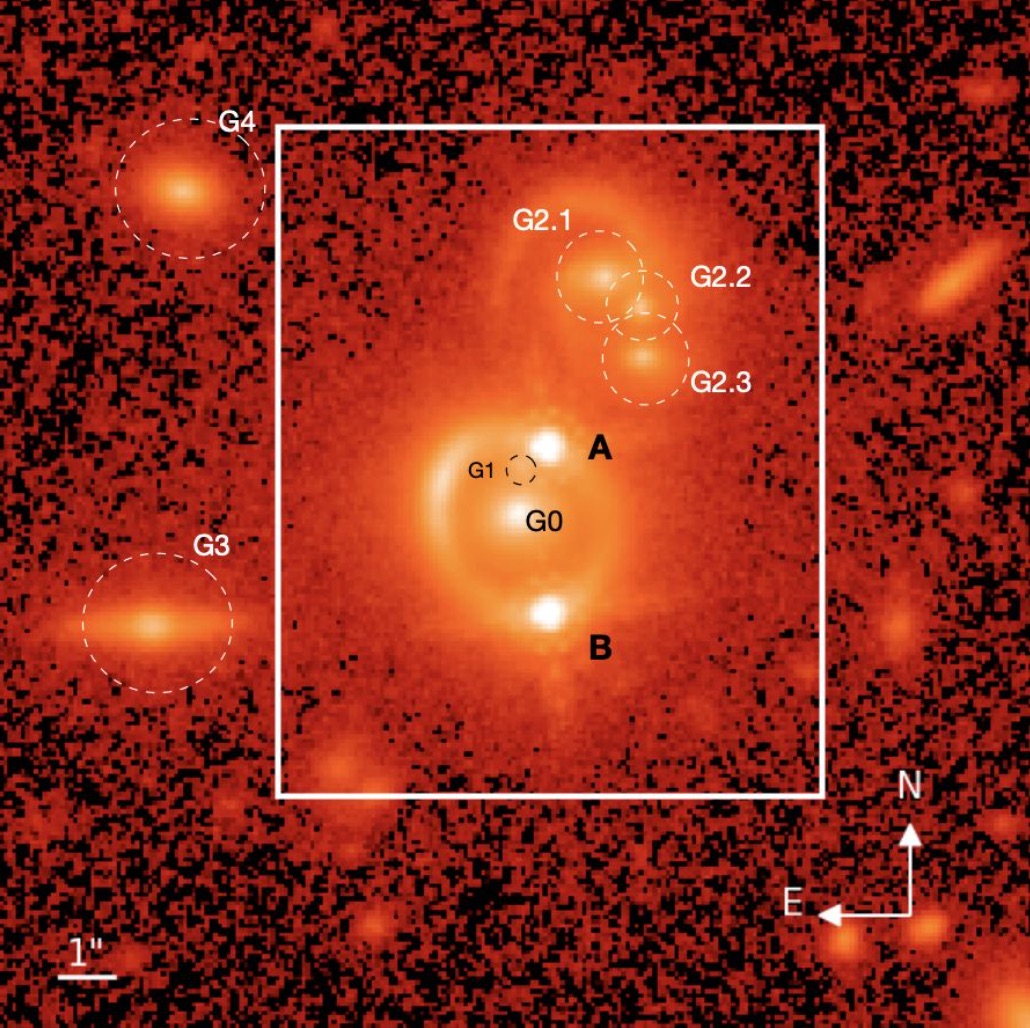}
\caption{The lensed quasar SDSS J1206+4332. The lensing galaxy G0 is located in the center of the image. A and B are the double quasar lensed sources. G2 is a triplet galaxy, and G3 and G4 are other nearby galaxies. The white rectangle marks the field of view of MAAT. Image Credit: Hubble Space Telescope. \citep{Birrer2019}.}
\label{fig:sciencelensqso}
\end{figure}

However, to turn the measured time-delays into distances and $H_0$, accurate lens models are required. IFS observations of lensed quasars can spatially resolve the kinematics of the deflector lens, and measure redshifts and stellar masses of other galaxies along the line-of-sight, enabling a large improvement in breaking the so-called ``mass-sheet degeneracies'' \citep{Falco1985}. This will provide a more accurate mass model for the lens, which will lead to a considerable improvement on the $H_0$ value \citep{Shajib2018,Birrer2018}. The primary distance measurement is given by the time-delay distance, which is a multiplicative factor that depends on three angular diameter distances, the distance to the observer, the lens, and their relative distance. The combination of these measurements with the spatial and spectral information coming from the deflector lens (its stellar velocity dispersion, a proxy for the lens potential) will provide the angular diameter distance to the lens in a unique and independent way, which can then be used to constrain the cosmological parameters in the context of the standard (and non standard) cosmological models \citep{Paraficz2009,Jee2015,Shajib2018}. 

\begin{figure}[htb]
\centering
\includegraphics[width=0.9\linewidth]{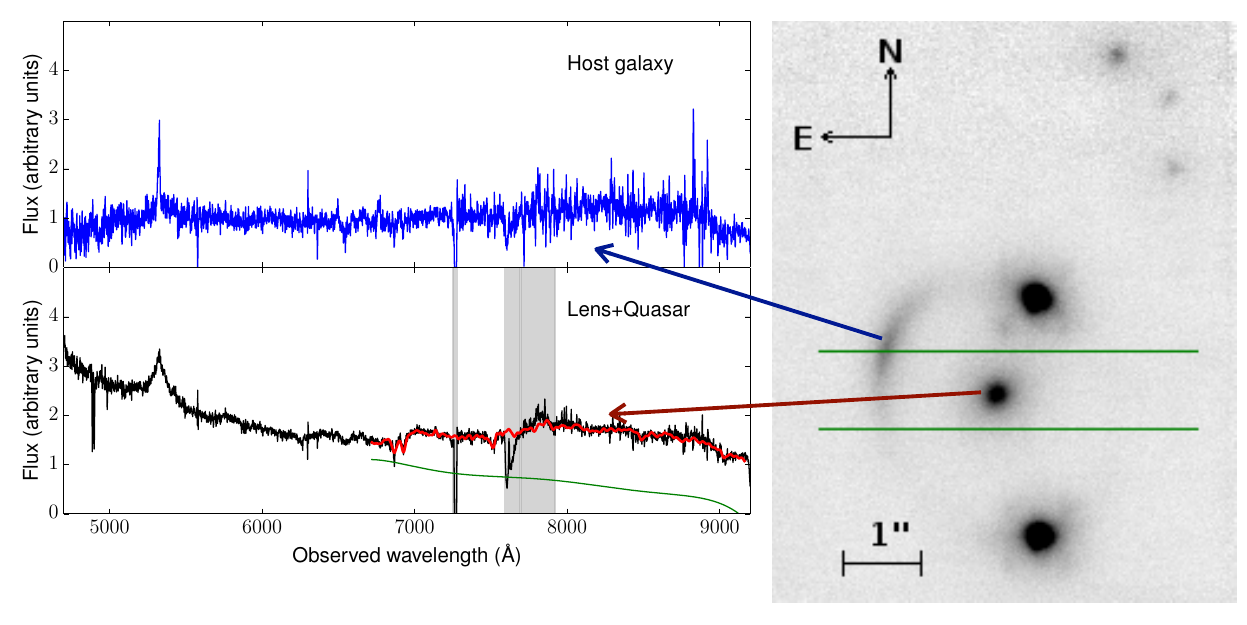}
\caption{Right: Imaging of SDSS J1206+4332 with LGSAO-NIRC2. The two quasar images, the deflector galaxy and the lensed host galaxy are all clearly visible. The lensed host traces a typical fold configuration. Three faint objects are visible to the NW, at $\sim4.5''$ from the deflector. The green horizontal lines mark the extent of the DEIMOS long-slit used for spectroscopy. Left: DEIMOS slit spectra. Top panel shows the 1D extracted spectrum of the quasar host galaxy. In the bottom the 1D extracted spectrum of the quasar (northern point source) and the deflector galaxy. The red line is the best-fitting spectrum obtained from the velocity dispersion fitting \citep[see][]{Agnello2016}). The MAAT@GTC FoV
will allow to obtain resolved 3D-spectroscopy of the entire image shown in the right panel.}
\label{fig:figrs}
\end{figure}

Therefore, highly-multiplexed spectroscopy is key to accurate measurements. For lenses visible from the South, VLT-MUSE has been pivotal in reaching this aim: \citet{Shajib2020} have constrained the value of $H_0$ to within 4\% from just one lens, and the combination of six lenses (of which three with MUSE data) yielded a measurement of $H_0$ within 2.4\%. In the foreseen sample of 12 lenses discovered by STRIDES and followed up by TDCOSMO, $H_0$ can be brought to 2\% uncertainties with ancillary data of sufficient quality.

The number of lenses found in the SDSS is larger than expected from the most optimistic theoretical predictions \citep{Oguri2005}. The total number of lensed quasars depends on the limiting magnitude of the observations: assuming a value of $r = 20$ mag, we have a total number of $\sim$ 200 sources visible from La Palma. However, we need a statistically significant number (e.g. 40-50) of quasars in order to get a very good precision on the Hubble constant and then make important constraints on the cosmological parameters and models. We will focus on the quasars that will fill the MAAT field of view and that are bright enough to follow-up in time the variable light emitted by the lensed quasars: as an example, the quasar SDSS 1206+4332 (\citet{Birrer2019}, see Figure~\ref{fig:sciencelensqso}) extends over a region that can be completely covered by MAAT in a single shot. Multiple dithered exposures will deliver a better resolved and high spectral quality dataset that, combined with the MAAT spectral and spatial resolution, can provide an exquisite measurement for the spatial kinematic distribution and the velocity dispersion of the lens potential, while observations in time will give an estimate for the quasar emission time-delay. Taking the SDSS1206+4332 source as model, characterised by $g=17.6$ mag and $i=17.4$ mag, a single exposure of 300s, using the R1000R grism, will provide a S/N value of $\sim30$ for the quasar spectral emission. Although for the time-delay measurements we can also use much shorter time exposures, longer exposures are preferred in order to enhance the faint signal from the lens sources and the fainter background galaxies that can belong to the cluster; a cluster lens requires $\sim10\%$ accurate velocity dispersion of the deflectors.

Based on the experience with MUSE we already know that between 3 and 5 hours on target are needed, depending on the lens brightness. Seeing better than $1\arcsec$ is required - the median seeing at GTC is $0.8\arcsec$ - in order to separate efficiently the quasar image spectra from the lens galaxy spectra, which in turn minimizes shot-noise from the quasar spectra and maximizes $S/N$ for the measurement of lens velocity dispersion (see Figure~\ref{fig:figrs}). Lensed QSO MUSE cubes typically have a seeing  around $0.8\arcsec$ \citep{Sluse2019,Shajib2020}, see also the data taken with MUSE at VLT by \citet{Caminha2017,Caminha2019,Bergamini2019}.

The lack of an instrument like MUSE at northern latitudes, with the only exception of KWCI at Keck (which only covers the blue spectral range), will render MAAT@GTC a very competitive instrument for this type of science, furthermore because the extended blue sensitivity of MAAT with respect to MUSE (BlueMUSE at VLT will only become operational by 2030; \citet{Richard2019}. In summary, 40 lenses with measured stellar kinematics can bring $H_0$ to within $1\%$ and build a distance ladder that is entirely based on time-delay lensing distances over the $0.3<z<2.5$ redshift range.


As a spin-off, MAAT can also provide a super-resolved study of distant quasars and their host galaxies, using the same gravitational lens systems. As a matter of fact, there are even some known ``coronograph lenses", where the narrow-line region around the central engine of the source is stretched over wide-arcs and even mapped in a different location from the broad-line region \citep{Koopmans2002,Agnello2016}.

\subsubsection{Strongly lensed supernovae}

The recent discoveries of strongly lensed supernovae \citep{Kelly2015,Goobar2017} have opened yet a new road to study the high-redshift universe as well as the physics of supernovae at large distances, when the Universe was much younger. In particular, with lensed SNe Ia, it is possible to combine them as standard candles through the well-known light curve correlations \citep{Phillips1993,Perlmutter1995} and at the same time via the lensing time-delay between multiple images, which was recently used to determine the Hubble constant \citep{Grillo2018}. The current Young Supernova Experiment survey \citep{Jones2019}, using the Pan-STARRS telescopes  to catch the light of SNe at the very early stages of their explosion, is expected to observe four lensed SNe in 2 years, with the exact rate depending on the time cadence of each field of view and  the lensed sources inside them covered by the survey. Looking forward, projects like ZTF-2, and in just a few years LSST, will find large numbers of lensed SNe. Simulations by \citet{Goldstein2019} predict that LSST will find hundreds of lensed SNe, at which point follow-up facilities will be in very high demand. 

The spectral capabilities provided by MAAT will allow us to study the SN light-curves, as well as the host galaxy environments and lensing galaxies of these lensed SNe at high redshifts, given the large signal-to-noise provided by the large magnifications. Unlike strongly lensed quasars that require multiple and continuous monitoring in time, supernova time-delays can be measured in just weeks, as shown in \citet{Dhawan2019}. Furthermore, as the supernovae fade on time-scales of months, these systems lend to unique measurements of the lens system and SN host  galaxy which are unfeasible for quasars \citep{Mortsell2019} and allow for exciting measurements of substructures through microlensing effects \citep{Goobar2017,Dhawan2019,Mortsell2019}. Intriguingly, \citet{2020arXiv200410164J} showed that spectroscopic measurements alone can provide competitive measurements of time-delays for SNe~Ia, an exciting new avenue that can be exploited with MAAT.

The spatial and spectral information on the lens and the time-delay signal as provided by MAAT, in a single shot repeated in a fixed number of exposures - the 4D spectroscopy –, will allow to constrain the Hubble constant through the measurement of time-delay effects, as described for quasars, and at the same time infer the distance through the use of luminosity relationships, as successfully applied in  recent years to SNe Ia \citep{1995ApJ...450...14G}, which have lead to the discovery of the acceleration of the Universe \citep{Riess1999,Perlmutter1999}. Finally, MAAT observations will also probe evolutionary effects of supernovae and their environments through spectroscopic observations of highly magnified high-redshift cases, unveiling the luminosity function distribution up to high redshifts. 

This type of observations should be requested in Target-Of-Opportunity mode, given that when and where a lensed SN will be discovered is unknown in advance. Moreover, we need additional observations in order to monitor the appearance of multiple deflected emission by the lens source. Consequently, we will focus, initially, on a limited number of targets (2-3) per semester, asking for a total number of 5-6 epochs per SN.

The observation itself of lensed SNe, in particular SNe Ia at redshift $z=1.0$, is not an easy task. However, the lensing brightness amplification  will improve the detectability for high-z SNe, which presents us with a unique occasion to study the distant Universe with these sources. Assuming a spatial binning of $2 \times 2$ and R300R grism, for one hour exposure time (4$\times$900s) we expect to observe with MAAT a SN of $V=23.0$ mag with a $S/N~\sim10$, for the region around 4000--4500 \AA\ where SN features of \ionn{Fe}{ii}, \ionn{Ca}{ii} and \ionn{Mg}{ii} are generally observed. Following epochs will require longer exposures, unless we further bin the spectrum, which however will make us lose only limited information on the spectral emission. Again, this type of observations cannot be performed with MUSE at the VLT due to its lack of sensitivity below 4800 \AA.\footnote{BlueMUSE at VLT will only become operational by 2030; \citet{Richard2019}.}

\subsection{Identification and characterization of EM-GW counterparts}
\label{sec:maatgw}

The detection of gravitational radiation from the binary neutron star (BNS) merger, GW170817 \citep{Abbott2017b}, along with a short gamma-ray burst (sGRB) just 1.7 seconds later \citep{Goldstein2017}, followed by the identification of an associated kilonova AT2017gfo in the  nearby galaxy NGC 4993, at just 40 Mpc distance \citep{Arcavi2017}, opened a new multi-messenger window in astrophysics and cosmology. The forthcoming observational runs by the enhanced-sensitivity interferometers Advanced LIGO/VIRGO and KAGRA will extend the reach of possible additional kilonova detections to about 200 Mpc distance. These events will become the targets of the most intense observational campaigns in time-domain astronomy. Our knowledge about kilonova physics currently relies on the sole detection of GW170817/AT2017gfo \citep{Abbott2017b}, which has raised a number of very important questions about the connection of BNS mergers and short GRBs, the production of heavy elements in the Universe, as well as the potential use of BNS-mergers as “standard sirens” for cosmology. Furthermore, important open questions remain about the history of the stellar populations leading to the BNS system progenitors. MAAT@GTC will shed light on the formation and evolution of compact objects, emission processes and the expansion rate of the universe by addressing a set of these open questions through an accurate analysis of high-resolution and IFS data of the kilonovae and their host galaxies. Using a set of apposite techniques, developed to study the stellar population, structure and dynamics of gamma-ray bursts and superluminous supernovae host galaxies through IFS, we can study, with an unprecedented accuracy, the properties of the kilonova progenitors parent stellar populations, such as the star formation history, metallicities in the immediate circum-burst environments, and use the dynamical information to determine the evolutionary history of the neutron star binary system, as well as to determine the host galaxy distance and sharpen the measurement of the local expansion rate of Universe, $H_0$.

This information can be obtained with the use of new powerful synthesis population codes for the analysis of single and binary stars (with or without stellar rotation implemented, \citealp{Fernandes2013,Eldridge2016}). The use of advanced methods coming from machine learning techniques, aimed to find the best stellar templates, will also provide the nebular component of the immediate environment, obtained with a simple subtraction of the best stellar template, and consequently allow the information of the gas component around the kilonova source \citealp{Levan2017}). 

\begin{figure}[htb]
\centering
\includegraphics[width=0.95\linewidth]{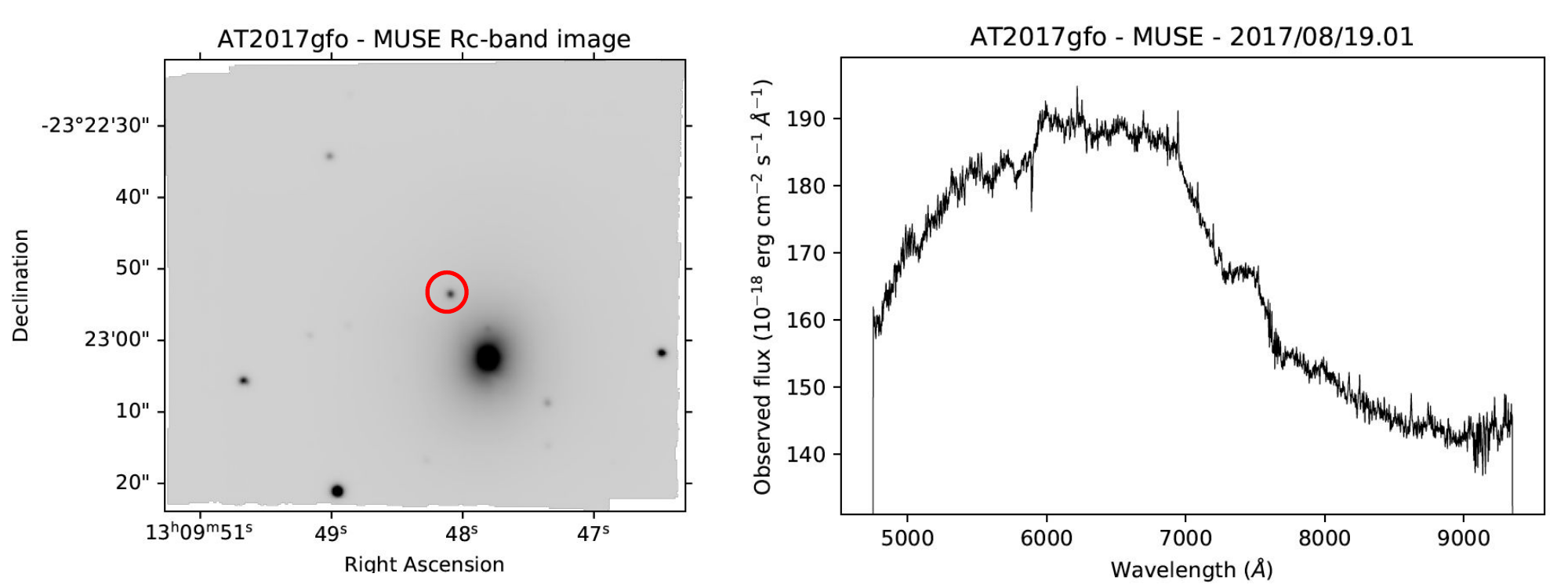}
\caption{
(Left panel) The equivalent Rc-band image extracted from the MUSE-VLT data of the KN AT2017gfo obtained ~ 1.5 days after the GW detection. The KN is marked with a red circle. (Right panel) The spectrum of AT 2017gfo as extracted from the MUSE-VLT data cube.} 
\label{fig:science4993}
\end{figure}

\begin{figure}[htb]
\centering
\includegraphics[width=0.9\linewidth]{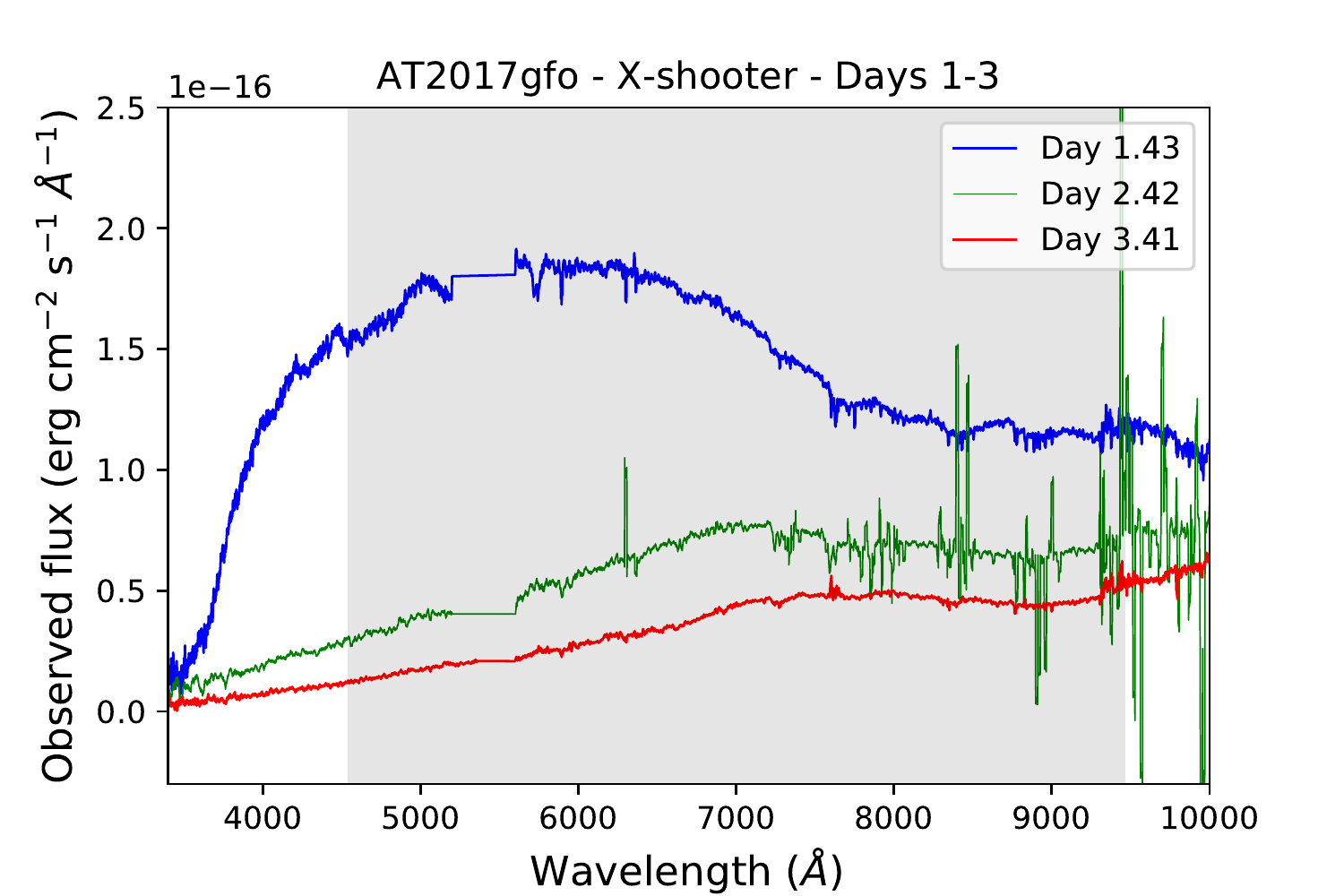}
\caption{
The spectral time series of kilonova AT2017gfo during the first 4 days of its emission \citep{Pian2017,Smartt2017}. The wavelength range plotted shows the broader  coverage that can be observed with OSIRIS+MAAT, using a combination of the R1000B and R1000R grisms, as compared to MUSE (gray area). This advantage of MAAT bluer spectral range will be maintained for almost a decade before BlueMUSE is operational \citep{Richard2019}.}
\label{fig:scienceKilonovaspectra}
\end{figure}

The IFS observations that will be carried out with  MAAT@GTC will also allow to follow-up the electromagnetic counterparts of newly-detected gravitational signal events and will help in building multi-epoch spectral energy distributions. Among the many advantages provided by MAAT with respect to other similar operating IFUs in 10m-class telescopes, like MUSE at VLT/ESO, the coverage of the ``blue’’ region of the optical spectrum represents one of the most important additions; this MAAT advantage will be maintained for almost a decade before BlueMUSE at VLT becomes operational \citep{Richard2019}. The early emission from kilonovae, believed to originate from radioactive decay of unstable r-process elements, is characterised by a very hot black body spectrum, with the peak of the thermal emission in the earliest instances is located at near-UV / blue wavelengths. In the days that follow, the emission originates from a lanthanide rich region, the kilonova ejecta cools down implying a redder thermal emission peaking in the near-IR. Consequently, the early peak of the kilonova emission, containing information of the viewing angle and geometry of element composition, is more intense at blue wavelengths, where MAAT is particularly sensitive. In Figure~\ref{fig:scienceKilonovaspectra} we show an example drawn from the X-shooter spectral series of the kilonova AT2017gfo \citep{Pian2017}, which shows the potential of investigating the blue region of the wavelength spectrum as provided by MAAT: several transitions of r-process elements fall in this spectral region (we mention the possible resonance line of \ionn{Sr}{ii} 4077,4215 \AA), which points out to an important role for MAAT@GTC for kilonova science in the years to come.

The observational strategy strongly depends on the number of detections of strong candidates by the LIGO-VIRGO Collaboration. The expected start of MAAT observations will coincide with the next observing run (namely the O4 run, expected to start observations in late 2021) at the LIGO-Virgo inteferometers, which will be characterised by an improved detector sensitivity and then a larger distance threshold for BNS detections. At the same time, the LSST will start to observe the full-sky visible from Chile, providing potential candidates for each GW detection at very deep magnitudes (current estimates give limits of $r \sim$ 24.5 mag for a single exposure with LSST). Similarly, optical surveys in the Northern hemisphere will scan the error region of each GW detection (we mention the GOTO telescope, currently operating at La Palma whose synergy with MAAT@GTC is fundamental for the success of the program) providing then possible candidates for each BNS merger detected by LIGO-Virgo. According to the detection rate for binary neutron stars (BNS) during the current O3 run (five BNS detections in eight months of observations) and considering that GTC will cover $\sim$ 60\% of the entire sky (excluding southern latitudes and Sun-constrained regions of the sky) we expect to observe two BNS candidates per semester. 

According to the above considerations, the strategy that we intend to adopt in a single semester of observations, will consist in obtaining a first observation in Target-of-Opportunity mode as quickly as possible once the error region is down to the size of the MAAT FoV - providing immediate information about the transient. Given the rapid evolution of kilonovae (they fade very quickly after two weeks from the BNS merger) we will ask for additional epochs very close in time: we plan three or four more epochs to be distributed in the first 10 days of the KN emission. A late observation, to be executed any time after the source has faded, 
is necessary in order to provide direct information about the stellar population underlying the source position. Each single BNS event will then require a set of 1 hour, which corresponds to a total requested time of 20 hours.

The kilonova AT2017gfo at the distance of 40 Mpc represents our best reference event to be used in our analysis to quantify the number of expected events within the reach of MAAT. According to the observed evolution and the distance of 40 Mpc of AT2017gfo, and given the sensitivity of MAAT, we expect to observe the peak brightness emission of kilonovae at the distance of 200 Mpc (expected magnitude V$\sim$19.5 mag) with a signal-to-noise of $\sim$10 per spectral bin assuming standard 1$\times$1 binning and the use of the R1000B grism. This estimate will improve if we integrate over the entire PSF of the source, permitting us to follow the evolution of the KN with similar signal-to-noise values below 2-3 magnitudes from the peak, which is equivalent to cover a week in terms of kilonova evolution (see Figure~\ref{fig:scienceKilonovaspectra}).

Kilonovae are also promising accurate distance indicators. As demonstrated by \citet{Dhawan2019}, the $H_0$ estimate using the “standard siren” measurement of GW170817 could be significantly improved by using the wavelength dependent intensity of the EM counterpart, AT2017gfo. The reason for this is that the time dependent SED of the kilonova is sensitive to the viewing angle towards the BNS merger plane. Since the latter is degenerate with the distance estimate from the GW signal, MAAT@GTC would allow us to do the same “standardization” for BNS merger GW distance out to 200 Mpc, i.e., the volume probed by the interferometers, and beyond the distance to which one can safely expect viewing  angle constraints from radio data when a GRB is observed (about $10\%$ of the cases). Furthermore, the latter requires high-density interstellar medium surrounding the GRB, unlikely to be the case for many of the BNS mergers. 
In BNS merger events there are also a number of intrinsic parameters characterizing the source that can impact the light curve and any observable steaming from it, including their potential use as distance indicators. It is already known that there is a relationship between ejecta mass and those of the NSs in the binary for a given equation of state (EOS) of matter. Matter in the interior of these ultra-dense compact stars can attain values beyond several times that of nuclear saturation density in terrestrial nuclei $ 2 \times 10^{14}$ $\rm g/cm^3$. It is believed that matter in the NS core forms a relativistic quantum nuclear (or hypothetically quark) system, as predicted by the theory of strong interaction. Different EOS models which predict different NS masses and radii in the BNS including internal quark deconfined phases are presented in the literature \citep{Baym2018,Ivanytskyi2019}. 
Recently, a coalescing event from a binary system composed of a black hole 23 times heavier than our sun and a much lighter object, of about 2.6 solar masses, has been detected by LIGO-Virgo as reported from GW190814 \citep{Abbott2020}. The unusual mass ratio adds yet more interest to the mass population of these individual compact objects, the lighter one being possibly a NS or a BH. However, an electromagnetic counterpart has yet to be observed. Constraining the EOS of dense matter is one of the current key problems in Physics. Numerical Relativity  simulations typically estimate the ejected mass, lepton fraction, and velocity of the ejecta \citep{Bauswein2013} and the expected light curve  with its associated uncertainties, finding correlations between the peak bolometric luminosity and the decline in the luminosity after a few days. MAAT will have the capability to scrutinize these KN light curves and use this information in distance measurement and hence the calibration of distance ladders, having a potential impact on the determination of $H_0$ and its associated uncertainty \citep{Coughlin2020}. At the same time it will indirectly  help shed light by constraining the EOS provided the number of events will be large enough.

In summary, kilonovae observations with MAAT offer unique possibilities in this thriving field.

\newpage 

\subsection{The host galaxy environment of supernovae}

\subsubsection{The environment of intermediate-z supernovae}
\label{sec:sngal}

Type Ia supernovae (SNe Ia) are the most mature and well-exploited probe of the accelerating universe, and their use as standardizable candles provides an immediate route to measure dark energy. This ability rests on empirical relations between SN Ia peak brightness and light-curve (LC) width \citep{1993ApJ...413L.105P}, and SN color \citep{1996ApJ...473...88R}, which standardize the absolute peak magnitude of SNe Ia with a dispersion of $\sim0.12$ mag ($\sim6\%$ in distance; \citealt{2014A&A...568A..22B}).  However, with over 1000 well measured SNe Ia \citep{2018ApJ...859..101S} each precisely photometrically calibrated to 2$\%$ \citep{2019MNRAS.485.5329L}, systematics, and in particular astrophysical systematics now dominate the SNe Ia error budget \citep{2014A&A...568A..22B,2019ApJ...874..150B}. 

To address this, recent studies have focused on detecting, characterizing and exploiting correlations between SNe Ia and their environments. Such measurements provide an indirect route into the progenitor physics of SNe Ia, allowing for an improved understanding of the diversity and potential evolution of these SN. Studies in both the local universe and at high redshift have shown that the rates and properties of SNe Ia depend of host galaxy measurements such as galaxy morphology \citep{2000AJ....120.1479H}, stellar mass \citep{2010MNRAS.406..782S}, specific star-formation rate \citep{2012ApJ...755...61S}, stellar population age \citep{2018arXiv180603849R} and metallicity ($Z$; \citealt{2011ApJ...743..172D}). Since all these galaxy parameters have been found to evolve with cosmic time, the properties of SNIa progenitors and, in turn, their ability to serve as a distance indicators may be affected by such evolution, which would add up to the systematic uncertainty budget.

Moreover, cosmological studies of SNIa have now firmly established a dependence of Hubble diagram residuals ($\approx$differences between distances estimated from SNIa peak brightness and those calculated assuming a fiducial cosmological model) on global host galaxy parameters, such as mass, age, and metallicity (e.g. \citealt{2010ApJ...715..743K,2010ApJ...722..566L,2011ApJ...743..172D,2011ApJ...740...92G,2010MNRAS.406..782S,2018MNRAS.476..307M}). The addition of a term in the standardization of SNIa absolute magnitudes in the optical that accounts for these environmental properties (e.g. the 
`mass step' or the $\gamma$-metallicity term; \citealt{2016MNRAS.462.1281M}) has proved to further reduce the scatter of the Hubble residuals.

Most of these studies are based on analyses of the integrated or central host galaxy spectra, and broad-band or narrow-band H$\alpha$ imaging. The effect of the local environment of SN Ia within galaxies in cosmological studies is almost unexplored. As an exception, \cite{2018A&A...615A..68R} presented an analysis of the dependence of SN light-curve parameters and Hubble residuals on their local environment using broad-band photometry. They show a significant dependence on the $U$-$V$ color, which is treated as a proxy for the stellar age of the underlying populations (bluer being younger). However, age derived from photometry (color) has several uncertainties, and degeneracies (with extinction and stellar metallicity), and it is not enough to determine precisely the cut-out in stellar ages of such dependence.

Recent studies of SNIa host galaxies observed with integral field spectroscopy (IFS) have opened a new window in the field. For instance, \cite{2013A&A...560A..66R,2018arXiv180603849R}, using observations of around a hundred very nearby host galaxies from the Nearby SN Factory \citep{2002SPIE.4836...61A}, showed that SNe Ia exploding at locations with higher star-formation intensity and higher specific star formation density could be more standardizable than those in passive local environments. 

In the last few years some efforts have been focused in compiling a statistical sample of SNIa host galaxies observed with IFS to probe this and other unexplored correlations with the local environment at low redshifts ($z\lesssim0.1$; \citealt{2013AJ....146...30K,2013AJ....146...31K,2014A&A...572A..38G,2016A&A...591A..48G}).

\begin{figure}[htb]
\centering
\includegraphics[trim=3.8cm 8cm 9.8cm 7.5cm, clip=true,width=0.7\columnwidth]{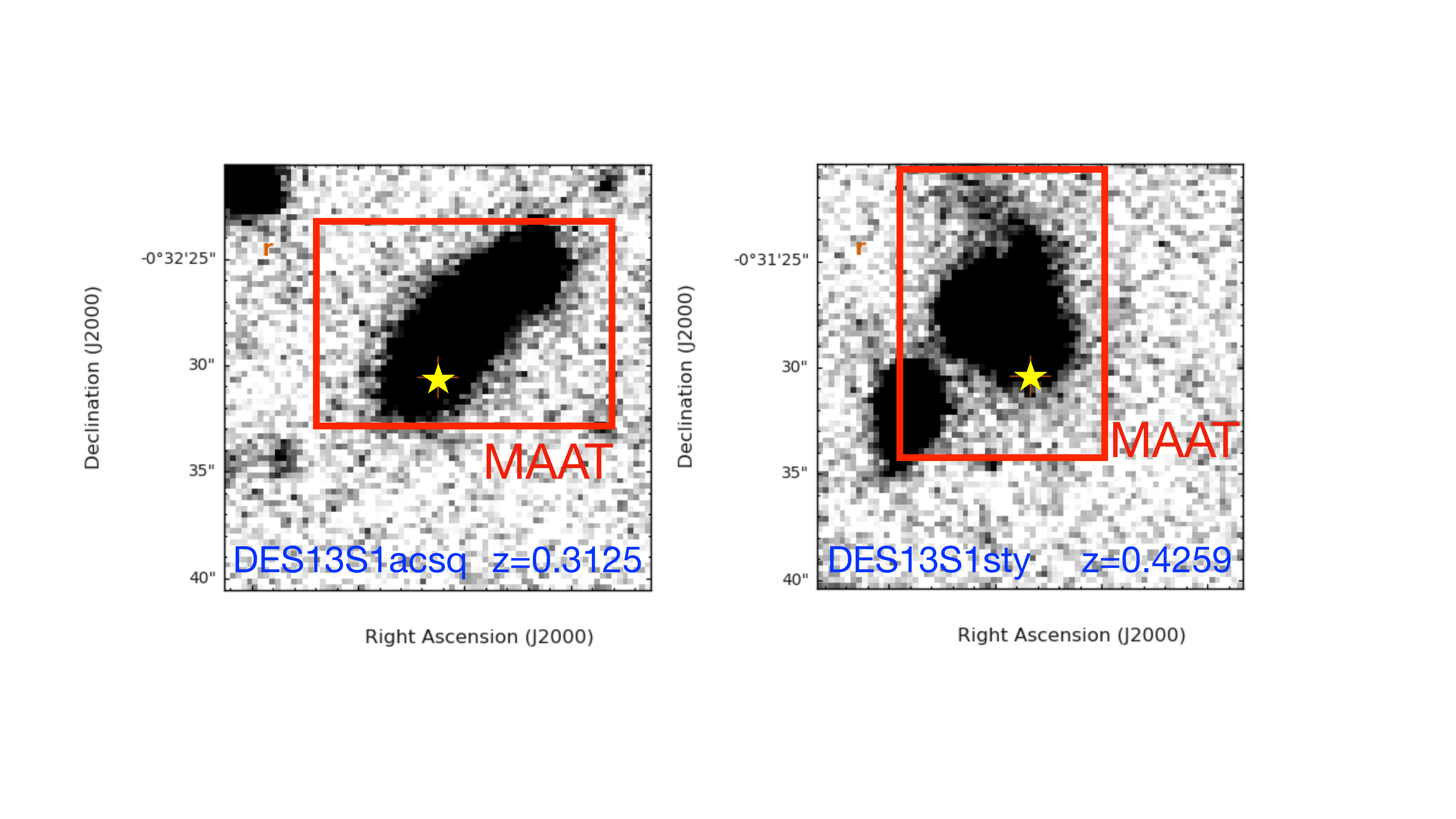}
\caption{Twenty arcsec squared cutouts of SNe Ia DES13S1acsq and DES13sty discovered by the Dark Energy Survey at redshifts $z=0.3125$ and 0.04259, respectively. The yellow star is located at the SN location, and the red rectangle corresponds to the projected field-of-view of MAAT on the sky. With the GTC+MAAT configuration we will be able to spatially-resolve the SN environment in the galaxy and, at the same time, provide both global and local parameters. Cutouts courtesy of Philip Wiseman (\textit{priv comm.}, but see \protect\citealt{2020MNRAS.tmp.1510W}).}
\label{fig:fig}
\end{figure}

There are currently a few instruments with the needed capabilities to perform such studies. While wide-field instruments allow to cover in full or at least large extents of host galaxies, high-spatial resolution is also needed to properly sample different regions of a galaxy including the location of the SNIa. 

In the Southern hemisphere, MUSE mounted to the 8.2 m Yepun UT4 Very Large Telescope (VLT) and KOALA at the 4 m AAT, are the best-suited instruments available. While MUSE is currently the only instrument combining the highest-spatial resolution ($0.2\arcsec$ spaxel) with the largest FoV (1 sq. arcmin), it lacks of coverage of wavelengths bluer than $\sim$4650 \AA\ (see e.g. \citealt{2016MNRAS.455.4087G} for an analysis of SN environments with MUSE); the blue range advantage of MAAT  will be maintained for about a decade before BlueMUSE becomes operational \citep{Richard2019}.
In the Northern hemisphere, PMAS mounted to the 3.5m telescope at the Calar Alto Observatory, is currently the only instrument available to perform such SN environmental studies. \cite{2012A&A...545A..58S} with their pilot work, and \cite{2018ApJ...855..107G} later compiling the largest sample of nearby SN host galaxies observed with IFS ($\sim$400; PISCO), have demonstrated the capabilities of this configuration.

At high-redshift, galaxies are both smaller in apparent size and fainter, so ideal instruments do not need large FoVs but to be attached to either a large-aperture telescope or a satellite. In addition, the usual features used to characterize the main properties of galaxies (e.g. strong emission lines, like H Balmer lines) have shifted to near-infrared (NIR) wavelengths. KMOS at the 8.2 m Antu UT1 VLT employs 24 small (8 sq. arcsec) integral field units with a configurable location within a field of $\sim$7 sq. arcmin and with a spatial resolution of $0.2\arcsec$ per spaxel, that provide IFS in the NIR. In the near future the James Webb Space Telescope (JWST) will provide IFS capabilities from space at near- and mid-infrared wavelengths. On the one hand, the NIRSpec instrument is equipped 
with an IFU  (FoV of 9 sq. arcsec, $0.1''$ per spaxel) that provides orders of magnitude gain in sensitivity with respect to ground-based facilities over 
the full 1 to 5 microns spectral range, and also takes full advantage of a very stable PSF. On the other hand the MIRI instrument provides the IFS 
capability over the 5-28 microns range, at somewhat larger sampling and FoV. MAAT@GTC will provide excellent complementarity in UV-optical wavelengths 
to the IFS studies carried out with the JWST.

Having the low-z ($z\lesssim0.1$) and the high-z ($z>0.5$) ends covered, there is currently no instrument that is suitable for studying intermediate redshift galaxies. Ideally, it would need to combine a FoV large enough to cover the extent of galaxies at this redshift range ($\sim$10-20 arcsec side), it would need to be mounted to a large-aperture telescope, and it would need to provide coverage in optical wavelengths up to 1 nm. MAAT@GTC would satisfy these three requirements. Figure~\ref{fig:fig} shows the locations (yellow stars) of two SNIa at intermediate redshifts from images obtained by the Dark Energy Survey \citep[DES;][]{Bernstein2012}. DES has compiled the largest, homogeneously identified sample of SNe Ia across cosmic time, with $\sim600$ spectroscopically classified SNe Ia and $\sim2000$ photometrically classified events to $z=1.2$. In the next few years, DES will produce the most stringent constraints on the nature of `dark energy' prior to LSST. The red rectangle represents the projected field covered by one MAAT pointing, which fits perfectly well the size of galaxies at these intermediate redshifts.

Combining past and current efforts to compile a large sample of low- and high-z SNIa host galaxies with PMAS/MUSE/KOALA and KMOS/JWST, we propose to use MAAT to obtain IFS observations of intermediate-z host galaxies of SNIa from the Dark Energy Survey.
This will allow us to fill the redshift desert that is unexplored with current instrumentation, and complete the studies of evolution of SNIa properties in a wide range of redshifts.

\subsubsection{Exploration of host galaxy environmental dependencies on energetic core-collapse supernovae}
  
Supernova is the most spectacular and dramatic phase during the life of a massive star. There exist two main flavors of SNe, namely the core-collapse (CC) and the thermo-nuclear SNe. The latter type has already been discussed in the previous section; here we discuss what can be the role of MAAT in unveiling the physical properties of the progenitor population of one of the sub-class of CC SNe, namely type-Ic SNe. This class of SNe is characterised by the absence of hydrogen and helium in their optical spectra \citep{Gal-Yam2017}. Moreover, the most energetic members of this class, e.g. type-Ic SNe showing broad lines (BL) and larger ejecta velocities, have been observed few days \citep{Galama1998,Hjorth2003} after the explosion of energetic long gamma-ray bursts (GRB), confirming the 
``Collapsar'' scenario as their origin \citep{Woosley2006}: their huge energy emitted in electromagnetic radiation is ascribed to the core-collapse explosion of a highly-rotating stripped-envelope massive star into a black-hole, which is able to produce in these late phases a relativistic collimated jet \citep{MacFadyen1999} giving rise to the observed prompt gamma-ray emission \citep{Gehrels2009}.

However, not all Ic-BL SNe are associated with a GRB: GRB-SNe represent only a tiny fraction of all Ic-BL SNe \citep{Cano2017}. This evidence cannot be explained with a jet pointing away to our line of sight: the relativistic jet should leave an imprint of its presence in the radio emission, but recent surveys \citep{Soderberg2006} (Marongiu et al. 2018) of nearby GRB-less Ic-BL SNe found no evidence for an associated off-axis relativistic component.
Interestingly, a recent analysis of well-studied GRB-less Ic-BL SNe show high-velocity components ($v_{ej}$ $\sim$ 30,000--40,000 km/s) in the very early optical spectra. This evidence was attributed to the presence of a ``choked-jet'', e.g. a jet that is not able to drill through the layers of the GRB progenitor star, but it is however able to generate a sub-relativistic cocoon emission that propagates laterally and inside the star until it then breaks-out once reaching the photosphere \citep{Piran2019}. 

We do not know yet why the majority of Ic-BL SNe do not provide enough fuel to the jet to escape from the stellar environment. A possible solution is provided by the lower
rotation of its progenitor final Fe-core, which points out to a higher metallicity of the progenitor star \citep{Maeder2001}.
This would imply a consistent mass-loss and consequently a final lower angular momentum for the Fe-core when compared with
GRB-SNe \citep{Woosley2006}, which instead represents the main ingredient to form a jet in broad-lined core-collapse
SNe. GRB-SNe are generally located very close to the the brightest regions of their host galaxies \citep{Fruchter2006}, similarly to type Ic SNe \citep{Kelly2008}. Moreover, the lifetime of Ic-BL SNe is relatively short, $\sim$ 10--20 Myr, and then these stars will die
very close to the region where they were born, which explains their proximity to bright \ionn{H}{ii} regions. Studying the local environment of these CC-SNe will then provide a direct information of the initial metallicity of their progenitor stars \citep{Galbany2016,Kuncarayakti2018}, which represents a crucial test to understanding if GRB-jets in SNe prefer low-metallicity and highly-rotating progenitors or this is not true, and then we should expect jet-like structure in all SN types.

An analysis of long-slit spectra of GRB-SNe and GRB-less Ic-BL SNe host galaxies has already revealed that there exists a distinction between the two subclasses: GRB-SNe are generally characterised by lower metallicity values (Modjaz et al. 2008), in line with theoretical expectation. Consequently, a systematic analysis, made with spatially-resolved detectors like MAAT@GTC, of the environment of GRB-SNe and type-Ic BL SN without an associated GRB will provide important clues on the stellar population that formed these SNe and then a crucial test for the above models.

\begin{figure}[htb]
\centering
\includegraphics[width=1.0\linewidth]{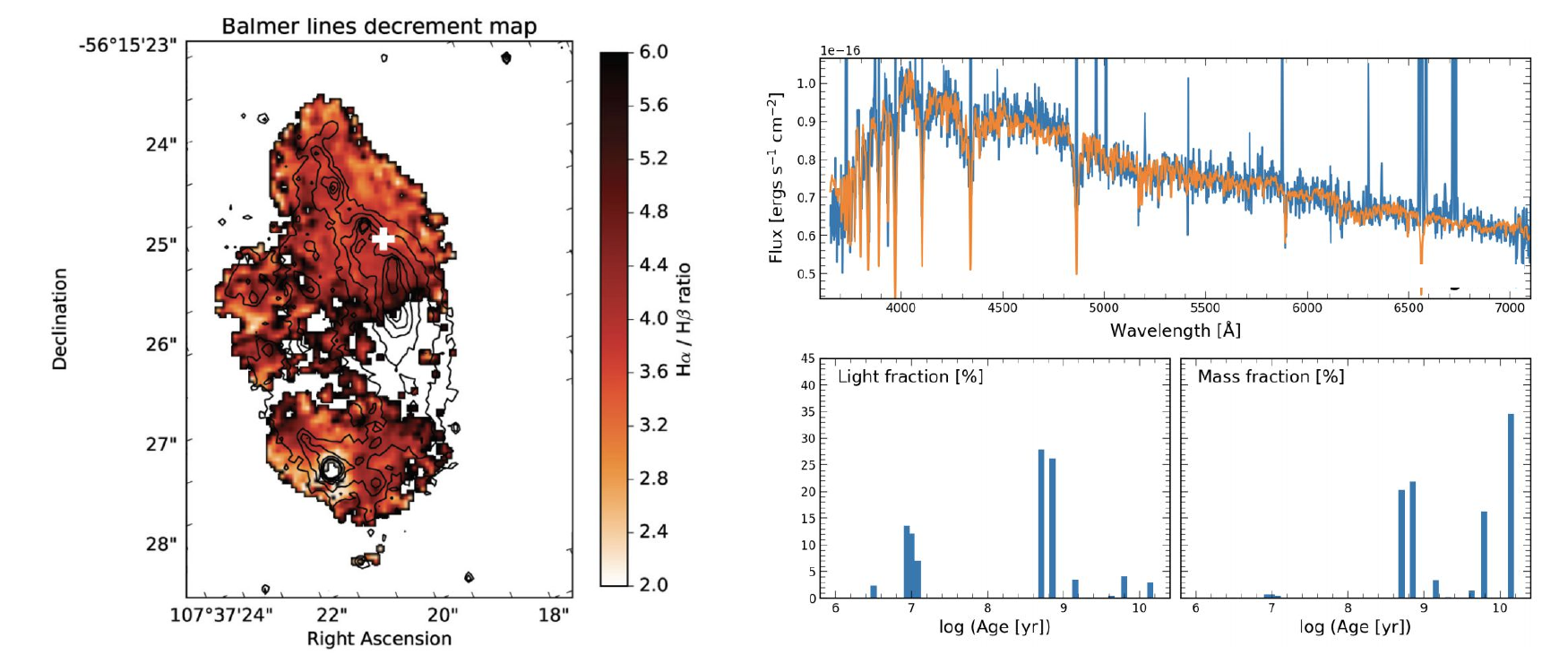}
\caption{
Left: The distribution of the color excess E(B-V) obtained from the Balmer decrement ratio for the case of the host galaxy of GRB 100316D \citep{Izzo2017}. The white cross corresponds to the location of the GRB. Right: The analysis of the region surrounding the SLSN 2017egm \citep{Izzo2018} using the STARLIGHT spectral synthesis code \citep{Fernandes2013}. The upper panels show the observed spectrum (blue curve) and the synthetic spectrum (orange curve) obtained from a combination of stellar spectra. The lower panel shows the stellar light fraction and the mass fraction as a function of the stellar age, as inferred from the spectral synthesis analysis at the position of the SLSN.}
\label{fig:sciencehostgal}
\end{figure}


The light from external galaxies is mainly composed by stars, in addition to the gaseous and dust components. However, the stellar populations responsible for the observed light are more mixed, which implies additional uncertainties on the real composition in terms of stars. In order to reveal the physical properties of the stars underlying a given region in a galaxy, which can be the location where a CC-SN was observed, we must deal with a larger set of star formation histories with composite populations and with different stellar evolution. IFS observations provide an enormous support in this research field, since we can study at high spatial and spectral resolution the immediate environments of SNe-Ia and the spatially-resolved global properties of the galaxy. This procedure is partly similar to what was proposed in section below about the study of kilonova environments. In the following, we describe the methods in  detail, focusing for this specific case on the estimate of the host galaxy extinction value.

\begin{figure}[htb]
\centering
\includegraphics[width=0.8\linewidth]{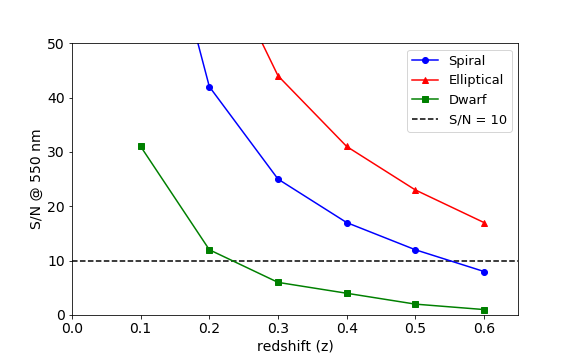}
\caption{Our simulation of the distribution of the signal-to-noise of spectral region, computed at the wavelength of 550 nm, for different type-Ic SN host galaxies as observed by MAAT@GTC. The black dashed line corresponds to our threshold for the signal-to-noise value useful to obtain reliable results from stellar synthesis population analysis.}
\label{fig:scienceSN}
\end{figure}

Each MAAT data cube will be analysed adopting a well-known strategy used for the study of stellar populations in distant galaxies. After defining a strategy for data binning, which can be based on optimising the signal-to-noise for each spectrum or on the study of single \ionn{H}{ii} regions for star-bursting galaxies, we will use dedicated codes capable of decomposing each single spectrum in terms of different ages and metallicities of the underlying stellar populations. This technique is based on the use of a pre-defined stellar libraries, with stellar spectra of different ages and metallicities  that  represent the starting “eigenvector” base for the final decomposition analysis. After finding the best solution for each spectrum/region, we can finally provide estimates for the main physical properties like the extinction. An example of the application of this method to a GRB-SN host galaxy and to a SLSN case is shown in Figure~\ref{fig:sciencehostgal}. The right panel shows an example of the decomposition of a spectrum through a pre-defined library of spectra with given ages and metallicities and how to obtain the residual nebular component; the left panel displays the 2D distribution of the color excess E(B-V) obtained from the Balmer decrement.

MAAT@GTC can observe GRB and type-Ic host galaxies at moderately high-redshifts, with the main goal of disentangling the stellar continuum (whereas possible) from the nebular component, thus resolving the main emission lines (\ionn{H}{i}, [\ionn{O}{ii}], [\ionn{O}{iii}], [\ionn{N}{ii}], [\ionn{S}{ii}], \ionn{He}{i}) in order to obtain estimates of different physical properties such as metallicity, star-formation rate, ionisation, stellar ages and  the extinction using both methods described previously. The target sample will be selected from the available sample of known long GRBs and type Ic SNe, as well as from future discovered SNe by operating surveys at intermediate redshifts ($0.1 < z < 0.6$).  Considering 1 hour of observing time per SN, we can split our program in four observing semesters; in each one we will observe $\sim$ 12 host galaxies in order to have a complete sample of $\sim$ 50 hosts after two years from the start of the program. This will allow a better understanding of the possible role of the environment and of the influence of the initial metallicity on the evolution of the progenitor star. 

Type Ic SNe are generally observed in late-type galaxies. GRB-SNe in particular have a ``preference'' for faint low-metallicity environments. We have simulated the signal-to-noise per spectral bin of MAAT, considering an average spatial sampling of $0.6''\times 0.6''$ that takes into account atmospheric effects (seeing), for different host galaxies of SNe-Ia and a total exposure time of 1 hour using a combination of the R1000B and R1000R grisms. In this specific case we have assumed three types of galaxies: 1) a late-type Sa spiral with absolute magnitude M$_V=-18$ mag, 2) an early-type E galaxy with absolute magnitude of M$_V=-20.5$ mag, 3) a dwarf starburst galaxy with absolute magnitude M$_V=-16$ mag. Results are shown in Figure~\ref{fig:scienceSN}, considering a signal-to-noise (SN) value estimated at 5500 \AA. Spectral decomposition can provide very good results if the spectra to be decomposed are characterised by a signal-to-noise of $\sim20$, but the analysis of the gas physical properties can be successfully done with lower values of the signal-to-noise, e.g. $\sim10$. We conclude that we can estimate physical parameters of the stellar continuum up to redshifts $0.6$ for an elliptical 
galaxy, $z=0.5$ for spiral hosts and $z=0.2$ for dwarf host galaxies, while we can study the gas properties (e.g. the analysis of the main emission lines) for all the types of galaxies in the redshift range considered. We further notice that we did not assume any spatial averaging, which will increase the value for the signal-to-noise for each single case. However, spatial averaging implies that the region covered by each single spectral region will be much larger, an effect that depends also on the distance and not only on the spatial sampling. This can have some drawbacks, given that for a larger spatial region inside the SN-Ia host galaxy we would observe a combination of multiple stellar populations, which can slightly affect the  value of the extinction at the location of the SN-Ia as inferred from stellar synthesis analysis.

\subsection{The abundance discrepancy in planetary nebulae with MAAT}

The ``abundance discrepancy'' problem is one of the major unresolved problems in nebular astrophysics, and it has been around for almost eighty years \citep{wyse42}. In photoionised nebulae ––both \ionn{H}{ii} regions and planetary nebulae (PNe)–– optical recombination lines (ORLs) provide abundance values that are systematically larger than those obtained using collisionally excited lines (CELs), the ratio between the two being the abundance discrepancy factor (ADF).
Solving this problem has obvious and deep implications for the measurement of the chemical content of nearby and distant galaxies, as this is most often done using CELs from their ionised interstellar medium.

\begin{figure}[htb]
\centering
\includegraphics[width=0.9\linewidth]{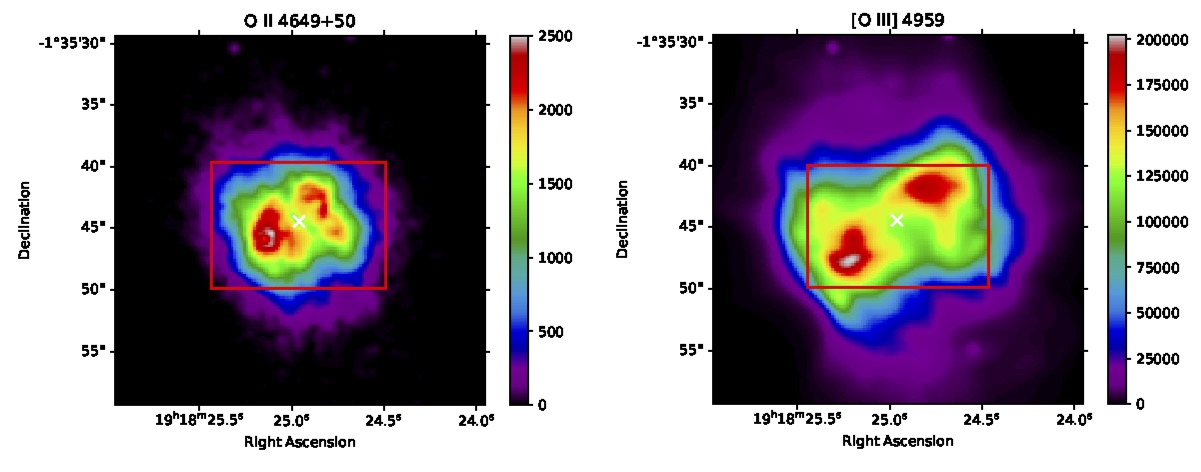}
\caption{MUSE-VLT emission line maps of planetary nebula NGC~6778 illustrating different spatial distribution of ORL and CEL emission in high ADF planetary nebulae (Garc\'ia-Rojas et al. \textit{in prep.}). Left: O~{\sc ii} $\lambda\lambda$4649+50. Right: [O~{\sc iii}] $\lambda$4959. The red box shows the FoV of OSIRIS+MAAT.}
\label{fig:nebulae}
\end{figure}

The reason of this discrepancy has long been a matter of debate \citep[see][]{garciarojasetal19} and no consensus has been reached today. 
Focusing on PNe, several spectroscopic studies have shown that the faint ORL emission is strongly enhanced in central regions of PNe with known close-binary central stars and high ADFs \citep[e.~g.][]{corradietal15, jonesetal16, wessonetal18}. These extreme ADFs have been associated with the presence of cold, metal-rich gaseous clumps in the nebula, which are very efficiently cooled by the heavy elements \citep{liuetal00}. The first clear evidence of two plasma components with a probable distinct origin was provided by \cite{garciarojasetal16}, who found, using OSIRIS-GTC tunable-filter imaging of a high-ADF PN, that ORL and CEL emission had clearly distinct geometries. Having spatially resolved information of both the emission of these lines as well as of the physical conditions can reveal crucial information on the mass ejection modes of both plasma components.
PNe with binary central stars and/or high ADFs are perfect targets to map the emission of both kind of lines (see Figure~\ref{fig:nebulae}), as well as the physical conditions obtained from several diagnostics, owing to the surface brightness of the usually very faint ORLs is large enough to be within the reach of large telescopes. 

MAAT@GTC offers the opportunity of accessing high-spatial resolution 2D spectroscopy even in relatively faint lines, such as metal ORLs, and additionally covering the blue ($\lambda <4600$ \AA) range of the optical spectra, where some important diagnostic on the abundance discrepancy lie and, therefore, address some interesting questions: i) to study if the spatial distribution of the electron temperature ($T_e$) sensitive [O~{\sc iii}] $\lambda$4363 auroral CEL correlates with O~{\sc ii} and/or O~{\sc iii} ORL emission and, if so,  quantify the effect on the determination of $T_e$ and the ADF  \citep{gomezllanosetal20}; ii) to map the spatial distribution of $T_e$ derived from the Balmer Jump at $\sim$3646 \AA\ that would give us an estimate of the spatial distribution of the $T_e$ of the cold plasma component; iii) to study the recombination contribution to the nebular [O~{\sc ii}] lines at $\lambda\lambda$3726+29 CELs in high-ADF PNe, which can make electronic density diagnostic from these lines unreliable \citep{wessonetal18}. It is worth mentioning that OSIRIS+MAAT would be the only instrument in a 10m-class telescope in the northern hemisphere providing simultaneously high-spatial sampling and a complete coverage of the optical wavelength range. 

Finally, these kind of observations combined with kinematics information obtained  from e.~g., MEGARA \citep{megara2016}, can provide strong constraints to 3D photoionization models and give crucial information to understand the physical mechanisms that are acting in the ejection of the metal-rich component. Summarizing, with the arriving of MAAT, the GTC would have a suite of instruments that can be of paramount importance for the topic of the abundance discrepancy problem.

\subsection{Accurate binary masses with MAAT}

One of the most important astrophysical parameters to be determined is the mass of an object, be it a star, a planet, a brown dwarf, or a stellar remnant (neutron star, black hole, white dwarf). For all but a handful of special cases, this is only attainable for objects belonging to binary systems, in which case it is essential to derive a spectroscopic orbit with the highest precision. For systems with very short orbits (with an orbital period of a few minutes to a few hours, i.e those that will likely merge, producing transients or gravitational waves), the need to sample well enough the orbit imposes tight constraints on the exposure time. When the primary is relatively faint (e.g., brown or white dwarfs, highly extincted or extragalactic systems), the use of \textit{\'echelle} spectrographs is not feasible and one needs to rely on low- and intermediate-resolution spectrographs attached to large telescopes. It is thus essential to ensure that the highest radial velocity accuracy can be achieved by such instruments.

A good example of the need for high-efficiency, intermediate-resolution spectroscopy on 10-m class telescopes comes in the form of the binary central star of the planetary nebula Henize 2-428. 
GTC-OSIRIS long-slit spectroscopy allowed \citet{2015Natur.519...63S} to derive a double-lined spectroscopic orbit of the star (while similar observations with FORS2 mounted on the VLT did not provide sufficient signal-to-noise, highlighting that these observation are on the limit of current GTC capabilities).  The radial velocity curve combined with photometric observations from smaller facilities indicated that the system was a double degenerate with total mass exceeding the Chandrasekhar limit and, thus, the strongest known candidate type Ia supernova progenitor.  Recent work has begun to indicate that at the resolution of GTC-OSIRIS the principal spectral lines used by \citet{2015Natur.519...63S} might be blended with absorption associated with diffuse interstellar bands \citep[DIBs;][]{2019cwdb.confE...7R}, hindering the accurate measurement of radial velocities -- in this case possibly leading to over-estimating the component masses.  The increased efficiency and resolution provided by MAAT would allow these lines to be resolved from their contaminants, and thus could prove crucial in unravelling the mystery of the origins of supernova type Ia \citep{2019MNRAS.482.3656R}.

Another important example of the possible contribution that MAAT would provide comes in the field of black hole binaries.  Black hole binaries are studied as single-lined spectroscopic binaries where the mass ratio is derived by measuring the radial velocity curve and projected
rotational broadening of the companion  star, the latter having typical values $\sim$40--100 km~s$^{-1}$ depending on orbital parameters.  The capabilities of OSIRIS have already been pushed to their limit in this field, via the use of extremely narrow long slits \citep{2020ApJ...893L..37T} where MAAT would offer a clear advantage in measuring both the radial velocity variations and rotational broadening, as well as allowing for measurements to be made for fainter, more distant systems.  The black hole mass distribution that will be obtained from Galactic binary systems using MAAT will improve our understanding of their formation mechanisms, as well as the origin of the black hole mergers detected with gravitational wave detectors.


According to \citet{1992ESOC...40...17G}, the uncertainty on a stellar radial velocity measurement ($\sigma_{rv}$) is inversely proportional to the  signal-to-noise ratio ($S$) of the spectrum as well as the resolving power ($R$) of the spectrograph to the power of $3/2$, i.e. $1/\sigma_{rv} \propto R^{3/2} S $.
As such, the increased efficiency of MAAT with respect to a typical OSIRIS long-slit (which, at 0.6\arcsec, generally implies significant slit losses) directly leads to an improvement in radial velocity precision. Furthermore, the increase in resolution again leads to a substantial improvement in radial velocity precision.

Using MAAT will typically lead to an increase in resolution by $\sim$60\% and increase in observed flux by $\gtrsim$20\% over OSIRIS with the $0.6''$ long-slit -- this results in a 2.2-fold gain in terms of radial velocity precision (i.e. $\sigma_\mathrm{MAAT}\sim0.45\sigma_\mathrm{OSIRIS}$).  Furthermore, this should be considered a \emph{minimum} gain as MAAT will maintain this precision even in poor seeing conditions (where OSIRIS slit-losses could become untenable), and MAAT removes any risk of losses or systematic effects due to inaccurate centering.

We also highlight that a number of higher resolution VPH grisms have been designed for OSIRIS but were never constructed -- each of which would push the MAAT resolving power up to 8,000.  With the designs available, the only associated costs would be for production.  The inclusion of the construction of one or more of these gratings in the MAAT budget would greatly aid the final exploitation of the instrument -- allowing for sub km~s$^{-1}$ precision measurements of the radial velocities of all but the faintest of sources (particularly around the strong H$\alpha$ absorption / emission line).

\subsection{Brown dwarfs and planetary mass objects with MAAT@GTC}

Brown dwarfs were first confirmed unambiguously 25 years ago \citep{Rebolo1995} and nowadays they constitute a well recognized population in open clusters, young moving groups, star-forming regions and the solar neighborhood. Their properties provide a natural link between stars and planets. More recently, planetary mass objects have been recognized free-floating in the field and also as companions to stars. Their masses overlap between those of brown dwarfs and planets, and their formation mechanism is not yet understood.

The Euclid and the Vera Rubin LSST surveys will provide the next major source of discoveries of brown dwarfs and planetary-mass objects during this decade. Spectroscopic reconnaissance of the candidates detected with those surveys will be in high demand. A strong synergy between Euclid, Vera Rubin LSST and MAAT@GTC presents itself to discover new \textit{bona fide} brown dwarfs and planetary-mass objects.

\begin{figure}[htb]
\centering
\includegraphics[width=0.45\linewidth]{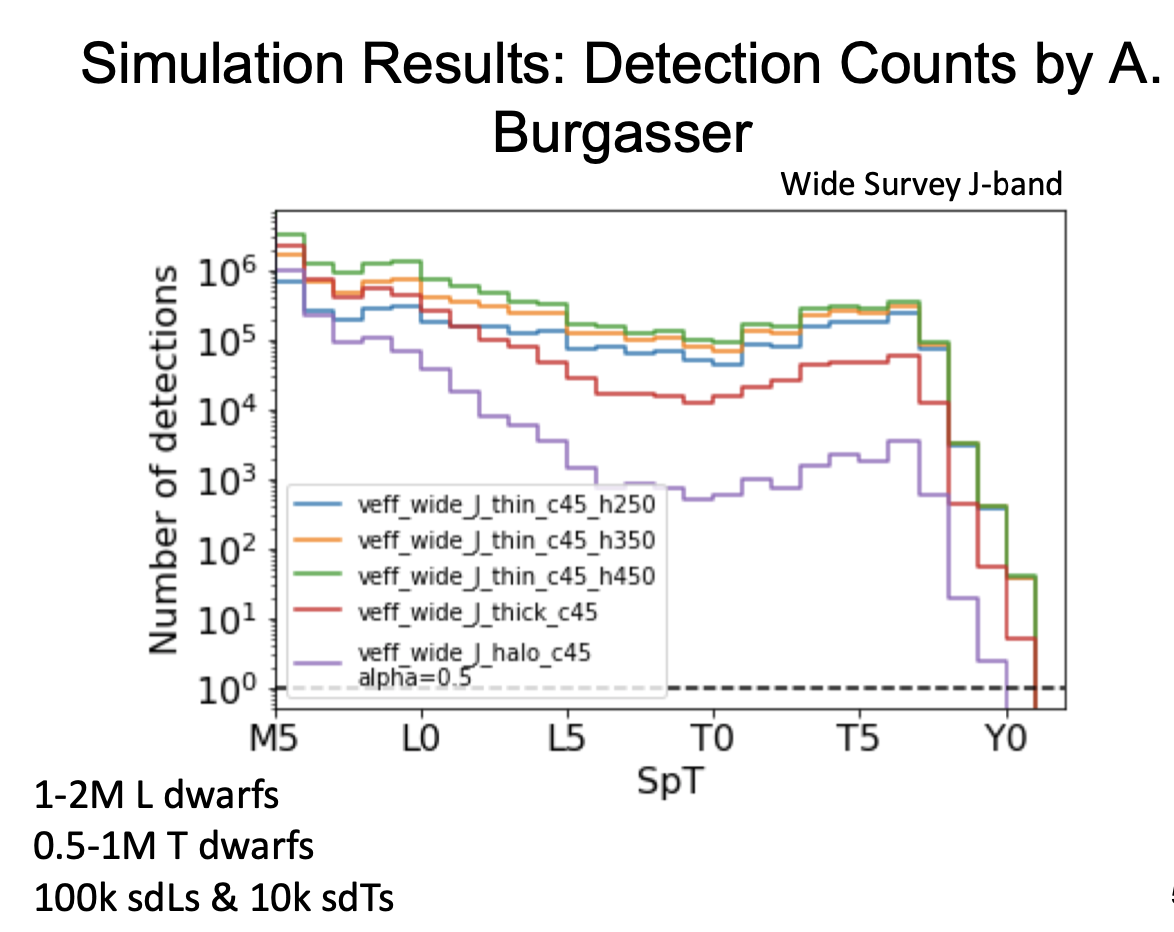}\quad\includegraphics[width=0.45\linewidth]{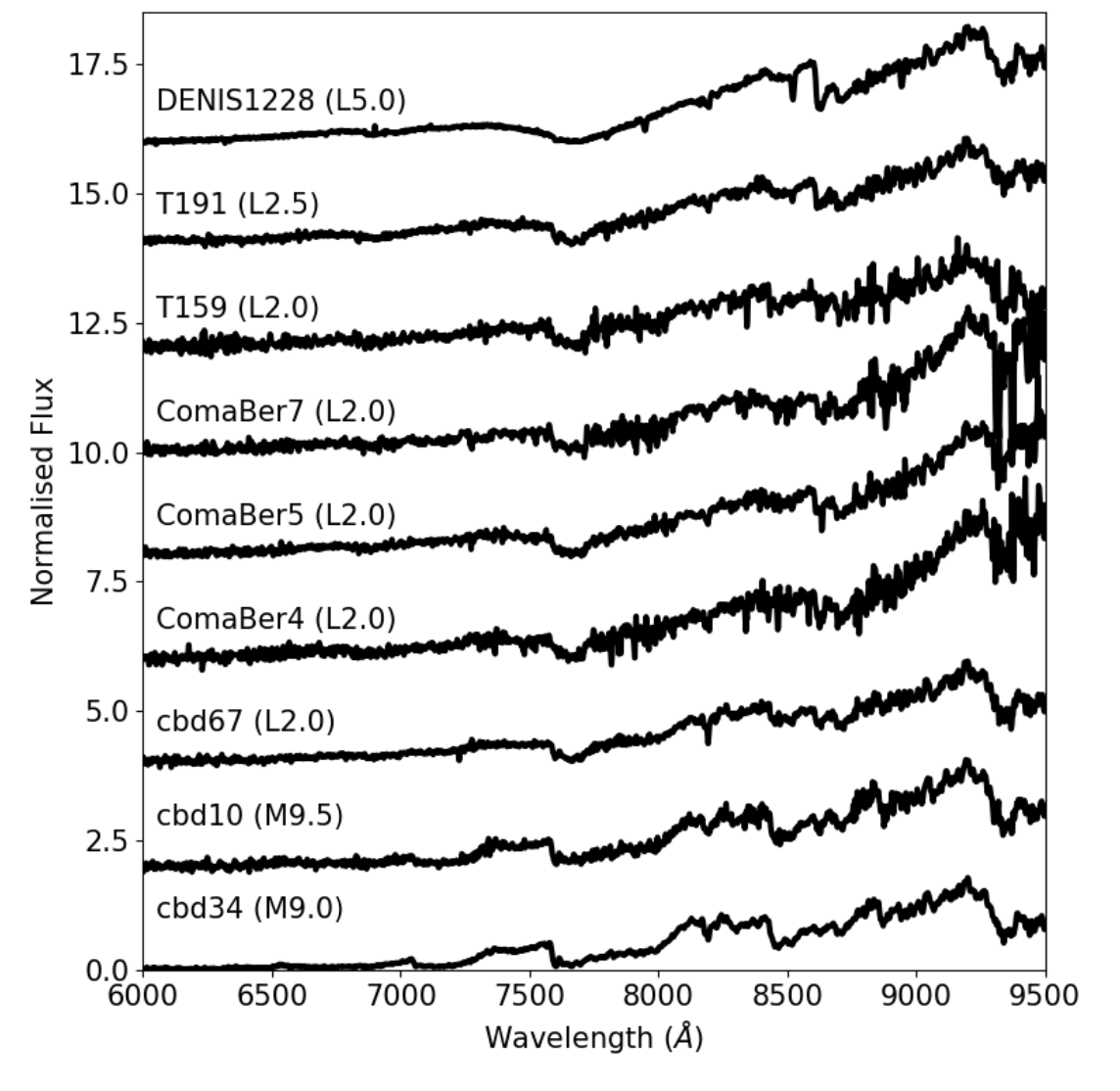}
\caption{Left: Simulated counts of L and T dwarfs detected in the Euclid wide survey in the $J$-band (Solano et al. 2020, MNRAS, \textit{submitted}). Right: OSIRIS@GTC spectra of brown dwarf candidates in the open star cluster Coma Berenices (Mart\'{\i}n et al. 2020, A\&A, \textit{in press}).
}
\label{fig:ege}
\end{figure}

In particular, our simulations of the Euclid performance indicate that over 1.5 millions L and T dwarfs will be detected in the $J$-band (Figure~\ref{fig:ege}, left panel). About 1\% of them could be young planetary-mass objects. Thus, Euclid can potentially increase the known number of brown dwarfs and planetary-mass objects by about 2 orders of magnitude, but it needs complementary observations to characterize them efficiently. We estimate that about 10,000 L and T dwarfs discovered by Euclid may be bright enough for follow-up observations with MAAT@GTC.
Among those, particular attention will be devoted to young objects of suspected planetary masses (about 100), and T subdwarfs (about 20), where we may be able to attempt the detection of primordial lithium abundance, a strong test of Big Bang nucleosynthesis.

A special science case that can be solved by the powerful combination of Euclid and MAAT@GTC data is the determination of the lithium depletion boundary among brown dwarf members in the Coma Berenices open cluster, the second nearest star cluster to the Sun. Using OSIRIS@GTC long-slit mode with the R1000R grating we have observed the faintest candidate L dwarfs in this cluster, investing over thirteen hours of observing time. The spectra allowed us to determine spectral types, measure radial velocities and search for the \ionn{Li}{i} resonance doublet at 6708 \AA\ (Figure~\ref{fig:ege}, right panel). We were able to confirm the cluster membership of four brown dwarfs, but we failed to detect lithium. Using evolutionary models for brown dwarfs we impose a lower limit of 550 Myr to the age of the cluster (Mart\'in et al. 2020, A\&A, \textit{in press}). In order to be able to find the resurgence of lithium in this open cluster we need to probe about one or two magnitudes deeper. Fainter brown dwarf candidates in Coma Berenices will be furnished by the Euclid wide survey, and we need to enhance the spectroscopic sensitivity of OSIRIS to search for lithium in those
objects in a reasonable amount of observing time. Given the increase in spectral resolution provided by MAAT@GTC, we estimate that lithium could be detected using the R300R grating instead of the R1000R grating with the one arc second slit. The gain in sensitivity needs to be quantified with detailed simulations.

The LDB might also be attempted in the Praesepe open cluster, with an age similar to Coma Berenices but further away and denser \citep{Wang2011A&A}. For example, we could test our recent discovery of an ultracool close binary member of Praesepe composed of a star close to the hydrogen-burning limit orbited by a brown dwarf at about 62 mas (Lodieu et al. 2020, \textit{submitted}). We could also envision an astrometric and radial velocity monitoring of some known brown dwarfs at different ages, for example in open clusters
(e.g. Pleiades, Praesepe, Hyades), star-forming regions, and young moving groups.

Our final remark is that MAAT@GTC is a very welcome addition to OSIRIS which would arrive in a very timely period when we expect an enormous increase in the number of new brown dwarf and planetary-mass candidates that will need spectroscopic confirmation.

\subsection{Synergy with GOTO and HiPERCAM@GTC on La Palma}

MAAT@GTC offers great synergy with the Gravitational Wave Optical Transient Observer (GOTO) on La Palma \citep{Dyer2018}. GOTO can instantaneously image a 40 square degree field of view to a depth of approximate 20th magnitude. The facility is fully robotic and optimised for fast response to LIGO-VIRGO alerts, with typically a few tens of seconds delay between the GW alert being issued and GOTO starting to take optical data of the field. The plan is eventually for GOTO to issue its own alerts of potential optical counterparts to GW transients, which could then be fed to OSIRIS for follow-up spectroscopy on the GTC. Since the typical positional accuracy of GOTO at the sensitivity limit is of order $1\arcsec$, MAAT@GTC is perfectly suited to this task, as attempting to acquire GOTO candidates on the existing OSIRIS long slit would waste valuable seconds and risk the transient fading below the detection threshold. An additional advantage of using MAAT for GOTO follow-up spectroscopy is that information would be obtained on the immediate environment of the GW source in its host galaxy, such as the stellar population, local star-formation rate and redshift (see Section~\ref{sec:maatgw}). It would also be highly desirable for the GTC to be equipped with a Rapid Response Mode (RRM), similar to that employed at the ESO VLT\footnote{\href{https://www.eso.org/sci/observing/phase2/SMSpecial/RRMObservation.html}{https://www.eso.org/sci/observing/phase2/SMSpecial/RRMObservation.html}}. The RRM at the GTC would automatically interrupt observations to slew to the GOTO candidate, with the capability to be on source and exposing within a minute of receipt of the GOTO alert. Preliminary discussions have already taken place between the GOTO team and GTC staff about implementing RRM at the telescope.

MAAT@GTC also offers great synergy with the high-speed, quintuple beam camera HiPERCAM\footnote{\href{http://www.gtc.iac.es/instruments/hipercam/hipercam.php}{http://www.gtc.iac.es/instruments/hipercam/hipercam.php}} on the GTC \citep{Dhillon2018}. The current plan is to mount HiPERCAM {\em permanently} on the GTC using a new mini-derotator on one of the unused Folded Cassegrain stations, making HiPERCAM a perfect tool to monitor known variable sources and perform follow-up photometry of transient sources discovered by surveys. MAAT would offer complementary spectroscopic observations at low-intermediate resolutions, perfect for spectral characterisation of the sources and performing radial-velocity curve studies, for example, at an increased resolution and sensitivity than with the current long-slit mode of OSIRIS.

\section{Instrument Overview}
\label{sec:tech}

In this section we describe MAAT in detail and the interfaces with OSIRIS. We provide its technical specifications and a complete description of the envelope, optics 
layout and parameters, overall throughput and performance, data simulations and pipeline overview, as well as the observation scheme with OSIRIS+MAAT. The engineering work 
presented in this proposal has been completed in close collaboration with the staff at GRANTECAN (see Section~\ref{sec:maatmanag}). The results of this study demonstrate that the construction of 
MAAT is feasible and meets the technical requirements for its installation on OSIRIS.

\subsection{MAAT on OSIRIS}
\label{sec:maatosiris}

\begin{figure}[htb]
\centering
\includegraphics[width=0.9\linewidth]{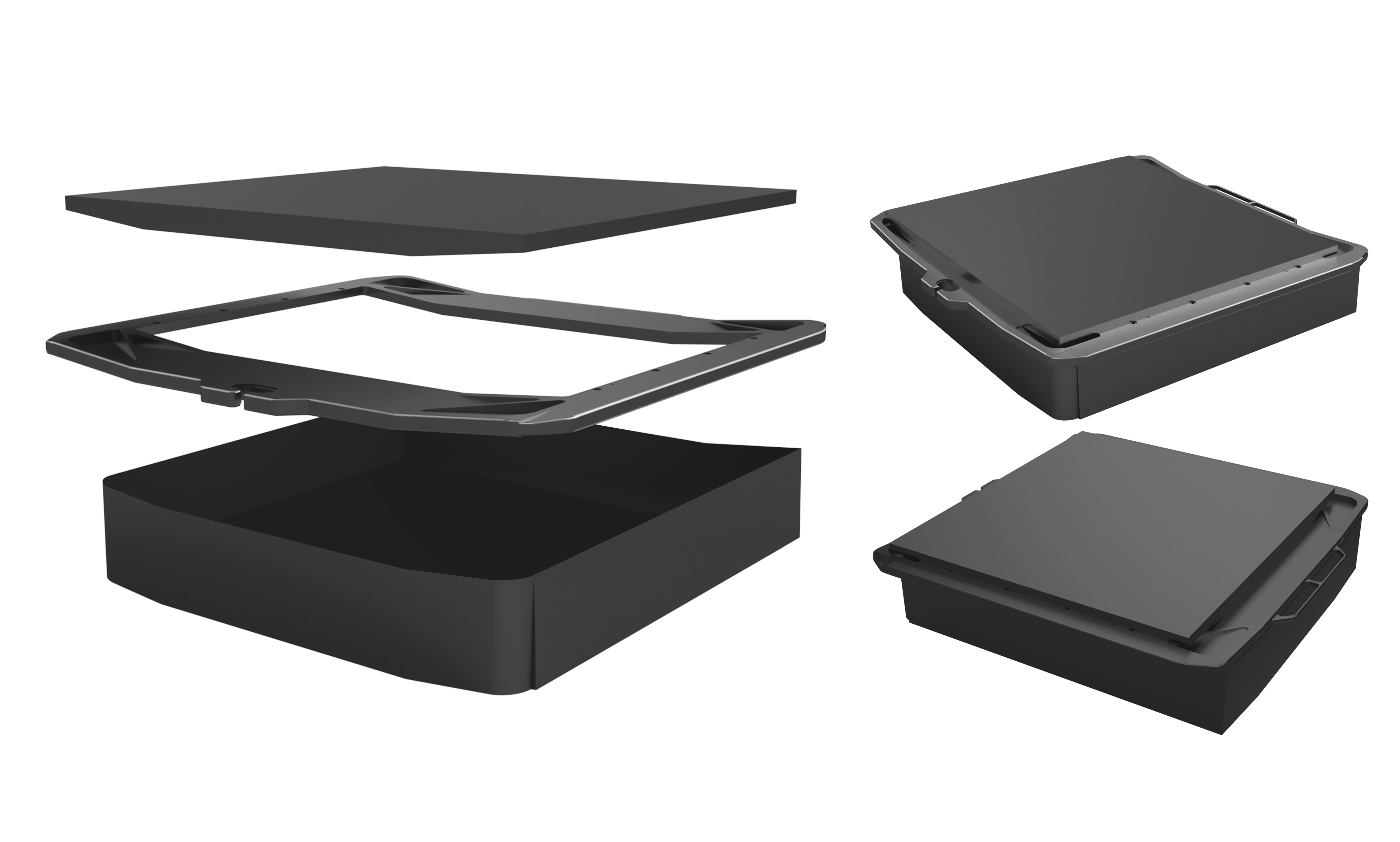}
\caption{A realistic overview of the MAAT module based on the exhaustive space mechanical and interfaces study done by the GTC staff. The edge of the MAAT module, accommodating the envelope, is identical as that of any OSIRIS mask. Here the OSIRIS MOS mask is shown, as an example, only as a reference of the overall shape of the MAAT module. Top and bottom directions point to the secondary M2 and to OSIRIS, respectively. The module has one entrance hole (not shown) for the IFU pick-off mirror on its surface pointing to M2. The surface pointing in the direction of OSIRIS will have the slit aperture. The module space dimensions are $x: 405$ mm, $y: 441$ mm, $z:$ 112 mm; see more details on the envelope study in Section~\ref{sec:maatenvelope}.} 
\label{fig:maatmodule}
\end{figure}

A realistic representation of the MAAT module is shown in Figure~\ref{fig:maatmodule}. This figure displays the space envelope of MAAT as a result of the exhaustive mechanical and interfaces study done by the GTC staff (see Section~\ref{sec:maatenvelope} for details and dimensions). The edge of the MAAT module, accommodating the envelope, is identical to that of any OSIRIS mask. All IFU optical elements are located inside the MAAT module, which will be located in the space equivalent to six mask-frame slots. Thus, when the IFS mode of OSIRIS is required for observing, the MAAT module will be inserted into the telescope beam as if it were a slit-mask. A pick-off mirror inside the module directs the light from the focal plane through fore-optics and onto the mirror slicer, and mirror elements, to reformat the focal plane into the pseudo-slit that passes the light into the rest of the OSIRIS spectrograph (see Section~\ref{sec:maatoptics} for the optics layout and parameters). We want to emphasise that there are no moving parts, neither cables nor electronics, inside the MAAT module, i.e. MAAT is an optical module that, to all effects, is \textit{seen} by the OSIRIS control system as another slit-mask frame. See Section~\ref{sec:maatobs} for the details of observing with OSIRIS+MAAT.

\begin{figure}[htb]
\centering
\includegraphics[width=0.7\linewidth]{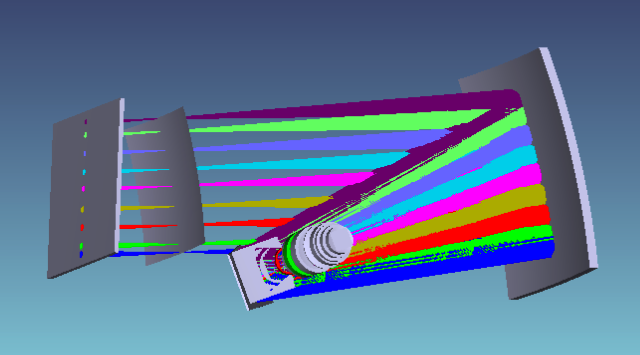}
\caption{OSIRIS optical layout in spectroscopy mode. The OSIRIS optical design is based on a focal reducer with a collimator-camera configuration, which provides an internal pupil of $\sim80$~mm in diameter. The OSIRIS collimator is an off-axis ellipsoidal mirror of conic constant $-0.94$ and a 1240~mm focal length. Its vertex is located 1290.64~mm away from the telescope focal surface and, displaced off axis by 228~mm relative to the telescope optical axis. An all-refractive camera is used to focus the collimated beam on a CCD detector. The camera consists on 9 lenses defining 6 optical elements, all spherical surfaces. The camera effective focal length of 181~mm. The last lens is the dewar window. Between the collimator mirror and the camera, a flat mirror is located to fold the optical path in order to fit the instrument to the envelope. See \citet{Cobos2003} for more details.}
\label{fig:osirisoptics}
\end{figure}

\begin{figure}[htb]
\centering
\includegraphics[width=0.7\linewidth]{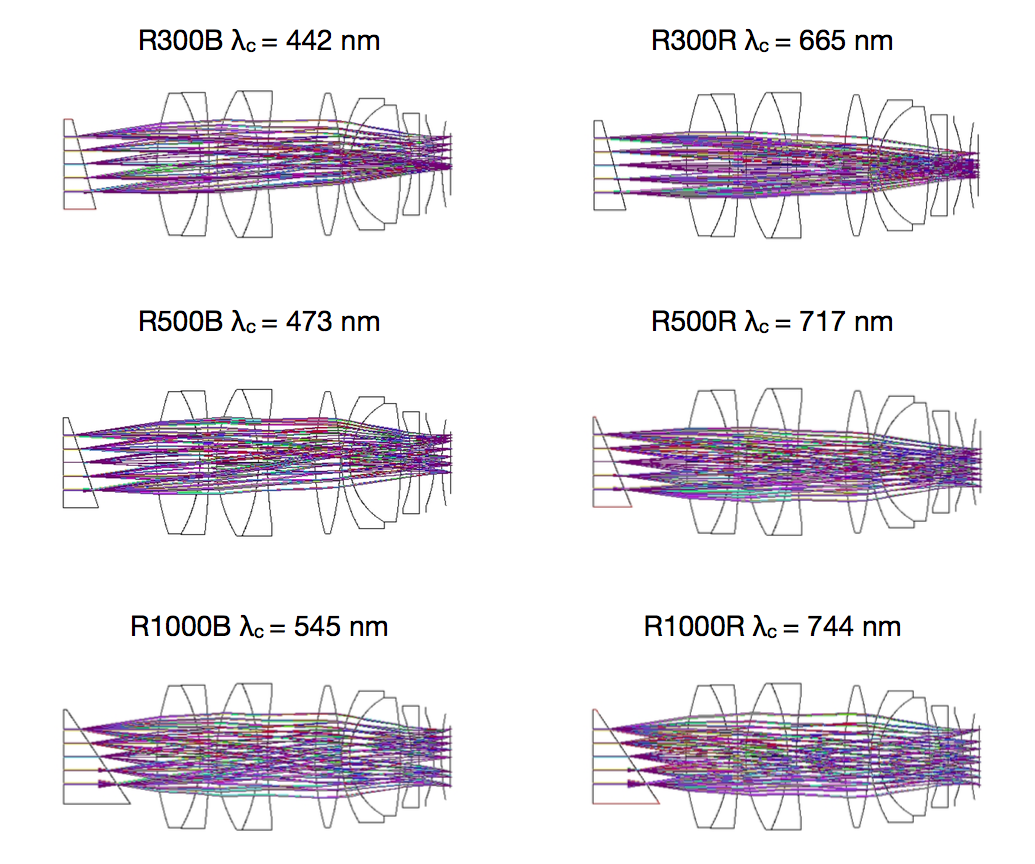}\quad\includegraphics[width=0.7\linewidth]{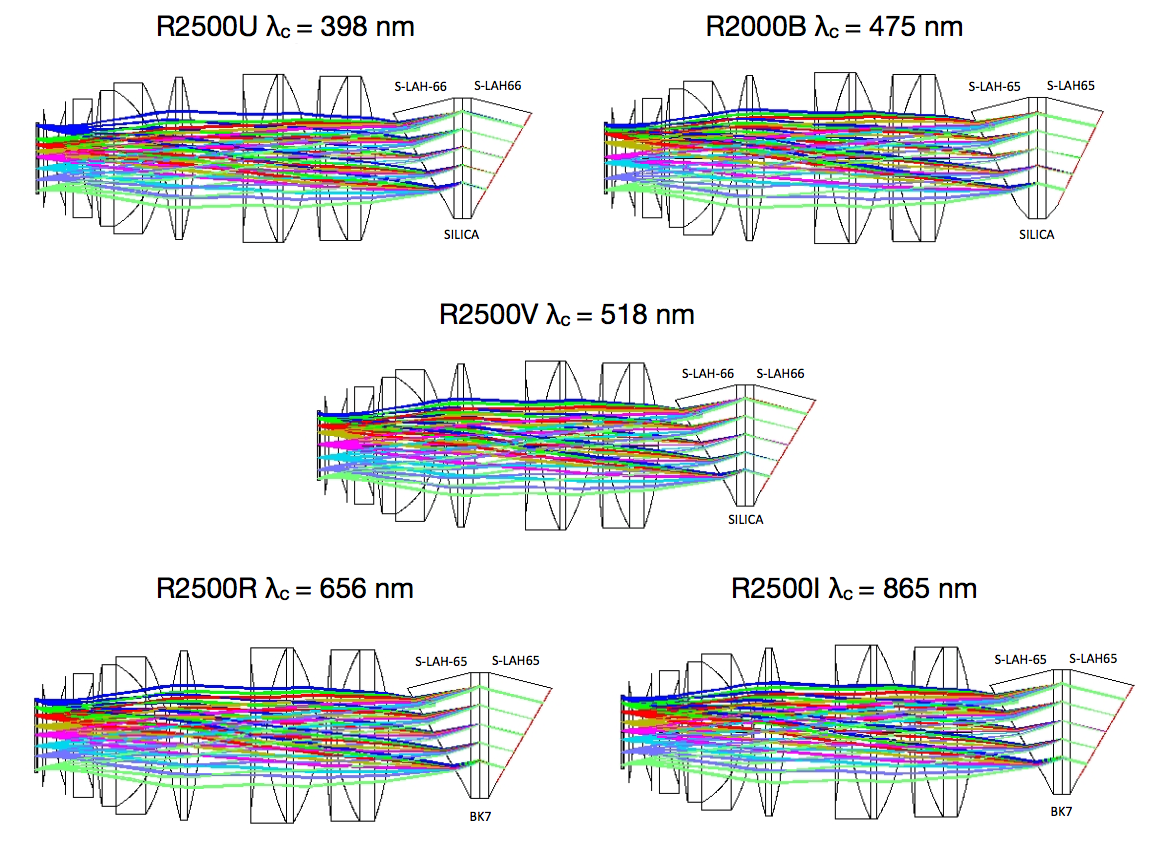}
\caption{Top: Zemax ray-tracing models of the OSIRIS suite of Grisms.
Bottom: The same for the suite of VPHs.}
\label{fig:osirisgrismsvphs}
\end{figure}

The concept of the IFU proposed for OSIRIS has been performed by our collaborator and optical scientist Robert Content. This has demanded a detailed study and in depth understanding of the OSIRIS spectrograph, both as designed (by studying its Zemax ray-tracing optical design, see Figure~\ref{fig:osirisoptics}) and as built (see below). We also had the collaboration of Ernesto S\'anchez-Blanco (OpticalDevelopment), who implemented in Zemax all OSIRIS Grisms and volume-phase holographic gratings (VPHs), since we initially had available only the OSIRIS optical design for imaging, but not for spectroscopy. Figure~\ref{fig:osirisgrismsvphs} shows the Zemax modeling of the OSIRIS suite of Grisms and VPHs respectively, which were integrated in the Zemax design model of OSIRIS for the MAAT optical preliminary conceptual study.

\begin{figure}[htb]
\centering
\includegraphics[width=0.8\linewidth]{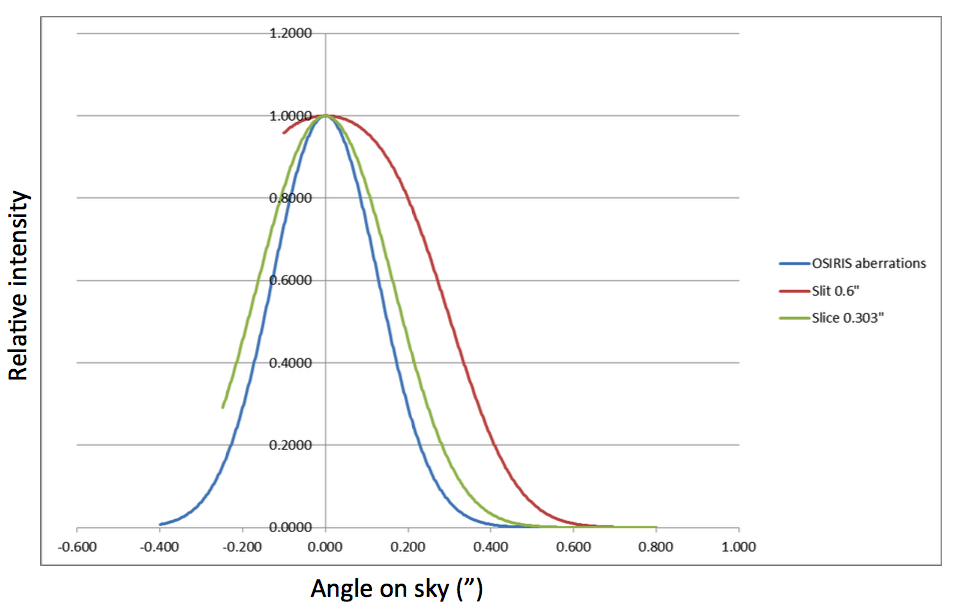}
\caption{PSF of OSIRIS for the MAAT slicer width of $0.303''$ as compared with the $0.6''$ wide long-slit. OSIRIS aberrations were assumed to be a Gaussian $0.3''$ FWHM. This is larger than the as-designed aberrations given by Zemax to take into account mechanical errors as alignments.}
\label{fig:osirispsf}
\end{figure}

MAAT will enhance the resolution power $R$ of OSIRIS by 1.6 times with respect to its $0.6''$ wide long-slit thanks to the $0.303''$ width of each mirror slice. This is a preliminary estimate and will depend on the correction of the spectrograph aberrations that can be corrected by the IFU re-imaging mirrors, especially the local defocus and astigmatism along the slit.
Figure~\ref{fig:osirispsf} shows the OSIRIS PSF for the slicer as compared to the long-slit. All the eleven OSIRIS Grisms and VPHs will be available to yield broad spectral coverage with moderate resolution (R=600 up to 4100) in the spectral range  360--1000 nm. Table~\ref{tab:maatresol} lists the main parameters of all Grisms and VPHs for the new IFS observing mode of OSIRIS+MAAT.

\begin{deluxetable*}{lcccccc} 
\small
\tablecaption{Resolutions and spectral ranges available with OSIRIS+MAAT for all available Grisms and VPHs\label{tab:maatresol}}
\tablehead{
\colhead{ID}  & \colhead{$\lambda_c$ (\AA)} & \colhead{$\Delta\lambda$ (\AA)} & \colhead{D (\AA/pix)} & \colhead{$R$ (LS$^1$)} & \colhead{Peak Efficiency} & \colhead{Type}
}
\startdata
R300B  &	4405  &	3600-7200  &	2.60  &	     575 (360)	  & 70$\%$ & Grism	 \\
R300R  &	6635  &	4800-10000  &	4.02  &	     560 (348)  & 70$\%$	   & Grism	 \\
R500B  &	4745  &	3600-7200  &	1.87  &	   	860 (537)  & 68$\%$	   & Grism	 \\
R500R  &	7165  &	4800-10000  &	2.58  &	   	940 (587)  & 67$\%$	   & Grism	 \\
R1000B  &	5455  &	3630-7500  &	1.13  &	   1630 (1018) & 65$\%$	   & Grism	 \\
R1000R  &	7430  &	5100-10000  &	1.40  &	   1795 (1122) &	65$\%$	   & Grism	 \\
R2000B  &	4755  &	3950-5700  &	0.46  &	   3465 (2165) &	87$\%$	   & VPH	 \\
R2500U  &	3975  &	3440-4610  &	0.33  &	   4090 (2555) &	70$\%$	   & VPH	 \\
R2500V  &	5185  &	4500-6000  &	0.44  &	   4025 (2515) &	80$\%$	   & VPH	 \\
R2500R  &	6560  &	5575-7685  &	1.56  &	   3960 (2475) &	80$\%$	   & VPH	 \\
R2500I  &	8650  &	7330-10000  &	1.73  &	   4005 (2503) &	80$\%$	   & VPH     \\
\enddata
\tablenotetext{1}{Resolving power for the OSIRIS $0.6\arcsec$ long-slit (LS) mode.}
\end{deluxetable*}

\begin{figure}[htb]
\centering
\includegraphics[width=0.25\linewidth]{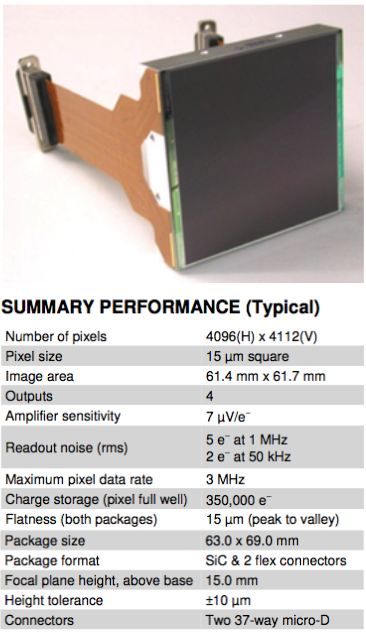}\quad\includegraphics[width=.6\linewidth]{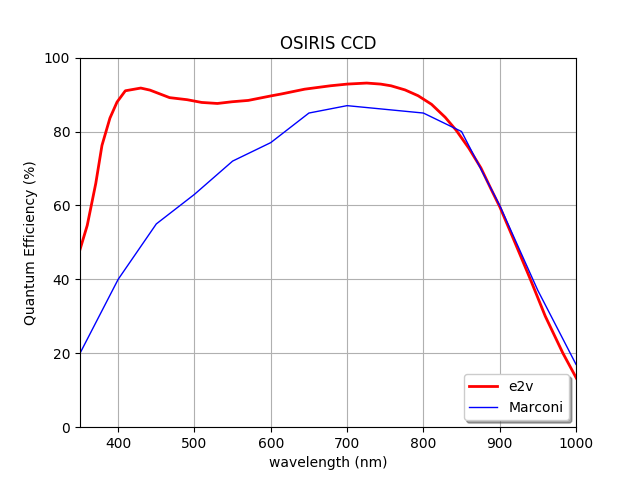}
\caption{Left: Characteristics of the new OSIRIS e2v CCD231-84 detector. Right: Spectral response of the new OSIRIS e2v CCD231-84 (red) compared to the current Marconi CCD44-82 (blue).} \label{fig:osirise2v}
\end{figure}

The proposed IFS observing mode for OSIRIS with MAAT will take advantage of the expected significant increase in the overall OSIRIS efficiency due to its relocation at the GTC Cassegrain focus (a net gain of $\sim10\%$ is expected in the total throughput by the removal of the tertiary M3), and to its new monolithic e2v $4k \times 4k$ detector. For our studies we have adopted the CCD231-84 Back-illuminated Scientific Sensor with $4096 \times 4112$ pixels, each of 15 $\mu$m. See Figure~\ref{fig:osirise2v} with the characteristics of this new OSIRIS e2v detector and 
its spectral response compared to the two Marconi CCD44-82 ($2048 \times 4096$ pixels) mosaic (with 37 pix binned gap between them) currently installed on OSIRIS. MAAT has been conceived and designed to take advantage of both actions, which will be completed by the end of 2021.

\subsection{MAAT envelope study}
\label{sec:maatenvelope}

\begin{figure}[htb]
\centering
\includegraphics[width=0.9\linewidth]{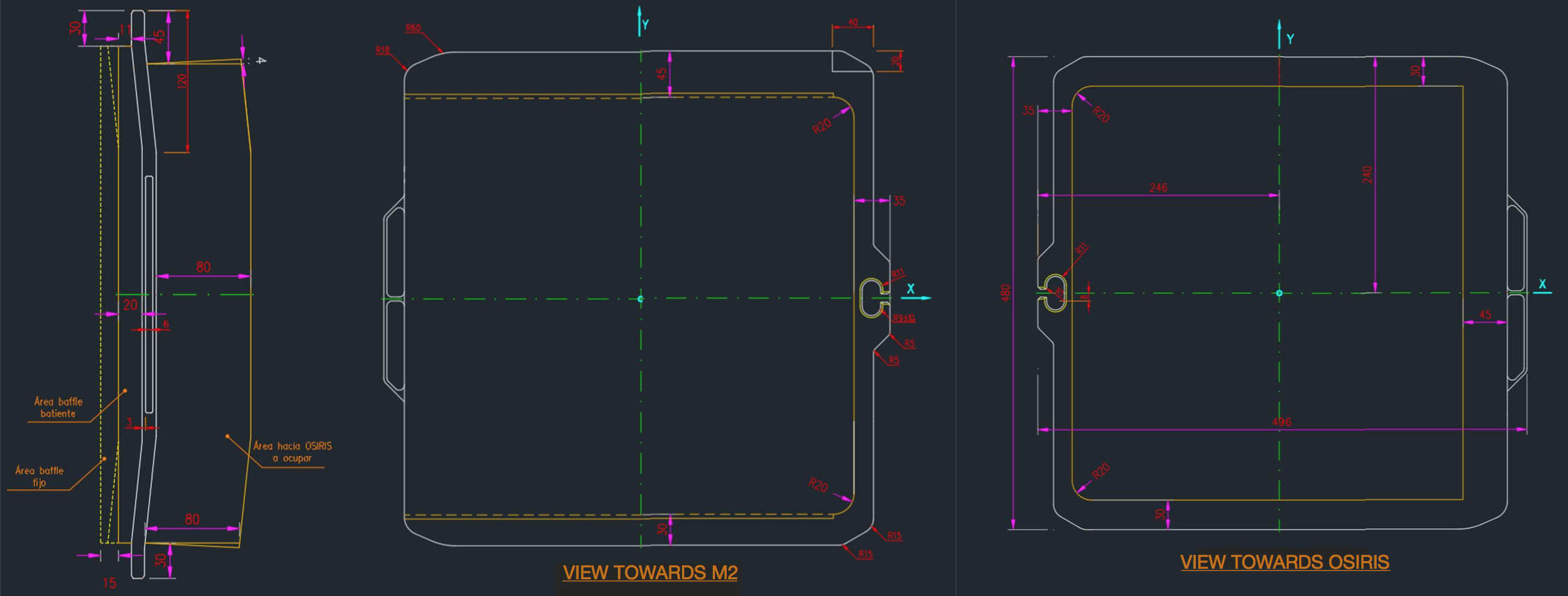}
\caption{Left: Side view of the MAAT module, as seen from the OSIRIS window, V2, that accesses the mask Loader mechanism. Middle: Back view of the MAAT module, as seen towards the secondary M2. Right: Front view of the MAAT module, as seen towards the OSIRIS collimator. (Credit: GTC).}
\label{fig:maatenvlopedim}
\end{figure}

We summarise in Table~\ref{tab:maatenvspec} the main MAAT envelope technical specifications. The 2D drawing with the detailed dimensions of the space envelope of the MAAT module is displayed in Figure~\ref{fig:maatenvlopedim}. All details are also available in the \textit{.step} file provided by GTC; Figure~\ref{fig:maatmodule} shows a 3D view.

\begin{figure}[htb]
\centering
\includegraphics[width=0.9\linewidth]{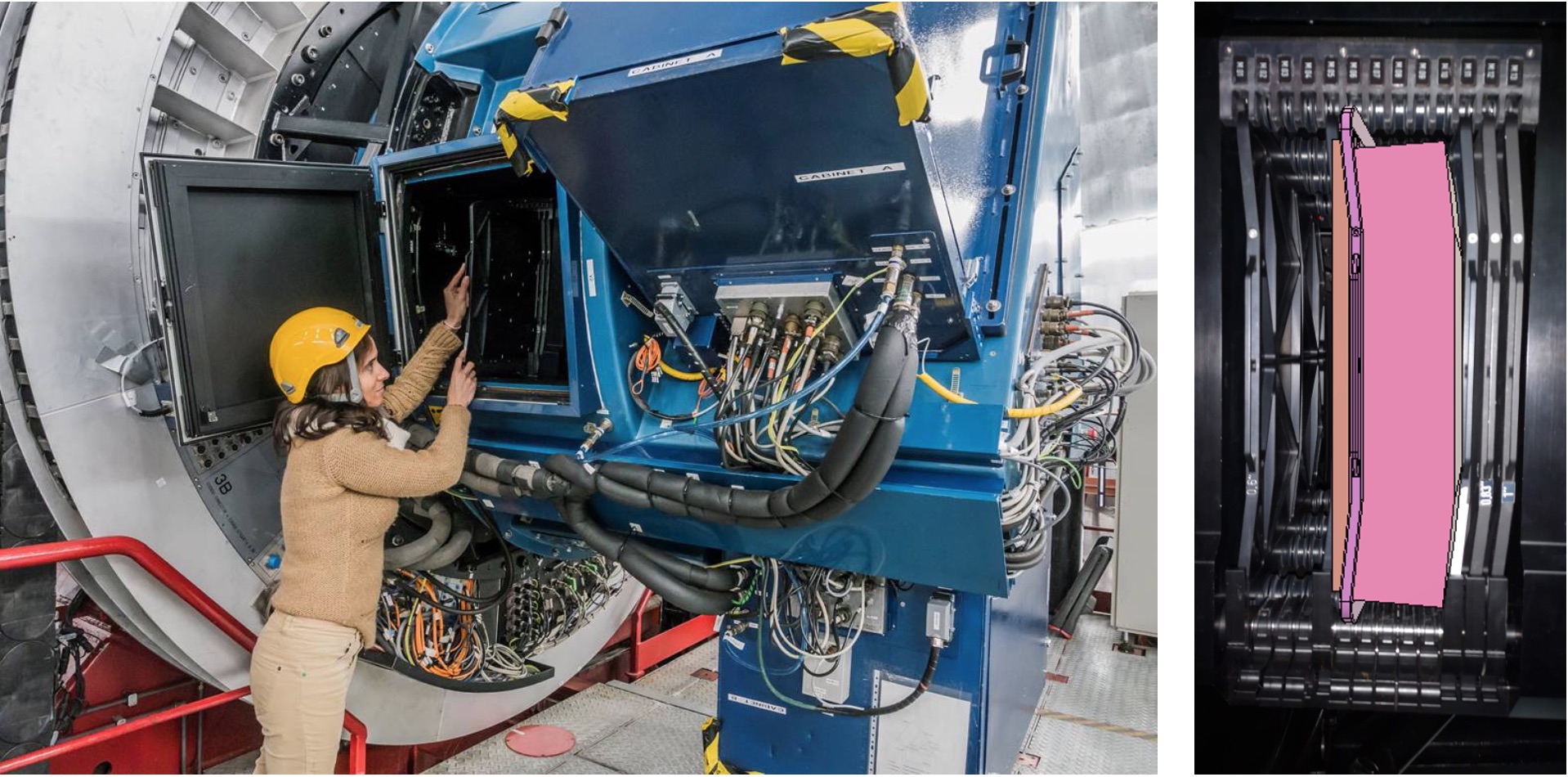}
\caption{Left: OSRIS mask exchanged operation as being done by a GTC staff member. Right: Mock-up
image of the MAAT module inserted in the OSIRIS Cassette structure. (Credit: GTC).}
\label{fig:maatgtcstaff}
\end{figure}

\begin{deluxetable}{rcc} 
\small
\tablecaption{MAAT envelope technical specifications\label{tab:maatenvspec}}
\tablehead{
\colhead{Parameter} & \colhead{Value} & \colhead{Note}
}
\startdata
Spectrograph focal plane radius\tablenotemark{1} & 1814 mm & Edges toward the telescope \\
Focal plane sag along slit\tablenotemark{2} & 11.07 mm & Edge to slit centre \\
Slit length\tablenotemark{3} & 400.2 mm & Ignoring curved focal plane \\
Angle slit vs. OSIRIS optical axis\tablenotemark{4}  & 4.16 $\deg$ & As in the OSIRIS Zemax \\
Space envelope & X: 405 mm & \\
               & Y: 441 mm & \\
               & Z: 112 mm & \\
Weight\tablenotemark{5} & 10 Kg & Maximum weight load of the MAAT module (TBC) \\
\enddata
\end{deluxetable}

\begin{figure}[htb]
\centering
\includegraphics[width=0.75\linewidth]{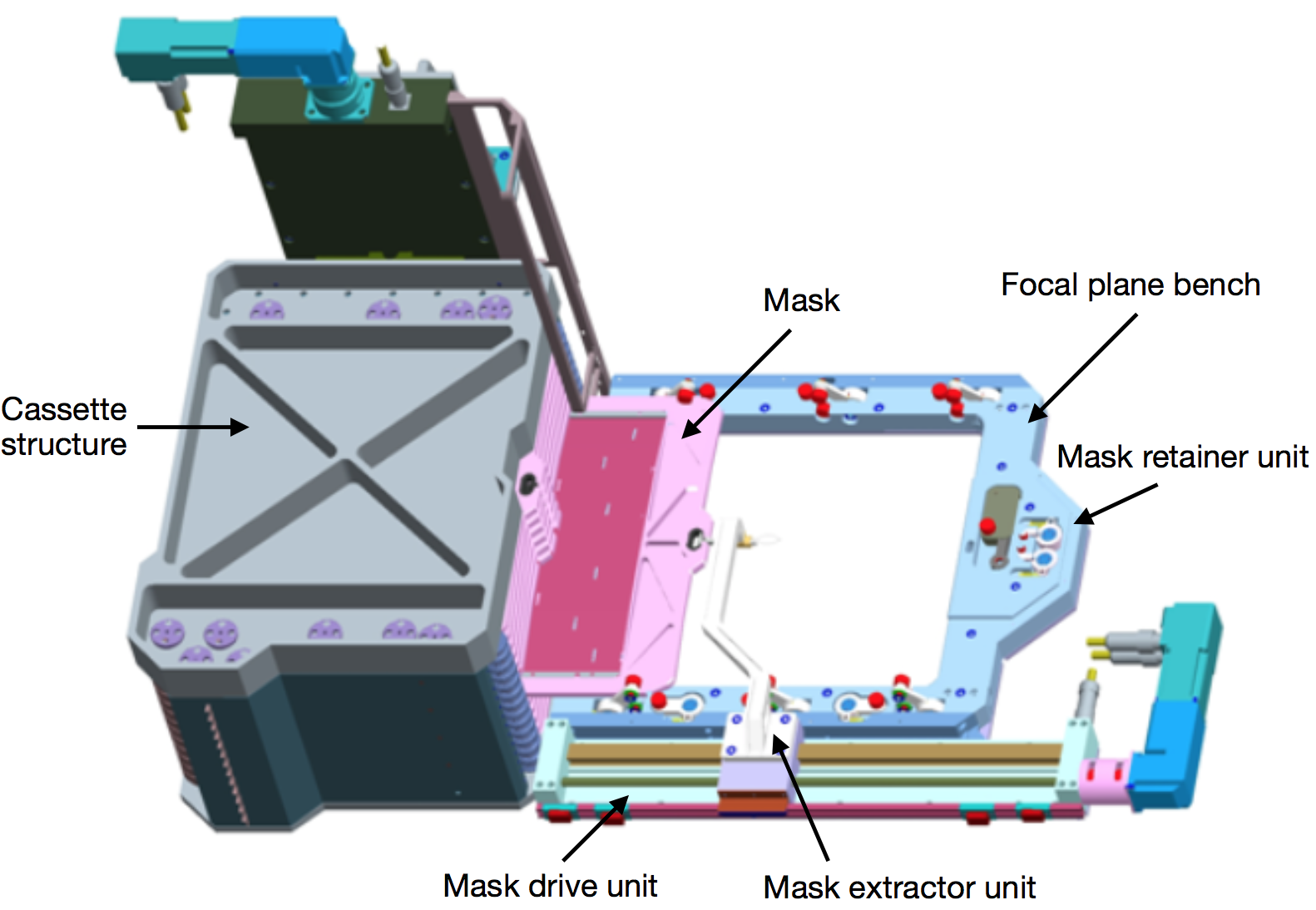}\quad\includegraphics[width=0.75\linewidth]{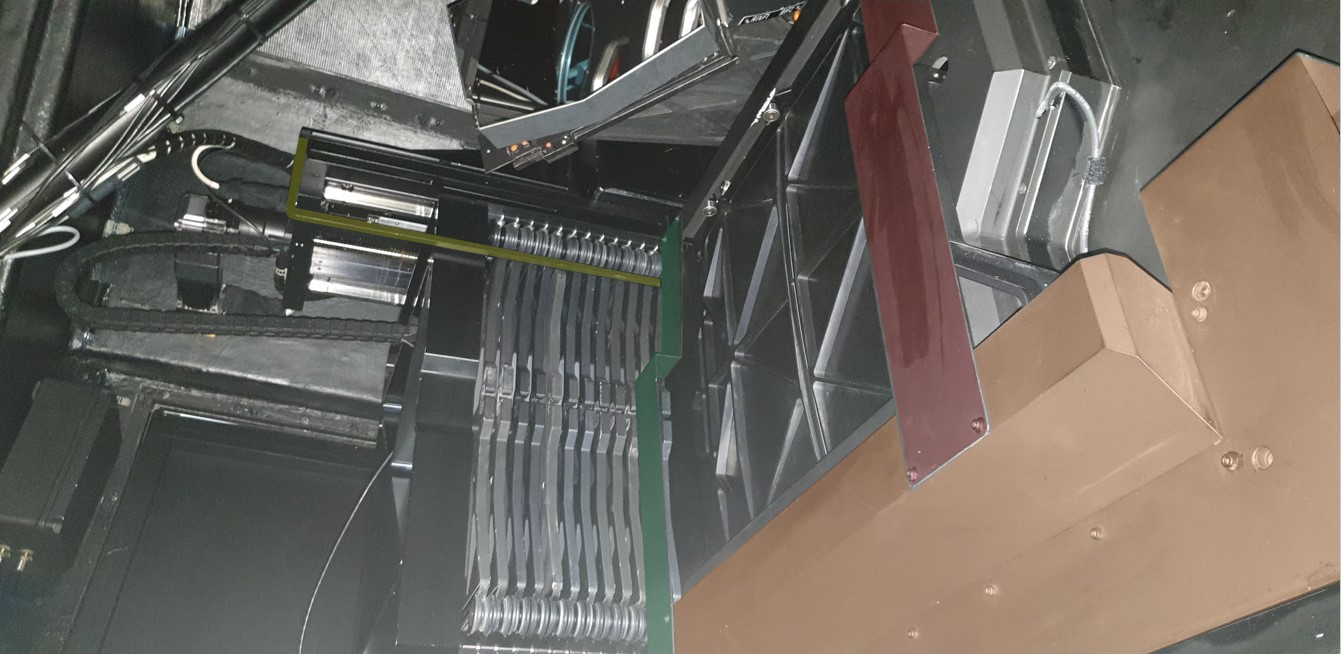}
\caption{Top: Model view of the OSIRIS Slit Subsystem without baffle, consisting of a Cassette structure (on the left in gray), with a capacity of 13 masks and a Charger, on the right in blue, which inserts the mask (in pink) selected by the control on the plane focal length of the telescope (Credit: GTC). The MAAT module will be inserted in the Cassette and occupies the equivalent to 6 masks (see text).
Bottom: Picture of the OSIRIS Cassette and focal plane assembly as built. We highlight in color different metal pieces: yellow for the masks edge protection, green and red 
baffle for mechanical support and as baffle (see text for more details). To all effects, the MAAT module will be inserted in the focal plane of OSIRIS when required as if it were another long-slit mask (Credit: GTC).}
\label{fig:maatmasksys}
\end{figure}

\begin{figure}[htb]
\centering
\includegraphics[width=0.4\linewidth]{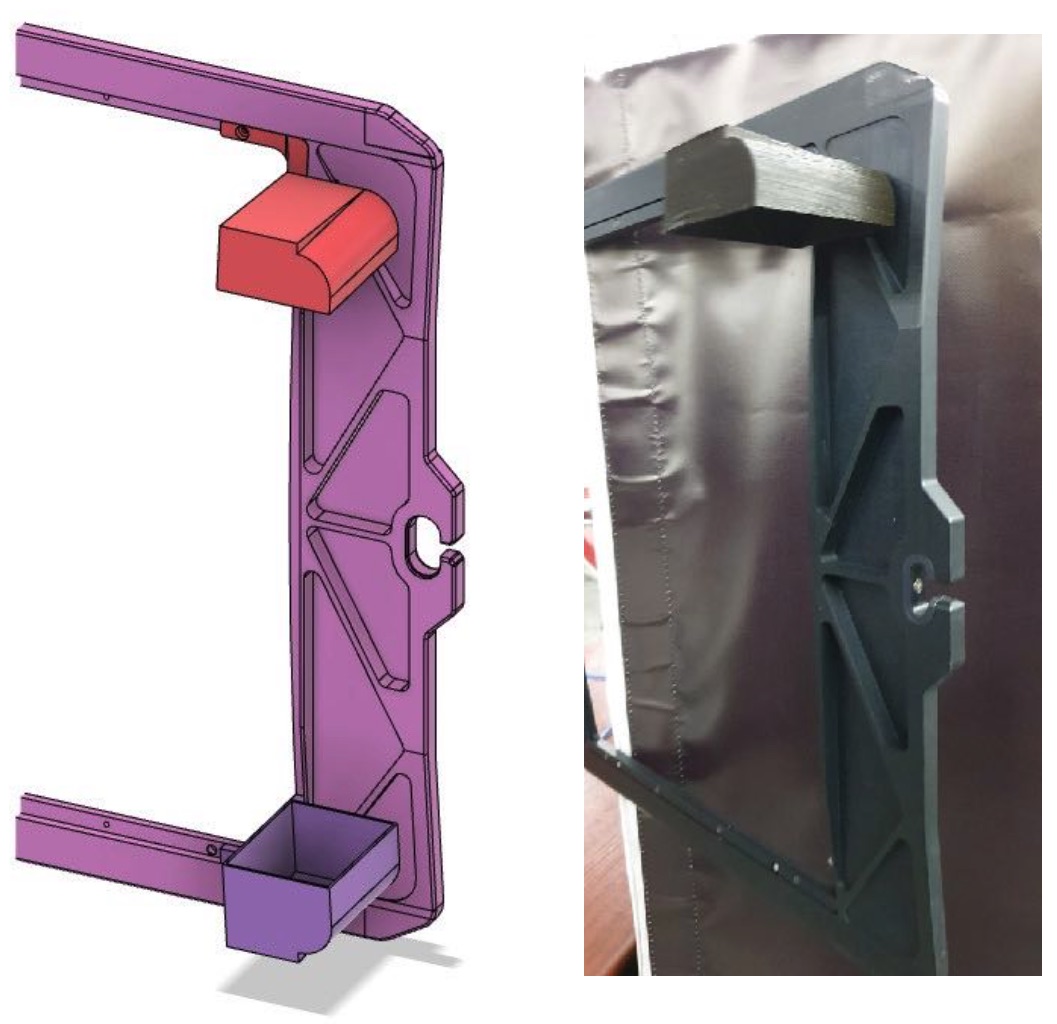}\quad\includegraphics[width=0.8\linewidth]{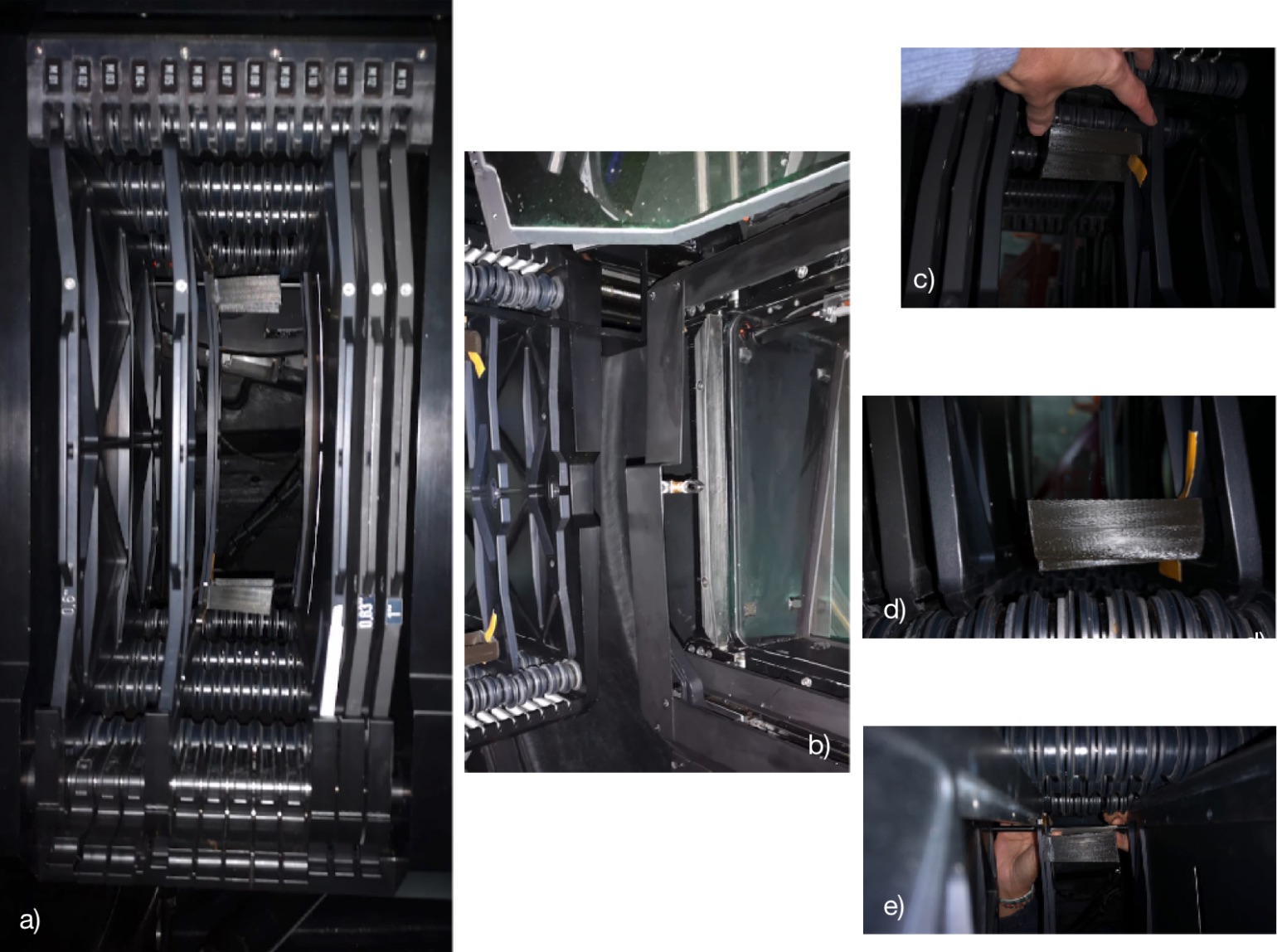}
\caption{Top: 3D printed dummies built for the \textit{in situ} testing of the MAAT envelope on OSIRIS (Credit: GTC).
Bottom: \textit{in situ} testing of the dummy of the MAAT module on OSIRIS: a) MAAT envelope dummy placed in the Loader mask, b) baffle to redesign in order to accommodate the subsystem mask frame plus MAAT envelope, c) width of the MAAT envelope, d) free space between the rollers and the lower part of the MAAT envelope (orientation $rotB = 80^\circ$), and e) free space between the upper part of the MAAT envelope and the masks edge protection (orientation $rotB = 80^\circ$). All these pictures have been taken with orientation of $rotB = 80^\circ$. (Credit: GTC).}
\label{fig:maatprint3dtests}
\end{figure}

The MAAT module will be inserted by the GTC staff (see Figure~\ref{fig:maatgtcstaff}) inside the OSIRIS Slit Subsystem shown in Figure~\ref{fig:maatmasksys}. The MAAT-module space envelope takes up the space equivalent to 6 masks in the OSIRIS masks Charger (see Figure~\ref{fig:maatgtcstaff}, right panel). The location of MAAT inside the OSIRIS 
Charger would be around the central position, leaving space to the left for 4 masks 
and to the right of MAAT for another 3 masks, i.e. a total of 7 masks will be reserved for OSIRIS standard observation modes (long-slit, MOS, calibrations). 
Figure~\ref{fig:maatmasksys} shows also a picture of the OSIRIS Cassette and focal plane assembly as built. The figure seems to show that in the upper part of the Mask 
Loader mechanism there is no range due to the proximity to the masks edge protection (in yellow color). This element is used to retain the masks when the 
instrument rotates. This hypothesis does not apparently become true when the MAAT envelope is introduced into the OSIRIS focal plane, since the masks edge protection 
seem not to interfere with the focal plane (to be confirmed when insertion tests could be done). Secondly, it appears that the metal baffle (grey colour) does not 
stop the MAAT envelope from being positioned into the focal plane. Finally, the third element is the mechanical support baffle (in green and red colour), which 
certainly avoid the insertion of the MAAT envelope into the focal plane, reason why there would be to redesign a wider structure baffle that would allow the MAAT 
envelope width fit the baffle dimensions. The final confirmation of these three elements is pending until the insertion masks test could be done, which means 
unmounting the mechanical support baffle (green/red colour).

\begin{figure}[htb]
\centering
\includegraphics[width=0.7\linewidth]{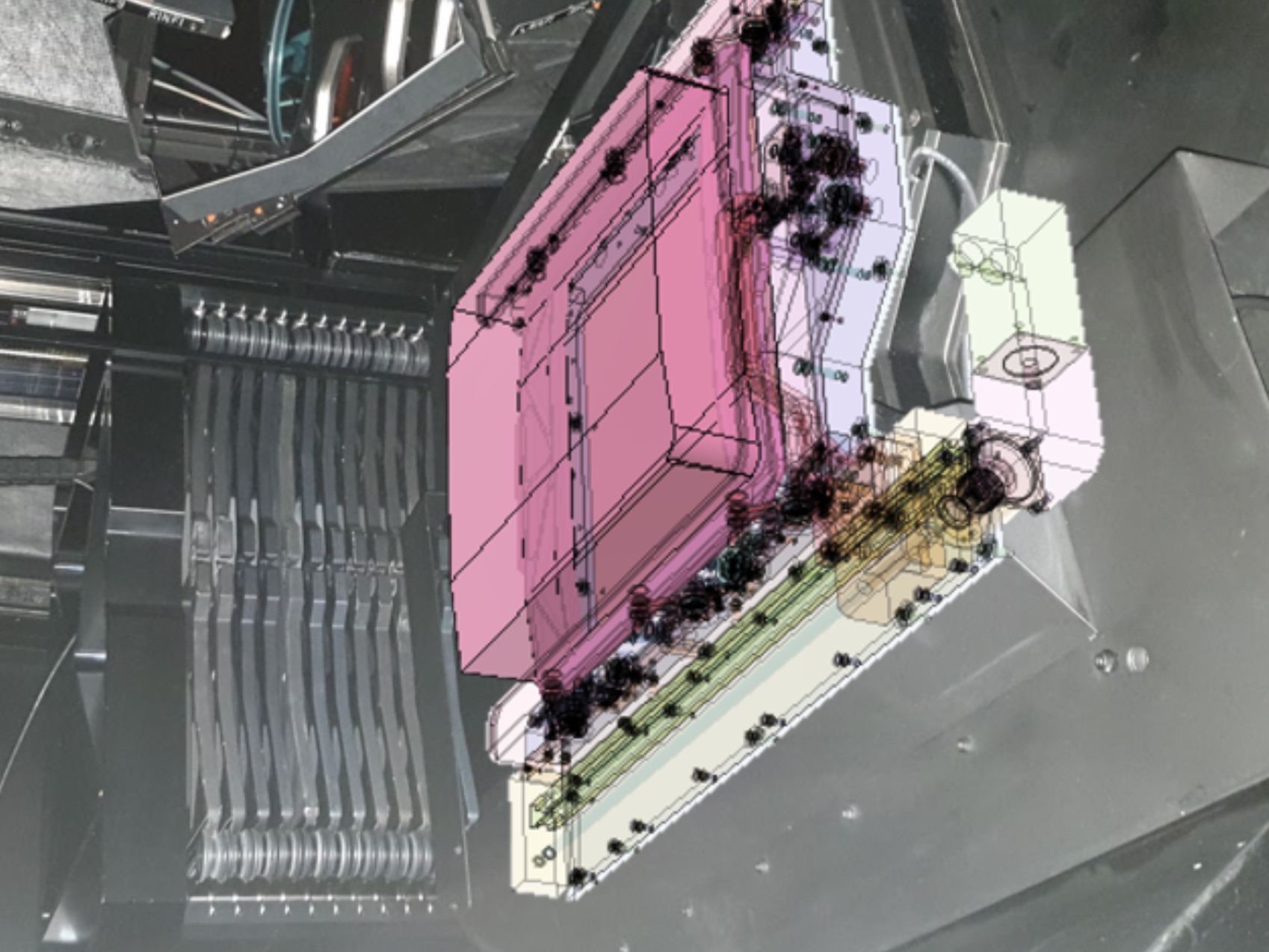}
\caption{The MAAT module has been placed on the OSIRIS focal plane. (Credit: GTC).}
\label{fig:maatinserted}
\end{figure}

The GTC staff has 3D printed two dummies to simulate the MAAT envelope (see Figure~\ref{fig:maatprint3dtests}) to be tested \textit{in situ} on OSIRIS. 
Figure~\ref{fig:maatprint3dtests} shows 
several pictures of the dummy of  MAAT as inserted in its nominal position inside the OSIRIS Cassette. These tests were performed on January 14, 2020 and have 
demonstrated that the MAAT assembly fits well into the Cassette, without interference with the lower and upper shaft cam rollers or the adjacent masks units. The tests need to be completed by inserting the MAAT module into the focal plane of OSIRIS. This initial design of the frontal envelope (towards the secondary M2) cannot be considered with the instrument in the current state (only the back envelope). After the tests which will be carried out at GTC, important modifications to the metallic baffle would be made in order to consider including the frontal envelope. These modifications will only take effect once the mechanical tests are performed \textit{in-situ} in OSIRIS. In a worst-case scenario, only the back envelope will be taken into account.

\begin{figure}[htb]
\centering
\includegraphics[width=0.7\linewidth]{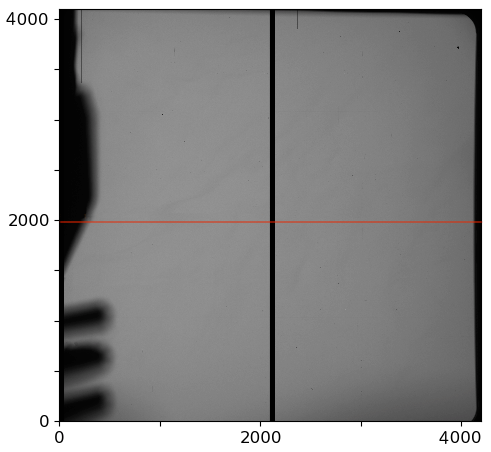}
\caption{Vignetting pattern as seen in an OSIRIS image. The red line marks the nominal position of the OSIRIS long-slit, which we adopted for the IFU pseudo-slit.}
\label{fig:maatvign}
\end{figure}

\begin{figure}[htb]
\centering
\includegraphics[width=0.75\linewidth]{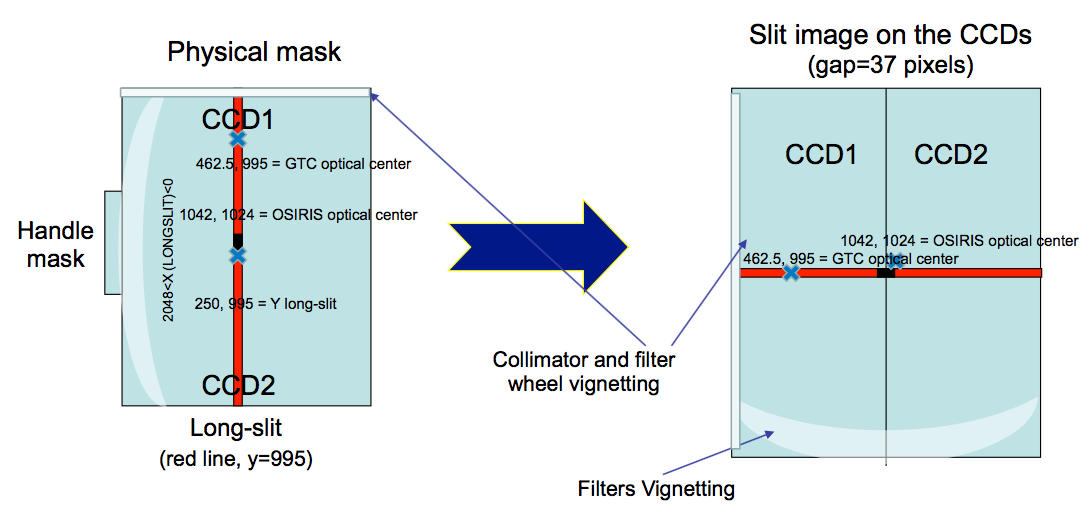}
\caption{Physical long-slit mask correspondence with the OSIRIS CCD (Credit: GTC).}
\label{fig:equivalencia}
\end{figure}

To all effects, the MAAT module will be placed in the focal plane of OSIRIS when required as if it were another mask (see Figure~\ref{fig:maatinserted}). The OSIRIS Mask Loader Mechanism allow us to remove the selected mask, in this case the MAAT module, from the Cassette and place it in the correct position of the Loader. It also allows to put it back in the Cassette when needed. It moves by means of a brushless motor, which coincides with the number 1 motor of the PMAC$\#1$ controller, assigned to the control of the $X-$axis movement. The motor (and drivers in the associated control part) is a Parker model SM232AD. Currently, the motor/encoder rotates at 2000 counts per lap. In principle, the motor should be able to retrieve up to 10 kg, which should be sufficient to handle the  MAAT module. Yet, this will have to be tested with a dummy to make sure that we do not put the motor under stress. Since the separation between two consecutive positions in the Cassette is 20 mm, it is necessary to move the encoder 80,000 beads to move from one discrete position to the next. Thus, the Table of discrete positions with respect to the “home” position will have to be updated to include the MAAT module in replacement for the equivalent to 6 OSIRIS masks.

Another critical aspect of the envelope study has been the understanding of the optical vignetting due to the OSIRIS collimator mirror support (and filter wheel). This feature is not present in the OSIRIS Zemax model as designed. However, it is a well known feature that our colleague Robert Content has studied very much in detail by looking at real 
OSIRIS images to map the vignetting pattern (see Figure~\ref{fig:maatvign}). In this figure we also mark the position of the OSIRIS long-slit which is at the same location as that of the pseudo-slit of the MAAT IFU. In practice, our design maximises the Field-of-View of the IFU, allowing a very minor vignetting in the top-left corner of the MAAT sky footprint (see Figure~\ref{fig:maatfootp}). More details are given in the section below. 
The mapping shown in Figure~\ref{fig:equivalencia} has been very useful 
for our understanding of the exact location of the OSIRIS and GTC optical-axis and the correspondence between the OSIRIS long-slit masks on the OSIRIS CCD,
and thus for the development of the IFU optical concept.

\subsection{MAAT optics layout and parameters}
\label{sec:maatoptics}

\begin{figure}[htb]
\centering
\includegraphics[width=0.9\linewidth]{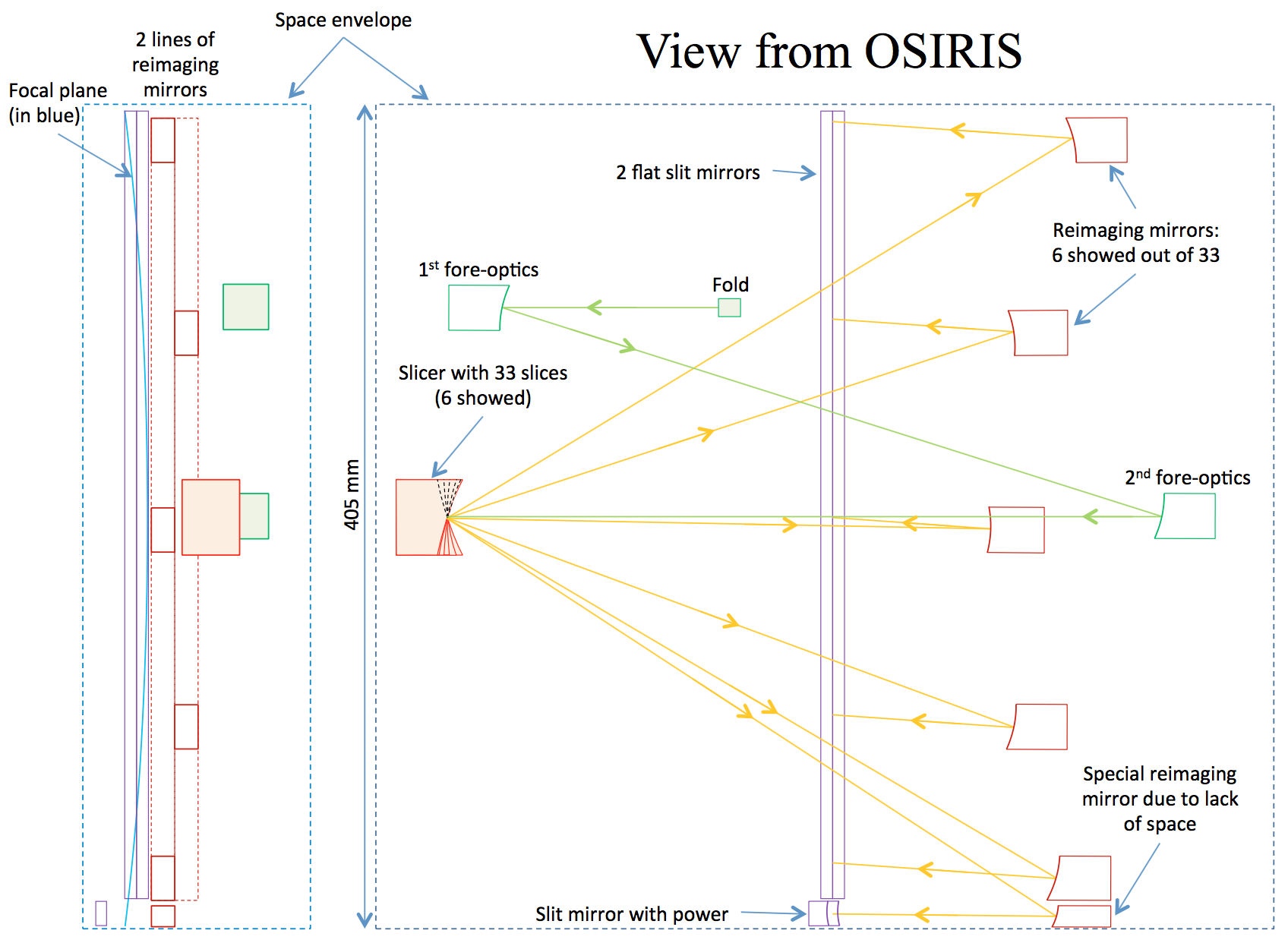}
\caption{Optics lay-out of the proposed IFU concept (only 6 out of 33 slices are shown). The IFU consists of a folded pick-off mirror, fore-optics, the slicing mirror array, re-imaging mirrors, and 2 flat slit mirrors. The positions and sizes of these elements are quite realistic (see left panel for a lateral view). The input field pick-off mirror folds the telescope beam into the fore-optics, which re-images the focal plane of the telescope on the slicing array with the proper magnification. The telescope focal plane is then divided into 33 slitlets by the slicing mirror array, similar to that shown in Figure~\ref{fig:KCWIslicer}. Each mirror in the slicer stack is tilted with a different angle so that the light is reflected in a different direction toward the re-imaging mirrors. Each beam from a slicing mirror hits a different re-imaging mirror, which images the slice on the slit. All the 33 slices are then imaged side-by-side on the slit with a gap between them of 4 pixels on the detector, avoiding cross-talk. The reflection on the 2 flat slit mirrors sends the light in the right direction into the OSIRIS spectrograph.}
\label{fig:maatopticslayout}
\end{figure}

\begin{deluxetable*}{rc} 
\small
\tablecaption{IFU optical parameters\label{tab:maatparm}}
\tablehead{
\colhead{Parameter} & \colhead{Value}
}
\startdata
Number of slices & 33    \\
Free space between slice images along slit & 0.42 mm \\
Slice image length on slit & 11.72 mm \\
Slice image pitch on slit  & 12.14 mm \\
Slice width & 0.8 mm \\
Slicer width  & $0.8 \times 33 =$ 26.4 mm  \\
Slice length  &  37.49 mm  \\
Focal length on slicer  &  544594 mm   \\
Focal ratio on slicer   &  f/52.365   \\
Slit mirror position vs slit focal plane centre (sag/2)     &  5.53 mm   \\
Distance re-imaging mirror to slit focal plane (centre)      &  83.25 mm  \\
Distance re-imaging mirror to slit mirror (centre)           &  $83.25 - 5.53 =$ 77.72 mm   \\
Distance re-imaging mirror to slicer (centre)               &   267.01 mm  \\
Footprint width on re-imaging (centre)                      &   5.27 mm  \\
Footprint length on re-imaging (centre)               &  $5.27 + 11.72 =$ 16.99 mm   \\
Distance re-imaging mirror to slit mirror (top edge) & 118.33 mm \\
Distance re-imaging mirror to slicer (top edge) & 363.8 mm \\
Footprint width on re-imaging (top edge, parallel to slit mirror) & 6.95 mm \\
Footprint length on re-imaging (top edge, parallel to slit mirror) & $6.95 + 11.72 =$ 18.67 mm \\
Minimum size largest re-imaging (2mm chip zone) & 22.67 mm $\times$ 10.95 mm \\
Pitch re-imaging on each row & $2 \times 12.14 =$ 24.28 mm \\
Re-imaging mirror surface shape &  toroidal \\
Typical re-imaging mirror radius of curvature (increase from centre to edge) &  $\sim$150 mm \\
Slice mirror surface shape & spherical \\
Rough estimate of typical slice radius of curvature & $\sim$300 mm \\
\enddata
\end{deluxetable*}

\begin{figure}[htb]
\centering
\includegraphics[width=0.4\linewidth]{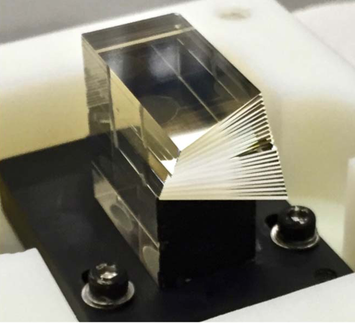}
\caption{A close-up view of the KCWI slicer built at Winlight. This element is made from optically contacted Zerodur. Each of the 24 slitlets is 0.5 mm thick and 14.8 mm tall. The reflective faces are convex with a radius of curvature of 5206 mm, see \citep{KCWI} for more details. The MAAT slicing mirror stack will be very similar but with 33 
slicers each 0.8 mm thick (see Table~\ref{tab:maatparm}).}
\label{fig:KCWIslicer}
\end{figure}

The proposed MAAT optical solution is based on the concept of Advanced Image Slicer (AIS) developed by Robert Content, who has designed the most relevant AIS-IFU optical systems in ground-based 10-m class telescopes and space projects to date. As examples, this includes the designs of the MUSE mirror-slicer IFU \citep{Henault2004}, the KMOS mirror-slicer system \citep{Content2006}, the Gemini GNIRS slicer \citep{Content1998}, and the JWST NIRSpec slicer \citep{Content2000ngst}. For completeness we should also mentioned his Gemini 
GMOS-IFU \citep{Content2000gemini} and Magellan IMACS-IFU \citep{Schmoll2004} designs using fiber-lenslets. We are also in close contact with the company Winlight in Marseille who are well recognised for the optical manufacturing (including opto-mechanics) of the MUSE and KCWI mirror-slicer IFUs (see Figure~\ref{fig:KCWIslicer}), among others, and complete spectrograph systems such as MUSE, DESI, and PFS. The results presented here support the feasibility of MAAT. Naturally this work will have to evolve into proper and formal phases of Conceptual, Preliminary and Final design for the eventual construction of MAAT (see Section \ref{sec:maatmanag}). At this stage of the project we do not address the optomechanical aspects of the IFU optical components and mechanical enclosure, these will be considered in the Preliminary phase of the design that will include the study of tolerances.  
 
The MAAT space envelope and interfaces presented and discussed in Section~\ref{sec:maatenvelope} have been closely taken into consideration to make sure that the entire optical system and their components will fit well into the available space. As mentioned above the MAAT box that contains the AIS-IFU optics will be placed into the input focal plane of OSIRIS, as in the case of the IFU-module for the GMOS spectrograph on Gemini \citep{Content2000gemini}. The IFU consists of an imaging slicer optical system with 33 slices each of $0.303'' \times 14.20''$ (with 4 pixels between slice images along the slit), and a slit length of 8.0 arcmin. The spatial sampling $0.303'' \times 0.127''$ is better than the typical seeing of $0.8''$, which guarantees an appropriate sampling even with $0.303'' \times 0.254''$ by adopting an $1 \times 2$ CCD binning. The slicer width $0.303''$ yields a spectral element of 3 pixels, i.e., a resolution 1.6 times larger than the standard long-slit width of $0.6''$, as discussed in Section~\ref{sec:maatosiris}. 

The IFU consists of a folded pick-off mirror, fore-optics, the slicing mirror array, re-imaging mirrors, and 2 flat slit mirrors. There are in total 6 reflections in the IFU system. The optics layout of the IFU system is shown in Figure~\ref{fig:maatopticslayout} for only 6 slicers. The positions and sizes of the slicer, slit, and re-imaging mirrors are quite realistic.
The input field pick-off mirror is fixed but it can be placed almost anywhere in the focal plane. It folds the telescope beam into the fore-optics which re-images the focal plane of the telescope on the slicing array with the proper magnification. The telescope focal plane is then divided into 33 slitlets by the slicing mirror array, similar to that shown in Figure~\ref{fig:KCWIslicer}. The proposed concept has similarities with that designed by Robert Content for OCTOCAM \citep{Content2018}. Each mirror in the slicer stack is tilted with a different angle so that the light is reflected in a different direction toward the re-imaging mirrors. Each beam from a slicing mirror hits a different re-imaging mirror which images the slice on the slit. All the 33 slices are then imaged side-by-side on the slit with a gap between them of 4 pixels (on the detector) to avoid cross-talks. The reflection 
on the 2 flat slit mirrors sends the light in the right direction into the OSIRIS spectrograph. Note that the 2 flat slit-mirrors in this drawing are incomplete, they have a more complex edge-shape than shown here. The slit beginning and end are determined by the vignetting of the collimator (as discussed in Section~\ref{sec:maatenvelope}). In Figure~\ref{fig:maatopticslayout}, the distance from the slit-end to the space-envelope-edge is about 3 mm at the top and 1.5 mm at the bottom. The input beam on the slit has an angle of about 4$^{\circ}$ due to the tilt of the frame (as that of the OSIRIS long-slit). The \textit{special} re-imaging mirror at the bottom does not need the beam tilt before its powered slit-mirror, which is tilted accordingly. The curved focal plane (pseudo-slit) on the left panel is part a real image and part a virtual image.

We provide in Table~\ref{tab:maatparm} the relevant parameters for all the IFU optical elements of MAAT: slicing mirror array, re-imaging mirrors and slit. The slicing mirrors are all spherical with the same radius of curvatuture. The re-imaging mirrors are toroidal in the proposed concept. Currently we are discussing with Winlight two extreme solutions: 
1) All re-imaging mirrors are spherical with the same RoC (positions of the mirrors give the focus). We expect that this would minimize cost but would degrade the performances both by increasing aberrations and changing the magnification on the detector (different slice images would have different $\arcsec/pixel$ so different width, thus different spectral 
resolution unless the slices themselves have different widths). Worst aberrations (at the edges of the slit) should be quite large; 2) Each re-imaging mirror is toroidal 
with the values of the radii and direction of the toroid different from the others, so there would be $2 \times 33 = 66$ different radii and 33 different directions. This 
would also permit to correct some of the aberrations of the spectrograph. This may increase the cost. The difference between the 2 radii of each 
mirror would be zero in the centre increasing roughly linearly to about $10\%$ at the edge. The average between the 2 radii should be around 130 mm in the centre increasing 
parabolically to about 170 mm at the edge. The goal here is to understand if having an identical shape on many mirrors does not change the costs or that 
it is the complexity of the shape that matters, not the number of identical mirrors. This is important because if the cost is the same for 33 spherical mirrors with 
different RoCs than with the same RoC, it would be a serious improvement to have them all different. Hopefully we will have very soon a detailed answer from Winlight with their trade-off study. This will help us to proceed with the current concept and start the formal Conceptual design phase. Finally, note that we have not studied in detail the
fore-optics: We are considering for now a mirror optical solution. In any case it seems a feasible component from previous experience.

As seen so far, the MAAT IFU slicer uses mirrors and hence we will require very good reflective coating for the entire spectral range $360-1000$ nm. Both, MUSE and KCWI have high efficiency in their slicer IFUs. 
The required transmission should be at least $80\%$ on average. Currently we are using the coating spectral efficiency data provided by Winlight to compute the expected efficiency of the IFU taking into account that there are 6 reflections in the IFU (see Figure~\ref{fig:maatopticslayout}). This gives an average reflectivity $>92\%$ over the whole spectral range. We should consider this a conservative value. Further investigation together with Winlight is going on to optimize the mirror coatings and increase the efficiency.

\subsection{Overall throughput and performance of OSIRIS+MAAT}
\label{sec:maatefficiency}

\begin{figure}[htb]
\centering
\includegraphics[width=0.8\linewidth]{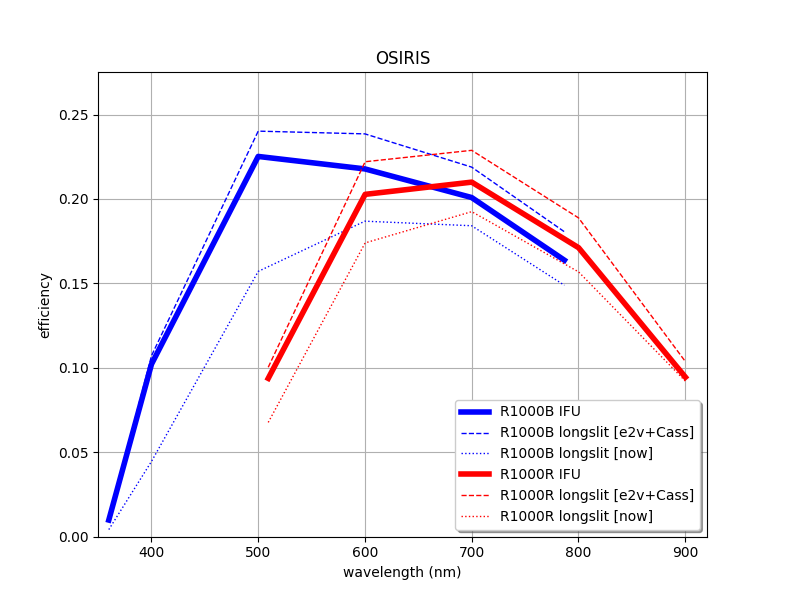}
\caption{Total throughput (atmosphere, telescope, and instrument) of the OSIRIS long-slit and IFU mode as currently mounted on the Nasmyth focus and as it will be on the Cassegrain, which will result in a significant gain in efficiency derived from the new e2v detector and the absence of M3. The curves are shown for the broadband R1000B and R1000R Grisms.}
\label{fig:maatefficiency}
\end{figure}

The overall efficiency of the OSIRIS+MAAT system (IFU mode including the atmosphere and the telescope) has been estimated starting from the results posted in the OSIRIS website measured from a spectrophotometric standard star through a wide slit, as a function of wavelength and for different Grisms and VPHs. The results are displayed in Figure~\ref{fig:maatefficiency} for the broadband R1000B and R1000R Grisms. We compare the expected total throughput of the OSIRIS IFU and long-lit modes  on the Cassegrain with that measured at its current location on the Nasmyth focus. The gain in sensitivity is significant below $\sim6000$ \AA, with a $\sim50\%$ enhancement in the blue domain $\sim3700-5000$ \AA, due mostly to the new e2v detector and the absence of the M3 tertiary. 

\begin{figure}[htb]
\centering
\includegraphics[width=0.8\linewidth]{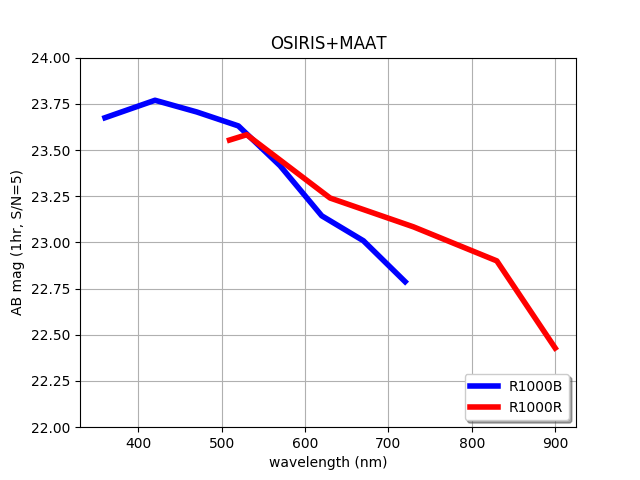}
\caption{OSIRIS+MAAT limiting magnitudes for the broadband R1000B and R1000R Grisms.}
\label{fig:maatlimmag}
\end{figure}

\begin{figure}[htb]
\centering
\includegraphics[width=0.7\linewidth]{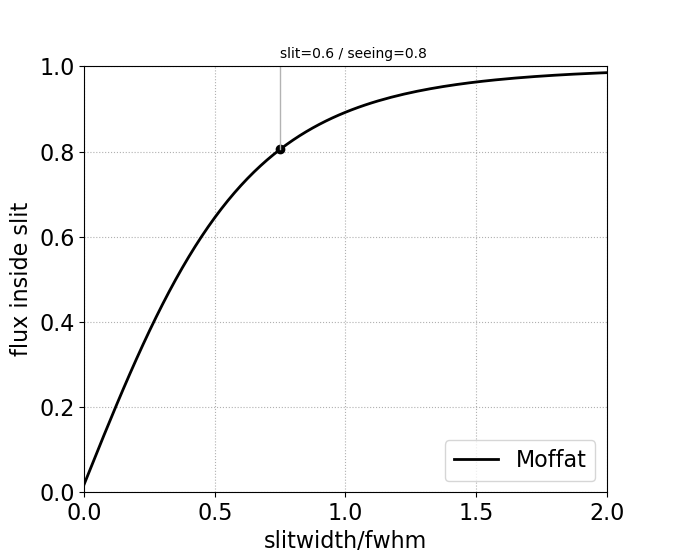}
\caption{Flux inside the slit vs. the slit-width to seeing (fwhm) ratio adopting a Moffatt PSF. For a typical seeing of $0.8\arcsec$ one expects an increased of $20\%$ in 
the total flux when the observation is done with OSIRIS+MAAT as compared to the standard OSIRIS $0.6\arcsec$ long-slit.}
\label{fig:maatgainflux}
\end{figure}

We have used the expected efficiency curve of OSIRIS+MAAT and the OSIRIS long-slit limiting magnitudes provided in the OSIRIS website to predict the performance of the OSIRIS IFU mode for spectral continuum observations of point sources. Figure~\ref{fig:maatlimmag} shows the limiting magnitudes (AB) to reach $S/N=5$ in 1 h of integration time with OSIRIS+MAAT for the broadband R1000B and R1000R (assuming $1\arcsec$ seeing, dark night, and airmass = 1.2). Note that these estimates can be obtained for any of the OSIRIS Grisms and VPHs.

In order to estimate the MAAT limiting flux we have used the MUSE@VLT exposure time calculator with the following assumptions to achieve a $S/N=5$ with 30 exposures of 1800s each: an extended Ly$\alpha$ emitting galaxy with a Sersic profile at redshift $\sim4$, a peak emission line flux of $3\times10^{-18}$ erg s$^{-1}$ cm$^{-2}$ arcsec$^{-2}$ \citep{wisotzki2018}, a line width of 600 km s$^{-1}$, spatial binning of $3\times3$ and integrated over the spectral line profile, seeing of $0.8\arcsec$, dark moon, and PWV=5 mmH$_2$O. 

We want to remark that a clear advantage of IFS as compared to long-slit is the fact that all the flux of the object (point sources or extended features inside the 
IFU FoV) can be collected (see Figure~\ref{fig:maatgainflux}), and hence enhancing the measured $S/N$ and allowing to perform absolute spectro-photometry.

\section{Data simulations}
\label{sec:maatdata}

We have carried out simulations to have an overview of the data reduction process. For this purpose we use MUSE@VLT data cubes as proxies for real targets on sky and then transform them through the instrumental specifications of OSIRIS+MAAT adopting the R1000R Grism to produce raw detector spectral images. The main limitation of this process is that MUSE@VLT does not provide information bluewards of 4750 \AA. 

We illustrate the process with four extragalactic astronomical targets: the kilonova GRB170817A in NGC~4993 as an example of time domain EM-GW counterpart and a point source, the galaxy IIZw40 as an example of a young low metallicity galaxy circumnuclear region, and two individual objects in the galaxy NGC~300: the giant \ionn{H}{ii} region De74 and the SNR S14. These last two cases are already showcased above (Section~\ref{sec:maat}).

\begin{figure}[htb]
\centering
\includegraphics[width=0.8\textwidth]{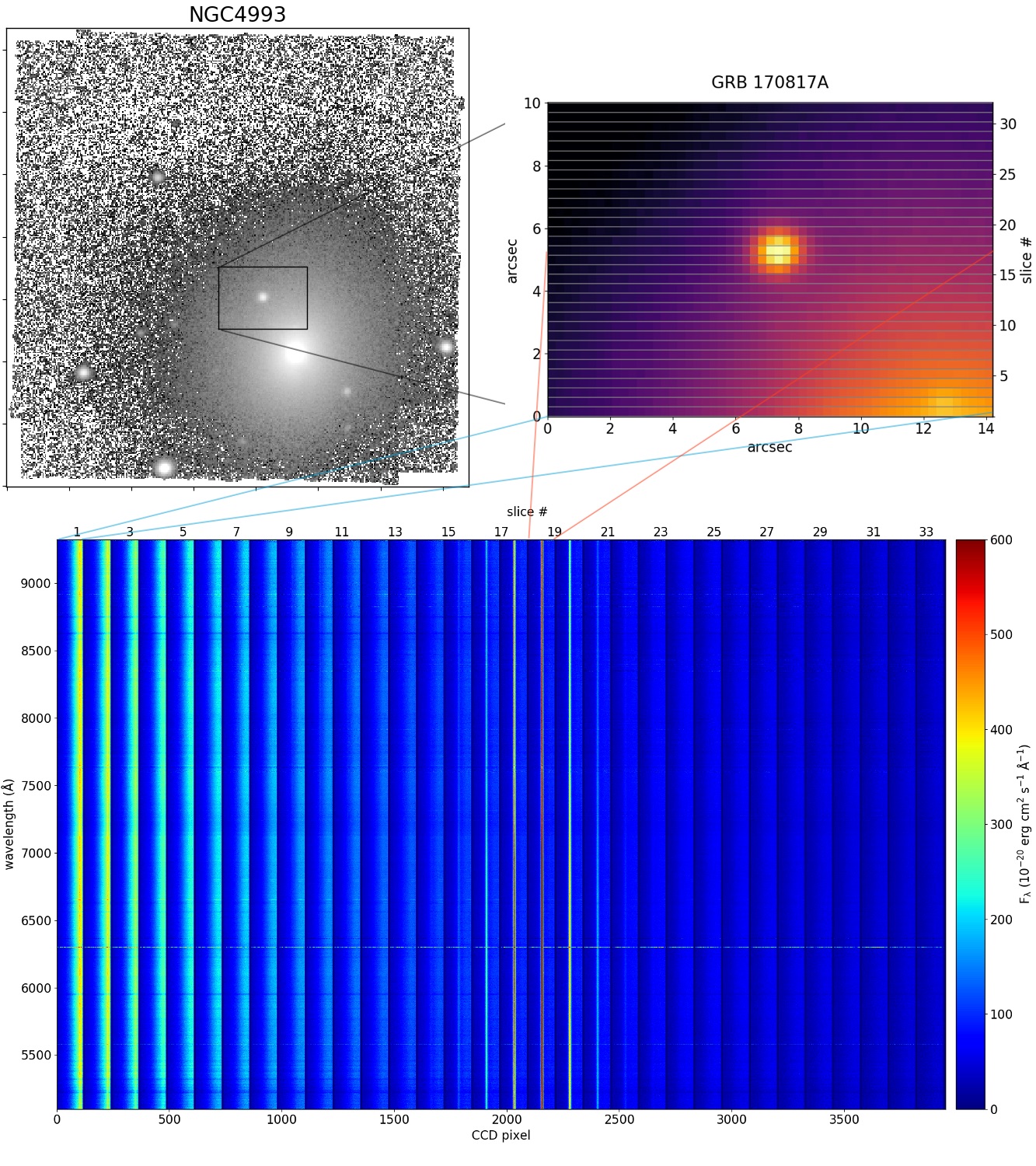}
\caption{MAAT simulation of the kilonova GRB170817A in NGC~4993. The simulations are based on a MUSE archival spectral cube of NGC~4993.}
\label{kiloMAAT}
\end{figure}

The top-left of Figure~\ref{kiloMAAT} shows an image of the galaxy NGC~4993 with the kilonova framed by a box of 14.2\arcsec$\times$10\arcsec\ corresponding to the MAAT FoV. On the right, the kilonova as viewed by MAAT, with 33 slices of 0.303\arcsec\ each in the vertical axis and 14\arcsec\ of 0.254\arcsec\ pixels in the horizontal axis. This input image is then dispersed by the slicer and re-imaged onto the CCD detector. This CCD spectral image is shown in the bottom, where the spectral direction is along the vertical axis and the spatial-slicer direction along the horizontal axis. (Notice that due to the limited blue cutoff of MUSE@VLT these simulations stop at 4750 \AA). In the spatial-slicer direction slice $\#1$, corresponding to the bottom of the image above, is located on the left end of the CCD, and all the slices follow to the right, each separated by 4 CCD pixels from the previous slice, so that the top slice in the image above is placed on the far right end of the horizontal axis in the CCD. This can be clearly followed in this field because the bulge of NGC~4993, seen in the bottom right side of the FoV, is spectrally detected in the left slices of the CCD frame (thin blue lines), while the spectrum of the kilonova centered in the central slices is seen as a point source continuum in the central slices of the CCD spectral frame (thin red lines).

\begin{figure}[htb]
\centering
\includegraphics[width=0.8\textwidth]{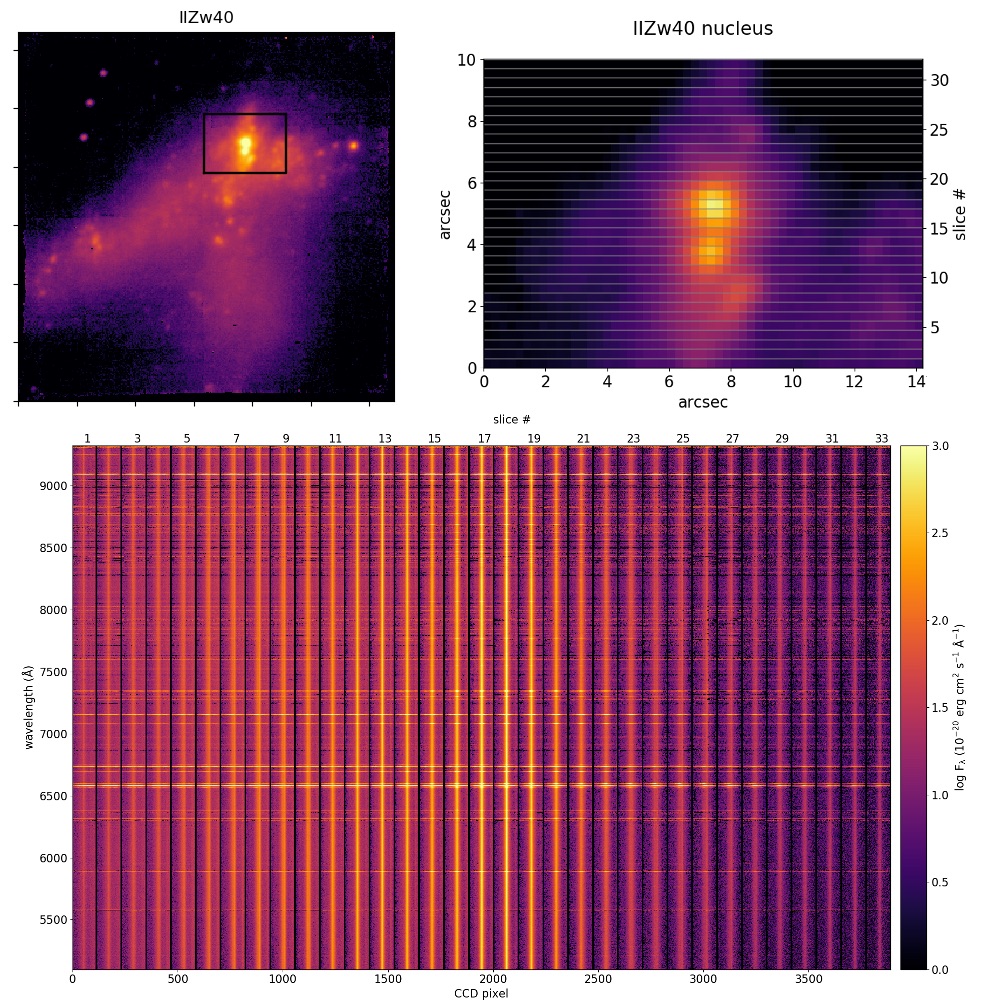}
\caption{MAAT simulation of the circumnuclear region in the low metallicity dwarf galaxy IIZw40. The simulations are based on a MUSE archival spectral cube of IIZw40.}
\label{iizw40MAAT}
\end{figure}

As in the case of the kilonova, Figure~\ref{iizw40MAAT} shows the corresponding images for the case of the low metallicity dwarf galaxy IIZw40. This case has been chosen to highlight the rich emission line spectrum.

The bottom images in Figures~\ref{kiloMAAT} and~\ref{iizw40MAAT} show a typical readout frame from the detector of OSIRIS+MAAT. This file has to go through the reduction pipeline to be calibrated in wavelength and flux and reassembled into a spectral cube fully corrected for instrumental effects.

\section{SESHAT: The reduction pipeline for MAAT}
\label{sec:maatpipe}

The aim is that SESHAT will be as automatic as possible, thus ensuring a smooth, consistent delivery of final reduced and calibrated datacubes ready for the scientific analysis. For this purpose the project already starts with the advantages of being part of a successful and stable instrument, OSIRIS, and those derived from its nature as a slicer, as compared to IFUs based on fibers. Furthermore, there is now a wealth of resources and knowledge available in the community for the calibration and analysis of IFS \citep[see e.g.,][]{delburgo2000,Husemann2013,RGB2015,Sanchez2016,Weilbacher2020}. The final public delivered data cube will be such that for each spaxel at least it will provide: flux, error (accounting for Poisson, readout, etc), and bad pixels flags (cosmic rays, bad CCD columns, vignetting, etc). These properties will be properly propagated from end to end of the pipeline.

Most IFS pipelines already available in the community have in common the main general tasks, and differ in some tasks specific of the associated instrument. Thus it is reasonable that SESHAT will have a ground layer adapted from one of the public available pipelines that have demonstrated its robustness and reliability for (semi-) automatic processing, and then a few tasks specific for MAAT in OSIRIS that will provide a customized catalog of characterization static files for all spectral range configurations available with the OSIRIS gratings.

The reduction includes the following steps: i) bias subtraction, ii) aperture definition and trace, iii) stray-light subtraction, iv) aperture extraction, v) throughput correction, vi) wavelength calibration: arc-line identification and dispersion correction, vii) sky subtraction, viii) cosmic ray rejection, ix) image reconstruction (generation of maps for the continuum, spectral indexes, velocity, etc), and x) differential atmospheric correction, as discussed in \citet{delburgo2000}.

Good wavelength mapping for all available gratings and spectral ranges need to be well studied during commissioning to provide the user with static calibration maps for each OSIRIS grating setup. This may require some user interaction during the reduction process. Cross-talk will be negligible. Absolute flux calibration will be much better than for fiber-based IFS, given the 2D spatial coverage of the slicer. For the sky subtraction one can decompose in continuum + emission lines (use per-slice Line Spread Function), and take into account slice-to-slice variations in flux calibration and the behaviour of OH line groups. This step is more uncertain in extended than in point-like sources because it implies an independent exposure for sky in the case of extended sources that fills the MAAT FoV. But see the ZAP work by \citet{Soto2016}.

All the relevant information will be included in the FITS header to help in automatizing the reduction process. Given that MAAT observing is akin to the long-slit mode of OSIRIS, the FITS header input to the reduction pipeline already has in place place all the keywords defined. Thus for the data frame read from OSIRIS we suggest that only two additional fields should be added: (a) ``MASKNAME = `MAAT' / Multi-object mask name", and (b) ``OBSMODE = `OsirisMAATspectroscopy' / Observation Mode". To this input header of the original file, the pipeline will append all the information relevant concerning all the steps carried out during the reduction and calibration.



We will also provide a \textit{quick-look} pipeline with basic functions to be used by the Support Astronomers during the observing runs. 

\section{Observing with OSIRIS+MAAT}
\label{sec:maatobs}

We propose the MAAT module to be permanently mounted on the OSIRIS Cassette from 2022 onwards, thereby enabling immediate response to transient targets and the execution of standard programs at any given time. Yet, it can be removed from the Cassette as any other OSIRIS mask if necessary since it uses the same mask-frame interface.

MAAT can be easily used by the GTC Support Astronomers via the OSIRIS control software by selecting the appropriate mask position of the Cassette where it is inserted. MAAT does not require from GTC staff any daily nor night technical engineering support or specific telescope operation. 

Overall, the observation procedure with MAAT is simple. To ensure 
accurate acquisition and centering the telescope must be pointed to the center of the IFU pick-off mirror (see Figure~\ref{fig:maatopticslayout}). This position is accurately known with respect to both the OSIRIS and telescope optical-axis positions. The $1\arcsec$ r.m.s. pointing error of GTC guarantees that the target will be placed right at the very center of the IFU FoV. The same acquisition procedure applied for MEGARA and EMIR MOS can be applied to MAAT, which will guarantee a target positioning better than $0.3\arcsec$ (a spaxel). Note that the main advantage of MAAT is that a broad-band image of the entire $10\arcsec \times 14.2\arcsec$ field could be generated from the 3D data cube, which will confirm the correct target acquisition, but at the same time will guarantee observations, and spectra, of targets whose position is known within an accuracy of few arcsecs. This represents another advantage to use MAAT, in particular for transient astrophysics. Focusing optimisation should be done even if one starts from a previous OSIRIS long-slit focus, but with the advantage that the slicer gives much more information than a long-slit, thanks to its imaging capability, thus a better focusing is possible. The Instrument Calibration Module at GTC with the three different calibration lamps (HgAr, Ne, and Xe) will be used to obtain the arc lines for the selected OSIRIS Grisms and VPHs. Any other calibration frames such as dark, bias, and flat-fields will be also necessary. MAAT generates similar data rates than any other typical OSIRIS long-slit observing runs. The FITS header from OSIRIS+MAAT will not require any new field; it will suffice to give appropriate values to the standard fields (see Section~\ref{sec:maatpipe}).

The MAAT team  will provide an on-line user-manual with detailed observing and calibration procedures, and guidelines for the public available data reduction pipeline. A quick-look pipeline will also be provided to assist the Support Astronomers during the observing runs, which will include acquisition and focusing scripts.

Our observing requirements for the science programs described in Section~\ref{sec:science} requires both rapid response to transient events as well as standard observations.

\section{Remarks and acknowledgements}
\label{sec:maatmanag}

This MAAT@GTC White Paper  fulfills the goals of presenting the different aspects of the project and collecting outstanding science cases that justify its need and the interest of the GTC community. This is based on the workshop ``Panoptic spectroscopy of our universe with OSIRIS+MAAT at GTC" held on May 5, 2020, to encourage and support the development of this new facility. Slides and videos of the workshop presentations can be found at the workshop website\footnote{\href{http://case.iaa.es}{http://case.iaa.es}}. A MAAT@GTC Proposal has been already submitted to the GTC Office for its consideration.

The MAAT@GTC Collaboration consists of a group of scientists and engineers in Spain (Centro de Astrobiolog\'ia CSIC-INTA, Departamento de F\'isica Fundamental Universidad de Salamanca, Departamento de F\'isica Te\'orica y del Cosmos Universidad de Granada, Instituto de Astrof\'\i sica de Andaluc\'\i a CSIC, Instituto de Astrof\'\i sica de Canarias, Instituto de F\'isica de Cantabria CSIC-UC), Australia (Australian Astronomical Optics - Macquarie University), Denmark (DARK, Niels Bohr Institute, University of Copenhagen), Mexico (Instituto Nacional de Astrofísica, Óptica y Electrónica), and Sweden (The Oskar Klein Centre for Cosmoparticle Physics, Stockholm University) led by Francisco Prada as MAAT Principal Investigator. We shall strive to integrate in the MAAT team other scientists and institutions from the GTC user community interested on this project.

The work presented here would have been impossible without the technical support and help of the GTC staff Manuela Abril (Optical Scientist and Contact person for MAAT), Kilian Henr\'{\i}quez-Hern\'andez (Mechanical Engineer), Andreas Gerarts (Mechanical Engineer), Luis A. Rodr\'{\i}guez-Garc\'{\i}a (Head of Engineering), Antonio Cabrera (Head of Astronomy), and Romano Corradi (Director). They have provided a detailed 3D space envelope study for MAAT and all the relevant documentation on the optics and mechanics of OSIRIS. We are grateful to Manuela Abril for her very efficient and instantaneous feedback along the feasibility study after our many requests and questions. Furthermore, their effort included 3D printed relevant pieces to test the envelope of MAAT inside OSIRIS. Proper credit to their work has been included in the main text and relevant figures.

We also acknowledge the help of Ernesto S\'anchez-Blanco (OpticalDevelopment, Spain) who implemented all OSIRIS Grisms and VPHs in Zemax. We thank the company Winlight in France that provided very useful feedback on the IFU optical manufacturing process and mirror coatings, and are grateful to Rub\'en Garc\'ia Benito and Jos\'e Miguel Iba\~nez for very useful comments and suggestions. 

F.P. is deeply grateful to everyone who is contributing to the MAAT@GTC development for generous and constructive support. Special thanks go to Prof. Jes\'us Marco (CSIC Vice-president for Scientific and Technical research), Prof. Antxon Alberdi (IAA Director), and all signatories of Letters of Support attached to our MAAT@GTC Proposal.

Based on data obtained from the ESO Science Archive Facility under requests number EPEREZ\#498200, EPEREZ\#514213, and EPEREZ\#522882.

\clearpage

\bibliography{main}

\end{document}